\documentclass[letterpaper]{mn2e}
\usepackage{amssymb}
\usepackage{graphicx}
\usepackage{color}

\title{Astroinformatics of galaxies and quasars: a new general method for photometric redshifts estimation}

\author[O. Laurino et al.]{O. Laurino$^{1, 2}$\thanks{E-mail: olaurino@head.cfa.harvard.edu}, R. D'Abrusco$^{2}$, G. Longo$^{3,4}$ and 
G. Riccio$^{3}$\\
$^{1}$Astronomical Observatory of Trieste - INAF, Trieste, Italy\\
$^{2}$Harvard-Smithsonian Center for Astrophysics - Cambridge (MA), US\\
$^{3}$Department of Physical Sciences - University of Naples, Naples, Italy\\
$^{4}$Visiting associate, Department of Astronomy, California Institute of Technology, Pasadena, 90125 CA, USA}
\begin{document}

\date{Accepted 2011 July 10. Received 2011 June 17; in original form 2011 March 09}

\pagerange{\pageref{firstpage}--\pageref{lastpage}} \pubyear{2002}

\maketitle
\label{firstpage}
\begin{abstract}

With the availability of the huge amounts of data produced by current and future large multi-band photometric 
surveys, photometric redshifts have become a crucial tool for extragalactic 
astronomy and cosmology. In this paper we present a novel method, called Weak Gated Experts (WGE), 
which allows to derive photometric redshifts through a combination of data mining techniques. 
\noindent The WGE, like many other machine learning techniques, is based 
on the exploitation of a spectroscopic knowledge base composed by sources for which a spectroscopic 
value of the redshift is available. This method achieves a variance $\sigma^2(\Delta z)\!=\!2.3\!\cdot\!10^{-4}$
($\sigma^2(\Delta z)\! =\!0.08$, where $\Delta z = x_{\mathrm{phot}} - z_{\mathrm{spec}}$) for the reconstruction of the photometric
redshifts for the optical galaxies from the SDSS and for the optical quasars respectively, while the Root Mean Square (RMS)
of the $\Delta z$ variable distributions for the two experiments is respectively equal to 0.021 and 0.35. 
The WGE provides also a mechanism for the estimation of the accuracy of each photometric redshift. We also present and 
discuss the catalogs obtained for the optical SDSS galaxies, for the optical candidate quasars extracted from the 
DR7 SDSS photometric dataset\footnote{The sample of SDSS sources on which the accuracy of the reconstruction has been 
assessed is composed of bright sources, for a subset of which spectroscopic redshifts have been measured.}, 
and for optical SDSS candidate quasars observed by GALEX in the UV range. The WGE  
method exploits the new technological paradigm provided by the Virtual Observatory and the emerging field 
of Astroinformatics.
\end{abstract}


\section{Introduction}
\label{sec:introduction}

The ever growing amount of astronomical data provided by the new large scale digital surveys in 
a wide range of the EM spectrum has been challenging the way astronomers carry out their everyday 
analysis of astronomical sources. These new data sets, for their sheer size and complexity, have extended beyond
the human ability to visualize and 
correlate complex data, thus triggering the birth of the new technological approach and methodology which
is often labeled as ``astroinformatics", a new discipline which lies at the intersection of many others: data mining,
parallel and distributed computing, advanced visualization, web 2.0 technology, etc. \cite{borne2009,ball2010}. X-informatics (where
the X stands for any data rich discipline), is growingly being recognized as the fourth leg of scientific research after
experiment, theory and simulations (see \emph{The Fourth Paradigm}, \cite{hey2009}). In this paper we shall 
present a new method for the estimation of photometric redshifts which fully take advantage of many of these 
new methodologies.

\noindent In the past, for many tasks such as, for instance, classifying different types of sources, 
determining the redshifts of galaxies, etc. astronomers had to rely mainly on 
spectroscopic observations which are still very demanding in terms of precious telescope time. 

Even though spectroscopy is still fundamental to gain insights into many physical processes, 
the unprecedented abundance of accurate photometric observations for very large samples of 
sources, has led to the development of what we can call {\it candidates astronomy}, i.e. the 
branch of astronomy which exploits photometry to accomplish tasks which in the past would have 
required spectroscopic data. This discipline stems from a long and rich tradition of
astronomical techniques based on the use of photometric information in low dimensional 
\emph{features} space (for example, colour-colour selection techniques). The main differences
relative to these classical methodologies reside in the statistical techniques and the size of the 
dataset considered (in terms of both the number of members and the dimensionality of the datasets).
In those cases where a very accurate evaluation of the uncertainties affecting the estimate
is possible, the loss of accuracy and effectiveness which is implicit in candidates astronomy, 
is compensated by the possibility to obtain very extensive samples with limited effort.
In the last few years, candidates astronomy has found many applications, 
such as, for instance, the determination of the spatial 
distribution of visible matter on very large scales through photometric redshifts \cite{arnaltemur2009}. 
In such cases, the statistical tools used to characterize the description of the distribution 
of the sources are specifically designed to trade-off between the lower accuracy 
of the derived quantities (e.g. photometric redshifts with respect to spectroscopic ones) and the 
increased statistics arising from the significantly larger size of the samples of sources. Another example is the study of the distribution of 
quasars through the use of photometric redshifts and reliable catalogs of candidate quasars 
selected on the basis of their photometric properties rather than through spectroscopic confirmations. 
The advantages of candidates astronomy over traditional astronomy are obvious: for instance, 
in the latter case, the number of quasars selected via spectroscopic identification in the Sloan Digital Sky Survey 
(SDSS) Data Release 7 (DR7) is $\sim7.5\!\cdot\!10^4$ \cite{abazajian2009}, while the number of candidate quasars 
extracted from photometric data with effective methods involving the modeling of the distribution 
of sources in the color space is almost an order of magnitude larger, ranging from $\sim\!10^6$ found 
by \cite{richards2009} to the higher $\sim\!2.1\cdot\!10^6$ in \cite{dabrusco2009}, the latter figures 
being much closer to the theoretically predicted number of quasars expected to lie 
within the limiting flux of the SDSS, $\sim\!1.3\!\cdot\!10^6$ reported in \cite{richards2009}.

Photometric redshifts are important for a large spectrum of cosmological
applications, such as, to quote just a few: weak lensing studies of galaxy clusters \cite{abdalla2008}, the determination 
of the galaxy luminosity function \cite{subbarao1996}, studies of specific types of cosmic structures like, for instance, 
the photometric redshifts derived in \cite{dabrusco2007} which were used to investigate 
the physical reality of the so-called Shakhbazhian groups, to derive their physical characteristics 
as well as their relations with other galaxy structures of different compactness and richness \cite{capozzi2009}. 

Many different methods for the evaluation of  photometric redshifts are available in literature. 
Without entering into much detail, it is worth reminding that all methods are based on the 
interpolation of \emph{a priori} knowledge available for more or less large sets of templates and 
differ among themselves only in one or both of the following aspects: i) the way in which the 
\emph{a priori} Knowledge Base (KB, for a detailed definition of KB, see section \ref{sec:datamining}) is 
constructed (higher accuracy spectroscopic redshifts or empirically or theoretically derived 
Spectral Energy Distributions (hereafter SEDs), and ii) the interpolation algorithm or method employed. 
In this context, modern wide-field mixed surveys combining multi-band photometry and fiber-based 
spectroscopy and thus providing 
both photometric data for a very large number of objects and spectroscopic information for a 
smaller but still significant subsample of the same population, provide all the information needed 
to constrain the fit of an interpolating function mapping the space of the photometric features. 
Most if not all photometric redshifts methods have been tested on the Sloan Digital Sky Survey (SDSS)
which is a remarkable example of these ``mixed surveys", which has allowed noticeable advancements 
in the field of extragalactic astronomy and, over the years, has also become a sort of standard benchmark 
to evaluate performances and biases of different methods. Nonetheless, it should be noticed that 
the SDSS spectroscopic sample is not unbiased, since limited to a bright subset of galaxies and 
quasars observed in the optical range and selected according to spectroscopic methods. The peculiar 
characteristics of the source samples for which both photometric and spectroscopic measurements are available 
should always borne in mind when considering the effectiveness of the machine learning methods tested.

\noindent As it will become evident in section \ref{sec:wge}, one of the main problems 
encountered in evaluating photometric redshifts is the critical dependence of the final accuracy on the 
parameters needed to fine-tune the method and the nature of the sources (i.e., galaxies or quasars). 
For example, in the template fitting methods, part of the degeneracy between the spectroscopic redshift 
and colors of the sources can be minimized by a wise choice 
of the SED templates \cite{bruzual2010}, at the cost of introducing
biases in the final estimates of the photometric redshifts. In other data mining applications the same
degeneracy can be minimized by applying priors derived from the distribution of the spectroscopic redshifts 
for the sources belonging to the KB, like in \cite{dabrusco2007}. 

In what follows, we shall just summarize some aspects which appear to be relevant for the 
class of the interpolative methods. Such methods differ in the way the interpolation 
is performed, and the main source of uncertainty is the fact that the fitting function is just an 
approximation of a more complex and unknown relation (if any) existing between colors and 
the redshift (for example, see \cite{csabai2003}). Moreover, due to different observational 
effects and emission mechanisms, a single approximation can hold only in a given range of 
redshifts or in a limited region of the \emph{features} space \cite{dabrusco2007}. In the last few
years, in order to overcome the effects of the oversimplification of the relation between observables
and spectroscopic redshifts, several methods based on statistical techniques for pattern recognition 
aimed at the accurate reconstruction of the photometric redshifts for both galaxies and quasars have 
been developed (and in most cases applied to SDSS data): polynomial fitting 
\cite{connolly1995, budavari2001, li2008}, nearest neighbors 
\cite{csabai2003, ball2008, budavari2001}, 
neural networks \cite{firth2003,collister2004,vanzella2004,collister2007,dabrusco2007,yeche2010}, support vector 
machines \cite{wadadekar2005}, regression trees \cite{carliles2010}, Gaussian processes 
\cite{way2006,bonfield2010} and diffusions maps \cite{freeman2009}.

\noindent These methods, when applied to SDSS galaxies in the local universe 
(i.e. $z\!<\!0.5$), lead to similar results, with a dispersion  
$\mathbf{RMS(\Delta z)\!\sim\!0.02}$. 
The extension of these methods to the intermediate redshifts range ($z\!<\!0.8$) is in theory possible 
for both quasars and galaxies and for the brightest sources in the SDSS dataset,  
by adding near infrared photometry to the Sloan optical photometry and by using 
as KB the large SDSS spectroscopic sample, sometimes combined 
with redshifts measured in other deeper surveys like, for instance, the 2SLAQ \cite{croom2004}. 
Even though in this redshift range the estimated photometric redshifts seem not to be affected 
by any peculiar systematic effect, all these methods suffer from strong degeneracies in specific 
regions of the photometric \emph{features} space when applied to sources, like quasars, which can be 
found at larger redshifts and whose spectra typically present strong emission and absorption features, 
because of several different effects, often depending on the specific method used: reduced statistics, 
strong evolutionary effects or observational effects, like peculiar spectroscopic features being shifted 
off and in the photometric filters adopted in a specific instrument (like shown for SDSS quasars in \cite{ball2008}). 

\noindent Such degeneracies manifest themselves mainly through high local fractions of catastrophic outliers, i.e. 
sources with photometric redshifts estimates differing dramatically from the spectroscopic value. 
Some of the previously mentioned techniques address the problem of the catastrophic outliers by 
providing probabilistic estimates of the photometric redshifts \cite{ball2008}, at the cost of an increased
computational burden, which may lead to an overall worse scalability. The Weak Gated Experts method 
(hereafter WGE) described in this paper
has been designed to be accurate, relatively fast when compared to the 
other approaches available in literature, and easily scalable in order to allow the processing 
of very large throughputs (like those that will be produced by the large synoptic surveys of the 
future). As it will be discussed in what follows, the WGE method is general and 
comprehensive since it adapts to different types of sources without requiring a specific fine tuning.
The WGE is the second step of an automated machine learning method whose ultimate
goal is to ease the exploitation of ongoing and planned multi-band extragalactic surveys. 
While not completely removing the catastrophic outliers (task which is impossible to achieve 
due to the physical limitations, as mentioned above), WGE achieves a fair characterization of the regions of the photometric 
feature space where the degeneracies happen and is consistently able, as discussed in section 
\ref{sec:errors}, to flag the photometric redshifts values which most likely are catastrophic outliers.
It is worth noticing that in our analysis we never take into account possible time variability as it is done,
for instance, in \cite{salvato2009}.

\noindent The paper is structured as follows: in section \ref{sec:datamining} the general 
features and design principles of the WGE method are discussed. In section \ref{sec:wge}, 
more details of the specific implementation of the WGE used for the problem of photometric 
redshift reconstruction and of the algorithms employed are provided. The description of the 
datasets used for the experiments\footnote{Throughout this paper, the word experiment will 
refer to a complete run of the WGE method, as customary in the data mining jargon.} and the 
\emph{feature} selection criteria can be found in section \ref{sec:boks}, 
while the experiments for the determination of photometric redshifts for galaxies and quasars 
are described in section \ref{sec:experiments}. The final catalogs of photometric redshifts for the
SDSS galaxies and candidate quasars are presented in sections \ref{sec:catgal}, \ref{sec:catqua} and \ref{sec:catquauv} 
respectively, together with details on the distribution of the catalogs to the community. 
A thorough discussion of the performances of the method for the reconstruction of the 
photometric redshifts can be found in section \ref{sec:accuracy}, while the determination of the errors
on the photometric redshifts estimates and the discussion of the catastrophic outliers 
are described in section \ref{sec:errors}. The conclusion and a summary of the results can be found in 
section \ref{sec:conclusions}.

\section{The underlying data mining methodology}
\label{sec:datamining}

The Weak Gated Expert (WGE) method is a supervised data mining (DM) model which aims at the 
reconstruction of a quantity, namely the \emph{target} (in this case the redshift 
of the astronomical sources) through a local reconstruction of an empirical relation between 
the observed \emph{features} of a sample of sources having otherwise measured \emph{targets}
(in this case, the spectroscopic redshifts). 
In the implementation discussed here, the WGE consists of a combination of clustering and regression 
techniques. 
In order to better explain how the WGE works, some DM concepts and definitions will 
be given in the paragraphs \ref{subsec:super}, \ref{subsec:clustering} and \ref{subsec:regression} 
respectively.

\subsection{Supervised vs unsupervised}
\label{subsec:super}

In the domain of Machine Learning (hereafter ML) methods, the problem of the extraction of 
knowledge from data can take place following two approaches: supervised and unsupervised 
learning. From this general point of view, the ML process can or cannot be derived from a set 
of well known examples. In the case of supervised learning, an unknown mapping function (the model) between the 
\emph{features} of a sample of sources and the corresponding \emph{targets}, can be determined using an 
\emph{a priori} Knowledge Base (KB). This approach 
is useful when the relation is either unknown or too complex to be treated analytically, as it is often 
the case with astronomical datasets. The usage of supervised ML algorithms 
requires three basic steps: 

\begin{enumerate} 
\item Training: in this phase, the algorithm is \emph{trained} by examples extracted from the Knowledge 
Base (KB) to derive a model.
\item Test: the model is tested against a set of data extracted from the KB but not used for 
training. Results are used to evaluate both the degree of generalization and the overall error 
on the reconstruction of the \emph{target} values. 
\item Run: the model is used to predict the values of the \emph{targets} for new input patterns.
\end{enumerate}

\noindent Optionally, a Validation phase may be implemented in order to avoid over-fitting on the
training set. Validation works exactly like the Test phase, the difference being that the model is chosen 
according to the minimum validation error instead of using the training error.

\noindent However, the extraction of knowledge can also take place without using 
any \emph{a priori} targets, i.e. using only the statistical properties of 
their \emph{features} distribution. In this case, the approach is said to be \emph{unsupervised}, 
and the knowledge extraction process is driven from the statistical properties of the data 
themselves.
In practice, all these techniques are not driven by hypotheses, as it happens in more classical approaches, but are driven 
solely by the data. This means that, while allowing a large set of unprecedented analysis methods, 
the DM approach leads to its own hypotheses, which may be then validated through, for instance, 
subsequent analysis or additional observations (in the case of photometric redshifts estimation, 
for example, spectroscopic follow-up observations aimed at confirming the estimated values of 
$z_{\mathrm{phot}}$).

\subsection{Clustering}
\label{subsec:clustering}

The most representative example of unsupervised analysis is the clustering of a population 
of data points associated to objects and defined by the so-called \emph{features} vector, obtained by partitioning the dataset 
into an arbitrary number of subsets. Each subset consists of objects that can be considered 
\emph{close} to each others by some metric definition, and are \emph{far} from objects belonging to other clusters. 
As before, clustering may be said to be supervised when the final number of clusters is assumed 
\emph{a priori}, while unsupervised clustering applies in the case the algorithm itself 
determines the optimal number of clusters representing the spatial features of a dataset in 
the \emph{features} space. Different clustering algorithms tend to produce different sets
of possible clusterings, 
associating each clustering with statistical figures so that the best or more efficient clustering 
can be determined off-line.

\subsection{Regression}
\label{subsec:regression}

Regression is defined as the supervised search for the mapping from a domain in $R^{n}$ to 
$R^{m}$, with $m<n$; a regressor is thus a model that performs a mapping from a \emph{features} space 
X to a target space Y. In order to find this mapping function without any prior assumption on its explicit form, one 
can \emph{train} a supervised method, providing it with a set of examples. The problem can be formally stated 
as follows: given a set of training data (training set) $\{(x_1, y_1),\dots,(x_n, y_n)\}$ 
a regressor $h\!:\!X\!\rightarrow\!Y$ maps a \emph{predictor variable} $\mathbf{x}\!\in\!X$ to the response variable 
$\mathbf{y}\!\in\!Y$.

\section{The Weak Gated Experts method}
\label{sec:wge}

The WGE method is an example of how a combination of different data mining techniques can prove very 
effective at overcoming some of the degeneracies that can be present in high dimensional datasets 
which are typical of astronomical observations. As it was stated before, in a supervised method the 
first step is obtaining a predictor by training a model on a training set. Since the WGE is itself a supervised 
method, in order to obtain a predictor it has to be trained (a general definition for the concept of training, 
validation and test for supervised machine learning algorithms can be found in section \ref{subsec:super}). 
At a high level of abstraction, the training of the WGE method (see paragraph \ref{sec:wgeimplementation} 
for a description of the actual implementation of the method), can be summarized in three distinct steps:

\begin{itemize} 
\item Partitioning of the \emph{features} space. 
\item For each partition of the feature space, a model for a predictor is determined (an \emph{expert}). 
This predictor maps each pattern of the \emph{features} space to the target space. 
The outputs of the predictors associated to the various regions of the partition define
a new \emph{features} space.
\item A new \emph{gate} predictor is trained to map the patterns extracted from the new 
feature space to the target values. This new space is an extension of the original one 
with the addition of the experts predictions.
\end{itemize}

\noindent Different partitions of the \emph{features} space need to be tried in order to increase 
the accuracy of the redshift reconstruction and reduce the uncertainties. However, in this case, the 
results must be validated against a {\it validation set} in order to assess data over-fitting (i.e. a particular 
decomposition of the \emph{features} space may lead to an accidental improvement in the reconstruction 
which depends solely on the dataset used to train the predictors). The whole model is then tested 
against the test set to measure the level of generalization achieved and to characterize the errors.

\noindent The gate predictor combines the responses from the experts in order to find patterns 
in the responses themselves, taking into account the input features as well. In 
this way, the gate predictor can resolve part of the degeneracies and provide better results. 


\noindent  The implementation of the WGE method which has been used for this work 
uses Multi Layer Perceptron (MLP) neural networks as experts and will be described in the 
following paragraphs where arguments justifying its application 
to the determination of the photometric redshifts for quasars with optical wide band photometry will be 
provided. It will also be shown that the WGE can be used to improve the overall performance of the 
reconstruction of the photometric redshifts as well. 
In this regard, the WGE improves over some of the caveats of the method proposed in \cite{dabrusco2007}, 
by providing a more robust approach, a large improvement in the accuracy of the redshifts determination
according to most statistical diagnostics 
and a substantial refinement in the characterization of the uncertainty on the $z_{\mathrm{phot}}$ estimation 
and the determination of the outliers. In conclusion, the WGE is general and can be applied without 
any differences to the problem of the estimation of the photometric redshifts of all types of the extragalactic 
sources.
The training, validation and test sets for the three different experiments with the WGE method have been 
randomly drawn from the KB, composed respectively by the 60\%, 20\% and 20\% of the total number of KB 
members.

\subsection{MLP predictors}
\label{subsect:mlppredictors}

Feed-forward neural networks provide a general framework for representing non linear functional 
mappings between a set of input variables and a set of output variables \cite{bishop1996}. 
This goal can be achieved by representing the non-linear function of many variables as the 
composition of non-linear \emph{activation} functions of one variable. A Multi-Layer Perceptron 
(MLP) may be schematically represented by a graph: the input layer is made of a number of 
perceptrons equal to the number of input variables, while the output layer will have as many 
neurons as the output variables (targets). The network may have an arbitrary number of hidden layers which 
in turn may have an arbitrary number of perceptrons. In a fully connected feed-forward network each 
node of a layer is connected to all the nodes in the adjacent layers. Each connection is represented 
by an adaptive \emph{weight} representing the \emph{strength} of the synaptic connection between neurons. 
In general, along with the regular units, a feed-forward network presents a \emph{bias} 
parameter for each layer. The bias parameter of the $k$-th layer is added to the activation function input of all the 
nodes in the $k\!+\!1$-th layer. We consider a generic feed-forward network with $d$ input units, $c$ output units 
and $M$ hidden units in a single hidden layer. This kind of network can also be defined as a two-layer network, 
counting the number of connection layers instead of the number of perceptron layers. The output of the $j$-th 
hidden unit of the $k$-th hidden layer is first obtained by calculating the weighted sum of the inputs:

\begin{equation}
a_j^{(k)}=\sum_{i=0}^dw_{ji}^{(k)}z_i^{(k-1)}
\end{equation}
 
\noindent where $w_{ji}^{(k)}$ indicates the weight associated to the connection from the $k\!-\!1$-th layer
to the $j$-th node of the $k$-th layer, and $z_i$ is the activation state of the unit, the sum running from $0$ to 
$d$ and including the bias parameter in the $k\!-\!1$-th units as $w_{j0}^{(k-1)}\!=\!b_{k-1}$ with a constant 
activation state $z_0^{(k-1)}\!=\!1$. 
\noindent Then, the output of the $j$-th unit of the $k$-th layer is:

\begin{equation}
z_j^{(k)}=g\left(a_j^{(k)}\right)
\end{equation}
 
\noindent where $g()$ is the activation function. In general, different nodes may have different activation 
functions even in the same layer. Most of the times, two distinct activation functions are set for the hidden 
layers and the output layer respectively. The output is obtained by the combination of these functions 
through the network. For the $k$-th output unit:

\begin{equation}
y_k=\tilde{g}\left(\sum_{j=0}^Mw_{kj}^{(2)}g\left(\sum_{i=0}^dw_{ji}^{(1)}x_i\right)\right).
\end{equation}
 
\noindent If the output activation function is linear ($\tilde{g}(a)=a$), the network output 
reduces to:

\begin{equation}
y_k=\sum_{j=0}^Mw_{kj}^{(2)}g\left(\sum_{i=0}^dw_{ji}^{(1)}x_i\right).
\end{equation}

\noindent  One of the most common differentiable activation functions, that is usually used to 
represent smooth mappings between continuous variables, is the logistic sigmoid function, defined as:
$$
g(a) = \frac{1}{1+e^{-\alpha}}
$$
\noindent where $\alpha$ is called steepness. The application of the logistic function requires the 
\emph{features} to be normalized in the interval $[-1, 1]$. In what follows we shall refer to the topology of a MLP and to the weights matrix 
of its connections as to the \emph{model}. In order to find the model that best fits the data, 
it is necessary to provide the network with a set of examples, i.e. the training set extracted 
from the KB. One of the methods for the determination of such model depends on the 
minimization of a cost function. The Back-Propagation (BP) is a common algorithm for cost 
function minimization implemented, in its simplest form, as an iterative gradient descent of the 
cost function itself. An important role in BP is played by the \emph{learning rate}, which can be viewed 
as the ``aggressiveness" with which the algorithm updates the weights matrix in each iteration (or epoch). The BP halts 
when either an error threshold is hit or a maximum number of iterations is reached.

\subsection{Regression with MLP}
\label{subsec:regressionmlp}

Photometric redshifts estimation is a regression problem. Regression, as already reminded in paragraph \ref{subsec:regression}, 
is defined as the task of predicting the dependent variable $y \in \mathbf{R}^N$ from the input vector $\mathbf{x} \in \mathbf{R}^M$
consisting of $M$ random variables. The input data $\mathbf{\{x_k|k=1,2,...,K\}}$ may be assumed to be selected independently 
with a probability density $P(x)$. The outputs $\mathbf{\{y_k|k=1,2,...,K\}}$ are generated following the
standard signal-plus-noise model:

\begin{equation}
\mathbf{y_k = f(x_k) + \epsilon_{(k)}}
\end{equation}

\noindent where $\mathbf{\{\epsilon_k|k=1,2,...,K\}}$ are zero-mean random variables with probability
density $P_{\epsilon}(\epsilon)$. The learning procedure of a neural network aims at minimizing
a cost function, for example the Mean Square Error (MSE) defined as:

\begin{equation}
\mathbf{MSE=\frac{1}{K}\sum_{k=1}^{K}(y_k-f(\mathbf{x_k}))^2}
\end{equation}

\noindent In this way, the best regressor is represented by $E(y|x)=\int yP(y|x)dy=f(x)$, where $E$ stands for {\it expectation}. 
Unbiased neural networks asymptotically ($K\rightarrow\infty$) converge to the regressor. Uncertainties in the independent 
variable can be accounted for by assuming that it is not possible to sample any $x$ directly, and by instead sampling the 
random vector $\mathbf{z}\in\mathbf{R}^M$ defined as:

\begin{equation}
\mathbf{z_k=x_k+\delta_k}
\end{equation}

\noindent where $\mathbf{\delta_k}$ are the independent random vectors extracted from the probability
distribution $P_{\delta}(\delta)$.
A neural network trained with data $\mathbf{\{(z_k,y_k)|k=1,2,...,K\}}$ approximates the function:

\begin{eqnarray}
E(y|z) = \frac{1}{P(z)}\int yP(y|x)P(z|x)P(x)dydx = \\
= \frac{1}{P(z)}\int f(x)P_{\delta}(z-x)P(x)dx
\nonumber
\end{eqnarray}
\noindent This means that, in general, $E(y|z)\neq f(z)$. The equality holds only when there is no noise.
If noise is assumed to be gaussian, it can be shown (\cite{tresp1994}) that, in some cases,
$E(y|z)$ is the convolution of $f$ with the noise process $P_{\delta}(z\!-\!x)$.

At a very high level of abstraction and in the light of the details of the approach 
discussed in the previous paragraphs, the WGE is a regressor trained to reproduce as accurately as possible 
the unknown correlation between \emph{features} and \emph{targets}.
Moreover, as it will be shown, the implementation discussed here is based on MLP algorithm. 
In the training phase, the WGE learns how to map the \emph{features} space into 
the \emph{target} space (i.e., the photometric \emph{feature} space to the redshift space):

\begin{equation}
\mathrm{WGE}_{train}: \mathbf{p} \rightarrow z_{\mathrm{spec}}
\label{eq:wgetrain}
\end{equation}  

\noindent where $\mathbf{p}$ is the vector representing a position in the photometric 
\emph{feature} space and $z_{\mathrm{phot}}$ is the corresponding value of the photometric redshift. 
Once trained, the WGE is used to evaluate photometric redshifts:

\begin{equation}
z_{\mathrm{phot}} = \mathrm{WGE}(\mathbf{p})
\label{eq:wgerun}
\end{equation}

\subsection{The Gated Experts}
\label{subsec:gatedexperts}

In most cases involving the determination of photometric redshifts, there is not a continuous mapping 
function from the \emph{features} space to the \emph{target} space and, therefore, a single MLP 
cannot produce an accurate reconstruction of the color-redshift relation \cite{dabrusco2007}\footnote{Although 
a single global model can, at least in principle, approximate any function even if piecewise defined, in real 
world problems it is very difficult or impossible to extract such global model from the data. In these cases 
the error function is very complex and the back-propagation process is likely to end in a local minimum.}. 
Also, there is not a single global noise regime throughout the \emph{features} space. Degeneracies are 
an example of how the noise regime changes in different colors intervals. Moreover, the input and target 
noises depend also on the measured magnitude of the sources and, in turn, on their distance from the 
observers which is the information encoded in the redshift itself. Since the colors distribution of the sources depends on the 
distance, the noise will depend on the input as well. Finally, in the case of statistically under-sampled 
populations of sources like, for instance, high redshift quasars, the sparseness of the KB itself varies 
with the value of the colors, i.e. over the regions of the \emph{features} space where the KB is defined. 
The attempt to learn the mapping function on different regions of the input space with different noise 
levels and different densities using a single network, is likely to fail since the network can either extract 
features that do not generalize well in some regions (local over-fitting), or cannot fully exploit all the 
information potentially contained in other regions (local under-fitting). 

\noindent In other terms, since the cost function is unique for a single network, a local 
overfitting in some regions may be compensated (in terms of contribution to the overall error)
by a local underfitting in other regions. For this reasons, a more complex architecture, following 
the \emph{mixture of experts} paradigm \cite{jordan1994} turns out to be more effective.

\noindent The basic idea behind \emph{experts} is in fact to learn 
different local models from data residing in different 
regions of the \emph{feature} space. These \emph{experts} are specialized over their subdomain and their 
outputs are linearly combined to form the overall output of the method. 
\noindent The \emph{gated experts} are somehow different since they non-linearly combine non-linear \emph{experts}. 
The input space is also non-linearly split into subspaces and one gating network
\footnote{The gating network is, as a matter of fact, acted by a committee of neural networks. This approach is 
necessary in order to find the best bias-variance trade-off \cite{krogh1995}.} 
is trained to learn both the partitioning of the input space and the input dependent coefficients $g_i(\mathbf{x})$ 
that are then combined to yield the system outputs $y_i(\mathbf{x})$:

\begin{equation}
y=\sum_{i=1}^M g_i(\mathbf{x})y_i(\mathbf{x})     
\end{equation}
 
\noindent where $M$ is the number of experts. 
%
%
This problem cannot be addressed by means of supervised learning only because in general it is 
not possible to infer any \emph{a priori} knowledge about the best partitioning of the input space. 
For this reason, a complex cost function has to be derived to take into account all the variables. 
The method for deriving this cost function is known \cite{weigend1995}, but it is necessary to 
bear in mind a few cautions:

\begin{itemize}
\item the cost function cannot be minimized with gradient descent but the problem itself can be reformulated 
and addressed by means of an Expectation Maximization (EM) algorithm; 
\item in order to find a consistent solution, it is necessary to assume that one and only one expert 
is responsible for each pattern. In other terms, it is necessary to make sure that 
there is a way of isolating different \emph{sub-processes} throughout the \emph{features} space. 
As it will be shown in paragraph \ref{subsec:partitioning}, for the reconstruction of the photometric 
redshifts of quasars, this assumption is false due to degeneracies.
\end{itemize}

\section{Weak Gated Experts implementation}
\label{sec:wgeimplementation}

In the implementation of the WGE used for the experiments described in this paper, each \emph{expert} is a standard neural network 
that learns a function $y_i(\mathbf{x})$ by means of a sigmoidal activation function hidden layer 
and a linear activation function output layer, as discussed in section \ref{subsect:mlppredictors}. 
The gating network, instead, has a classification flavor since its $K$ nodes in the output layer 
have a \emph{softmax} activation function:

\begin{equation}
g_j=\frac{e^{s_j}}{\sum_{i=1}^Ke^{s_i}}  
\end{equation}
 
\noindent where $s_{i}(\mathbf{x})$ is the output of the $i$-th node in the hidden layer. The outputs of the 
gating network are normalized to unity and their values express the competition 
among different \emph{experts}, which is meant to be a \emph{soft} competition since each input pattern 
has a non-null probability of being in the domain of each \emph{expert} (see section \ref{subsec:partitioning}). 

\begin{figure*} 
   \centering
   \includegraphics[width=4in, height=6in]{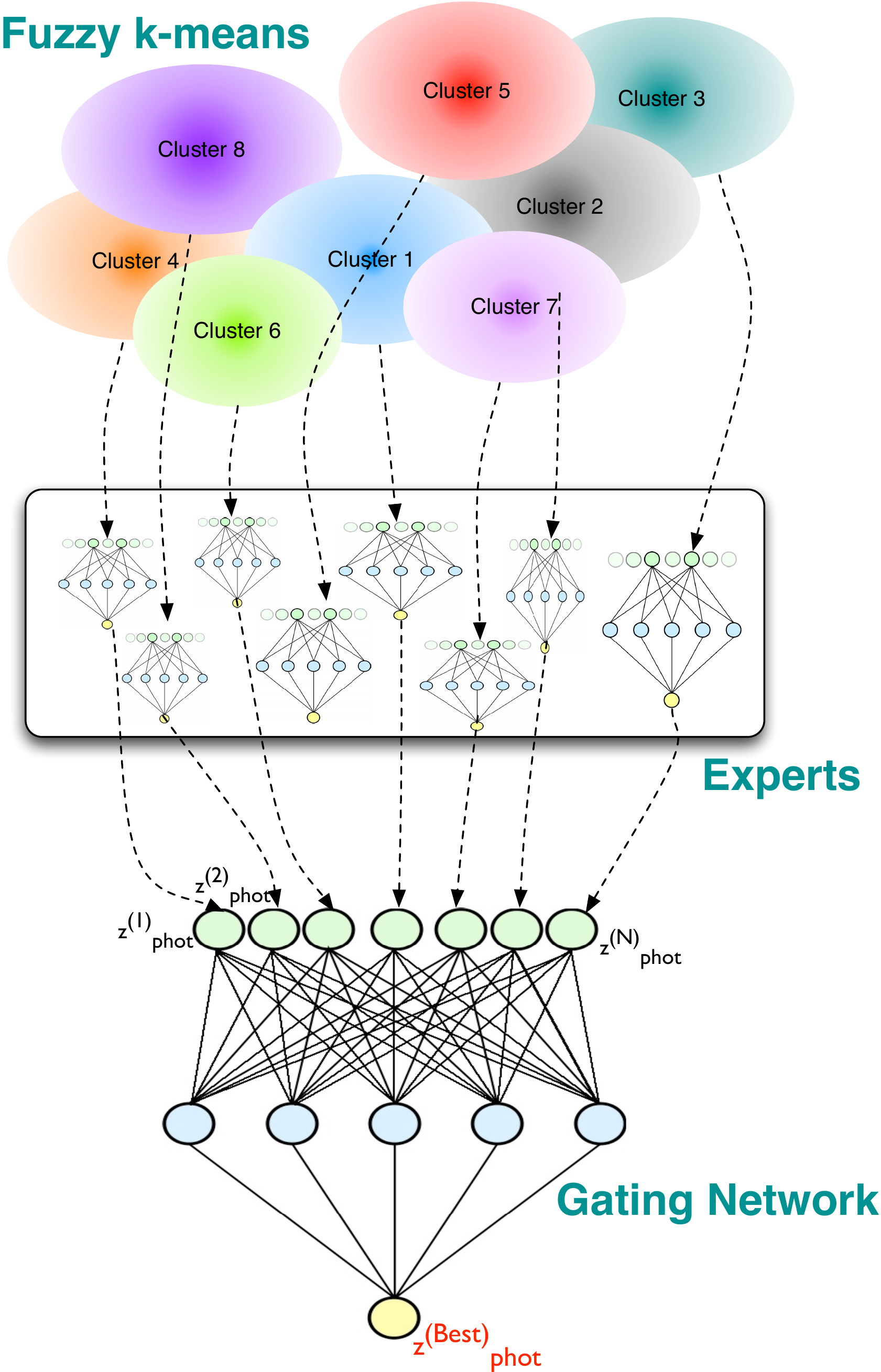} 
   \caption{Diagram of the implementation of the WGE method described in this paper. 
   		The ellipses associated to the clusters in the \emph{features} space have faded
		borders to stress the fuzzy nature of the clustering performed, while for the sake of 
		simplicity, only one gating network of the committee of experts is shown.}
   \label{plot:diagram}
\end{figure*}

\noindent The gated experts are combined through a non-linear superposition. This task, usually performed by an 
EM procedure together with the partition of the input space, in the WGE method is emulated by a ``weak" 
gating network, using a MLP network in a regression configuration and using the observed 
photometric \emph{features} and the outputs of the experts as \emph{features}. While trying to take advantage of the gated 
experts strengths, the WGE also takes into account the knowledge of the specific problem, from an 
astronomical point of view, as discussed in the following sections.
A diagram of the implementation of the WGE method used in the paper is shown in figure \ref{plot:diagram}.
In this plot, for the sake of simplicity, only one gating network is shown.

\subsection{Partitioning of the feature space}
\label{subsec:partitioning}

The \emph{gated experts} method requires an unsupervised approach to the partitioning of the input 
space. It is well known that the color distribution of extragalactic sources changes noticeably with 
the redshift, so that it is possible to determine distinct regions of the \emph{features} space 
where the color-redshift correlation follows different regimes.

\begin{figure*} 
   \centering
   \includegraphics[width=6in, height=6in]{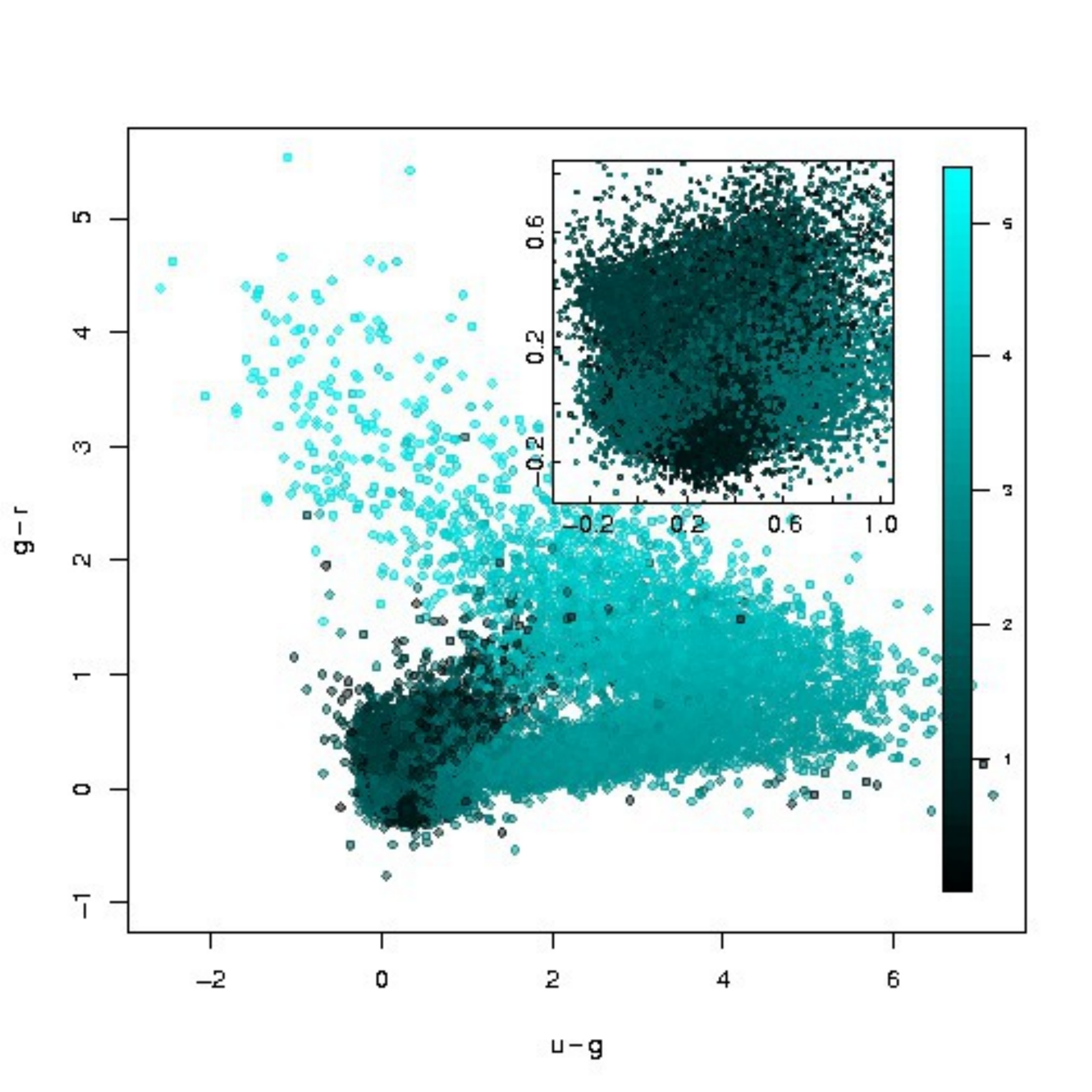} 
   \caption{Spectroscopically selected quasars in the SDSS DR7 dataset in the $u-g$ vs $g-r$ plot. The color 
   		of the symbols is associated to the spectroscopic redshift of the sources.}
   \label{plot:colvscol}
\end{figure*}

\noindent For instance, in figure \ref{plot:colvscol} it is shown the distribution of the sample of 
quasars observed spectroscopically by the SDSS in the DR7 in the $u\!-\!g$ vs $g\!-\!r$ color-color plot, 
where the color scale express the spectroscopic redshifts of the sources. Two main regions are clearly 
identified: a compact one where most of the objects lie, having redshifts in the interval 
$z_{\mathrm{spec}}\!=\![0, \sim\!2.5]$, and a vast region where the points are sparse and redshifts are larger 
than 2 with very few exceptions.  From an astrophysical standpoint, this can be explained with the 
fact that the Lyman break at redshift $\sim 3$ enters the optical SDSS $u$ filter, in turn yielding
larger values of the $u-g$ color.The inset zooms in the densest region of the plot, where most of 
the degeneracies arise. Although this plot shows a bi-dimensional projection of the 4-dimensional 
\emph{features} space (where it is possible that some of the degeneracies are resolved), this particular 
window is characterized by sources with similar colors and very different redshifts. These facts suggest 
that it is possible to divide the input space into different regions, two or more inside the window and 
one or more outside. Even if it is unknown \emph{a priori} whether the mapping function changes 
between these sub-domains, as it will be shown in paragraph \ref{subsec:features}, 
the error and noise regimes are different in such regions and, in particular, the densest ones are 
heavily affected by degeneracies while the others are mostly characterized by sparseness in the 
distribution of the points. In order to partition the input space, the implementation of the WGE method 
used for determination of the photometric redshifts employs a fuzzy version of a simple but effective clustering 
algorithm, namely the fuzzy $k$-means, or $c$-means \cite{dunn1973}. 
\noindent The classical $k$-means algorithm (hereafter ``sharp" k-means, opposed to the fuzzy counterpart), 
given the number of clusters $k$ and a metric definition, finds the centroids that minimize the distance with the 
objects belonging to their clusters while maximizing the distance among them by an iterative method. When convergence is reached, 
each point in the input space belongs to one and only one cluster. A different version of
the sharp $k$-means algorithm, namely the $c$-means, works 
exactly like its sharp counterpart for what finding cluster 
centroids is concerned, except that, in this case, each source belonging to the input sample
has a non-null probability of being a member of every cluster found by the algorithm,
even of very distant ones. In particular, each point $x$ belongs to the $k$-th cluster (identified with its centroid
$c_k$) with a membership degree $u_k(x)$ given by:

\begin{equation}
u_k(x)=\frac{1}{\sum_{j}\bigg[\frac{d(c_k, x)}{d(c_j, x)}\bigg]^{\frac{2}{(m-1)}}}  
\end{equation}

\noindent where $d(c_k, x)$ is the distance of the point $x$ from the $k$-th cluster and $m$ is 
a positive integer, which determines the normalization of the coefficients of the clustering. 
In this paper, the parameter has been fixed to $m=2$ so that the ``weights" associated to each
cluster are a linear function of the distance from the center of the cluster and the sum of the coefficients
is equal to 1. 
In practice, when partitioning the \emph{features} space, all the points with membership degree 
larger than an arbitrary threshold have been assigned to each cluster.
From a geometrical point of view, this allows to build clusters with soft boundaries, thus introducing
some redundancy in the datasets and translates, in the case of the determination of photometric 
redshifts, into the fact that the same pattern is allowed to belong to different clusters, so 
that part of the information contained in each pattern is shared the different experts trained on each 
of these clusters. For a discussion on the choice of the optimal set of \emph{features}, refer to paragraphs 
\ref{subsec:features} and \ref{sec:experiments}.
 
\subsection{The gating network}
\label{subsec:gating}

Although the WGE architecture addresses by itself the bias variance trade-off problem, a MLP used as a gated
network will introduce some variance and bias as well. This effect, mitigated by the WGE itself, is small but not
negligible.In order to address this problem, we modeled the gating network as a committee of $N$ identical MLPs trained on 
the same dataset. Each network will produce a slightly different result/ The final prediction is the average of all the predictions. 
The choice of the number of MLPs has been made by considering the bias and the variance of two randomly chosen 
distribution of photometric redshifts for each experiment and for several different numbers of trainings of the gating network.
The bias and variance for each couple of determinations of the photometric redshifts have been estimated 
using the mean and the standard deviation of the residual variable $\Delta z^{(phot)}$ between the two different 
determinations of the photometric redshifts:

\begin{equation}
\mathrm{bias}(z_{\mathrm{phot}}^{(1)}, z_{\mathrm{phot}}^{(2)}) =  <\!\Delta_{z}^{(\mathrm{phot})}\!> 
=  <\!(z_{\mathrm{photo}}^{(1)} - z_{\mathrm{photo}}^{(2)})\!>
\end{equation}

\begin{equation}
\mathrm{var}(z_{\mathrm{phot}}^{(1)}, z_{\mathrm{phot}}^{(2)}) = \sigma_{\Delta_{z}^{(\mathrm{phot})}} = 
\sigma_{(z_{\mathrm{phot}}^{(1)} - z_{\mathrm{phot}}^{(2)})}
\end{equation}

\begin{figure*}
   \centering
   \includegraphics[width=6in, height=6in]{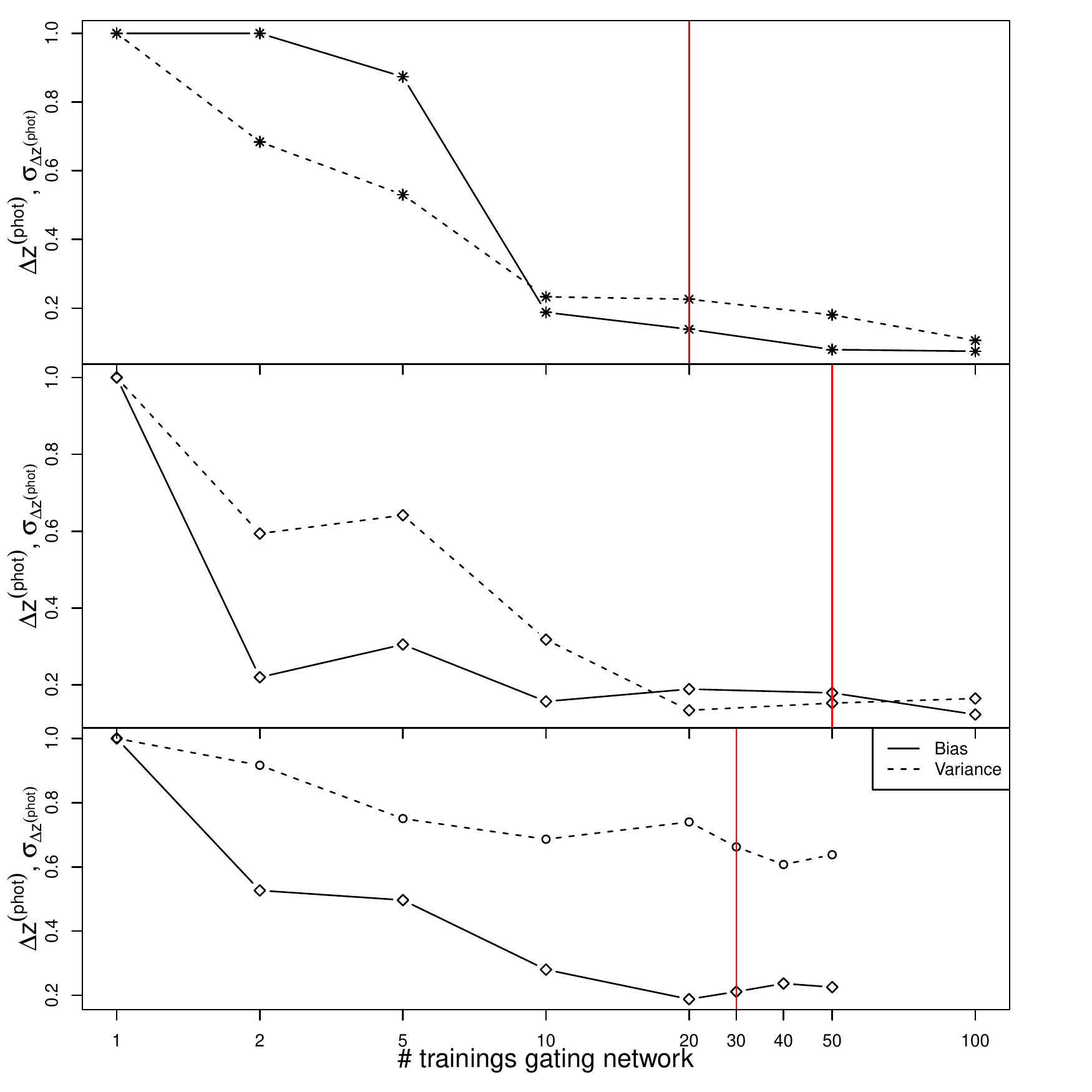} 
   \caption{Normalized bias and variances for two randomly produced realizations of the photometric redshifts
   distributions as a function of the number of the trainings of the gating network. The values of the this number
   used for the experiments and the production of the catalogs is indicated by red horizontal lines. From top to bottom, 
   optical quasars, optical+ultraviolet quasars and optical galaxies.}
   \label{plot:varbias}
\end{figure*}

\noindent These two variables (normalized to unity) are plotted against the number of trainings of
networks in the committee in plot \ref{plot:varbias}. The optimal numbers of networks for the three experiments has been chosen 
as those numbers for which for the variations of the bias and variance were lower than 5\% from the preceding 
realization, i.e. 30 gating network trainings for optical galaxies, 20 for optical quasars and 50 
gating networks for optical and ultraviolet quasars. The same procedure was used to determine the optimal
number of networks of the gating network for the determination of the errors on the photometric redshifts for each 
experiment. In this case, the threshold is reached is reached at 20 trainings for all experiments.

\section{The knowledge bases and features selection}
\label{sec:boks}

Three different KBs were employed during the training of the WGE method for the three classes 
of experiments performed, namely the evaluation of the photometric redshifts for:
\begin{itemize} 
\item optical galaxies with spectroscopic redshifts;
\item optical quasars with spectroscopic confirmation and redshift;
\item optical+ultraviolet quasars spectroscopically confirmed. 
\end{itemize}
\noindent The optical data for these three groups of experiments have all been extracted from the 
Sloan Digital Sky Survey (SDSS) DR7 database \cite{abazajian2009}.The confirmed spectroscopic 
quasars with both optical and ultraviolet photometry, used for the third class of experiments, have 
been retrieved from the dataset of crossmatched sources from the SDSS and GALEX surveys 
\cite{budavari2009}. A more detailed description of the three KBs can be found below:

\begin{itemize}
\item 1$^{\mathrm{st}}$ KB (optical galaxies). It includes all primary extended SDSS sources classified 
as galaxies according to the SDSS {\it specClass} classification flag (specClass == $\{2\}$), 
having clean measured photometry in all filters $(u, g, r, i, z)$, reliable spectroscopic redshifts 
estimates and brighter than the completeness limit of the SDSS spectroscopic survey, namely 19.7 
in the $r$ band. This sample, composed of $\sim 3.2\cdot 10^{5}$ sources, has been retrieved by querying the 
SDSS DR7 database for sources belonging to both $Galaxy$ and $SpecObjAll$ tables;
\item 2$^{\mathrm{nd}}$ KB (optical quasars): all spectroscopically confirmed SDSS quasars 
(specClass == $\{3, 4\}$), identified as point sources by any targeting program, with clean 
measured photometry in all filters $(u, g, r, i, z)$ and reliable spectroscopic redshifts 
estimates (this sample, composed of $\sim 7.5\!\cdot\!10^{4}$ sources, is a subset of the KB used 
for the extraction of candidate quasars described in \ref{subsec:candidates}). 
No specific cuts on the luminosity were performed. 
This sample has been retrieved by querying the SDSS DR7 database for sources belonging to the 
$SpecObjAll$ table;
\item 3$^{\mathrm{rd}}$ KB (optical+ultraviolet quasars): all spectroscopically confirmed optical SDSS quasars 
($\sim\!2.7\cdot\!10^{4}$ sources) associated to ultraviolet counterparts identified and observed by 
GALEX, with clean photometry in both optical $(u, g, r, i, z)$ and near and far ultraviolet bands 
$(nuv, fuv)$ and unambiguous positional cross-match (the sample of sources composing this KB 
is a proper subset of the second KB).
\end{itemize}
\noindent The queries used to extract the KBs from the SDSS and GALEX databases are reported 
in the appendix.

\subsection{Features selection}
\label{subsec:features}

\begin{figure*} 
   \centering
   \includegraphics[width=6in, height=6in]{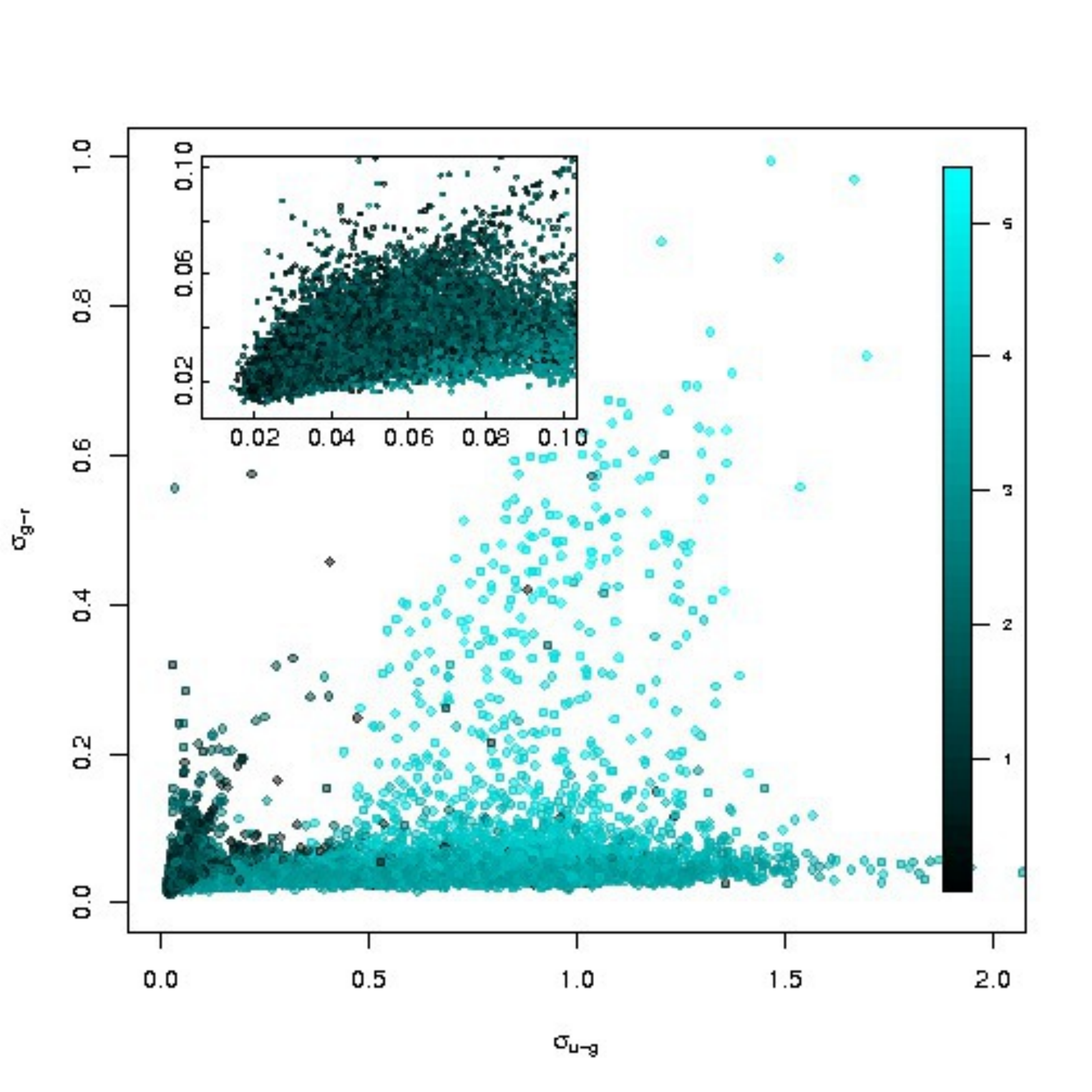} 
   \caption{Spectroscopically selected quasars in the SDSS DR7 dataset in the $\sigma_{u-g}$ vs $\sigma_{g-r}$ plot. 
   		The color of the symbols is associated to the spectroscopic redshift of the sources.}
   \label{plot:errcolvserrcol}
\end{figure*}

The selection process of the photometric \emph{features} used for the training of the WGE method 
(i.e. the \emph{features} of the experiment) was based on the assumption that most of the 
information needed to reconstruct the photometric redshifts of extragalactic sources is encoded 
in the observed magnitudes \cite{dabrusco2007}. However, since magnitudes are derived from 
fluxes, they tend to be correlated with each other and with the distance. Colors, instead, represent 
the ratio of fluxes measured in different filters and thus (once they have been corrected for extinction) they 
do not depend on the distance. Moreover, as it has already been discussed, the error regime changes 
with the redshift in the \emph{features} space defined by the colors, thus encoding some information 
on the redshift which can be used to partially remove the degeneracy in the unknown colors-redshift 
relation. In figure \ref{plot:errcolvserrcol}, the distribution of the same sample of quasars spectroscopically 
selected in the SDSS DR7 used in figure \ref{plot:colvscol}, is plotted in the plane generated by the 
errors on the colors $u\!-\!g$ and $g\!-\!r$ evaluated by propagating the uncertainty 
on the individual magnitudes. Even if the correlation between the error distribution and 
spectroscopic redshifts is not as clear 
as in the case of the color-color plot shown before, also in this case low redshift sources 
are almost completely contained in a window corresponding to errors generally smaller than $0.2$ 
in both colors (the inset of the plots zooms into the high density region located in the 
left bottom corner). The other points are instead distributed in an elongated feature corresponding to 
low and almost constant error on the $\sigma_{g\!-\!r}$ parameter and varying $\sigma_{u\!-\!g}$. 
Finally, only a small number of sources is spread all over the plot and has significantly higher 
redshifts.

\noindent In order to exploit all information contained in both the photometric \emph{features} and 
their uncertainties, the experiments discussed in this paper used the errors on the photometric 
colors to perform the clustering, and both colors and their uncertainties for the training of the 
experts. More combinations of \emph{features} and associated uncertainties were tested 
for each distinct experiment described here. All different combinations of \emph{features} produced 
less accurate reconstructions of the photometric redshifts. In particular, using as parameters of the 
clustering the colors only or the colors and their errors yielded, on average, 10\% larger MAD 
of the variable $\Delta z$ for all experiments.

\noindent For the first experiment involving the determination of the photometric redshifts for 
optical SDSS galaxies, the magnitudes used to derive the colors and their errors were the dereddened 
model magnitudes, i.e. the optimal estimates of the galaxy flux obtained by matching a spatial 
model to the source \cite{stoughton2002}. In this specific 
case, two different models are fitted to the two-dimensional images of each extended source in each 
band, namely a De Vaucouleurs profile and an exponential profile, and the best fitting model is used 
to calculate the model magnitude. The model magnitudes are then corrected for extinction according 
to the maps of galactic dust provided in \cite{schlegel1998}. For the samples of quasars used in the second 
an third experiments, the SDSS PSF magnitudes corrected for extinction were used to calculate optical 
colors and their uncertainties, while the remaining colors were calculated using the near and far ultraviolet 
magnitudes ($nuv$ and $fuv$ respectively) in the $PhotoObjAll$ table of the GALEX database 
\cite{budavari2009}, containing the photometric attributes measured for the sources detected in 
the GALEX imagery. 

\section{The experiments}
\label{sec:experiments}


For each KB a distinct class of experiments was performed by varying some of the parameters of the 
WGE method, and the ones yielding the best results for each of those classes, in terms of the accuracy of the 
reconstruction of the photometric redshifts (according to the statistical diagnostics used to characterize 
the accuracy of the reconstruction and discussed in section \ref{sec:accuracy}), are described in the 
next three paragraphs. The outputs of three the best experiments were also used to produce the catalogs
of photometric redshifts for SDSS galaxies and candidate quasars, described respectively in sections \ref{sec:catgal},
\ref{sec:catqua} and \ref{sec:catquauv}.  
In this section, the accuracy of the reconstruction of the photometric 
redshifts will be expressed by the robust estimates of the scattering 
of the variable $\Delta z \!=\!z_{\mathrm{phot}}\!-\!z_{\mathrm{spec}}$, evaluated through 
its median absolute deviation (hereafter MAD). Given a univariate set of variables
$\{\Delta z^{(1)}, \Delta z^{(2)},...,\Delta z^{(N)}\}$, the MAD of this sample is defined as:

\begin{equation}
\mathrm{MAD(\Delta z)} = \mathrm{median}(\Delta z - \mathrm{median}(\Delta z))
\label{eq:mad}
\end{equation} 

\noindent In other words, MAD is the median of the absolute deviation of the residuals from the median of the residuals itself. 
A modified version of the standard MAD statistics (hereafter MAD$'$) that can be used for the evaluation of the accuracy of 
the reconstruction of the photometric redshifts can be defined for the $\Delta z$ variable as follows:

\begin{equation}
\mathbf{\mathrm{MAD'(\mathrm{\Delta z})} = \mathrm{median}(\|\Delta z\|)}
\label{eq:mad2}
\end{equation} 

\noindent A summary of the \emph{features} used for the estimation of photometric redshifts and the errors on the photometric redshifts 
in these experiments are shown in tables \ref{table:experiments} and \ref{table:experimentserr} respectively, while the physical 
motivation behind the selection of the features used to train the WGE method has been given in the subsection \ref{subsec:features}. 
A more detailed characterization of the accuracy of the photometric redshifts reconstruction, obtained by means of distinct global and 
redshift-dependent statistical diagnostics, is discussed in paragraph \ref{sec:accuracy}. 

\noindent The criteria used for the choice of the best experiments for each class of experiments are the following, in order of 
decreasing priority:

\begin{itemize}
\item The total percentages of test-set sources with $\|\Delta z \| < 0.01$, $\|\Delta z \| < 0.02$ and $\|\Delta z \| < 0.03$ respectively
($<\!0.3$, $<\!0.2$ and $<\!0.1$ for the experiments involving quasars). These quantities, hereafter, will be referred to as $\Delta z_{1}$, 
$\Delta z_{2}$ and $\Delta z_{3}$ for galaxies and quasars as well; 
\item The value of the $MAD$ diagnostic of the $\Delta z$ variable as defined in equation \ref{eq:mad};
\item The value of the $MAD'$ diagnostic of the $\Delta z$ variable as defined in equation \ref{eq:mad2};
\end{itemize}

\begin{figure*} 
   \centering
   \includegraphics[width=6in, height=6in]{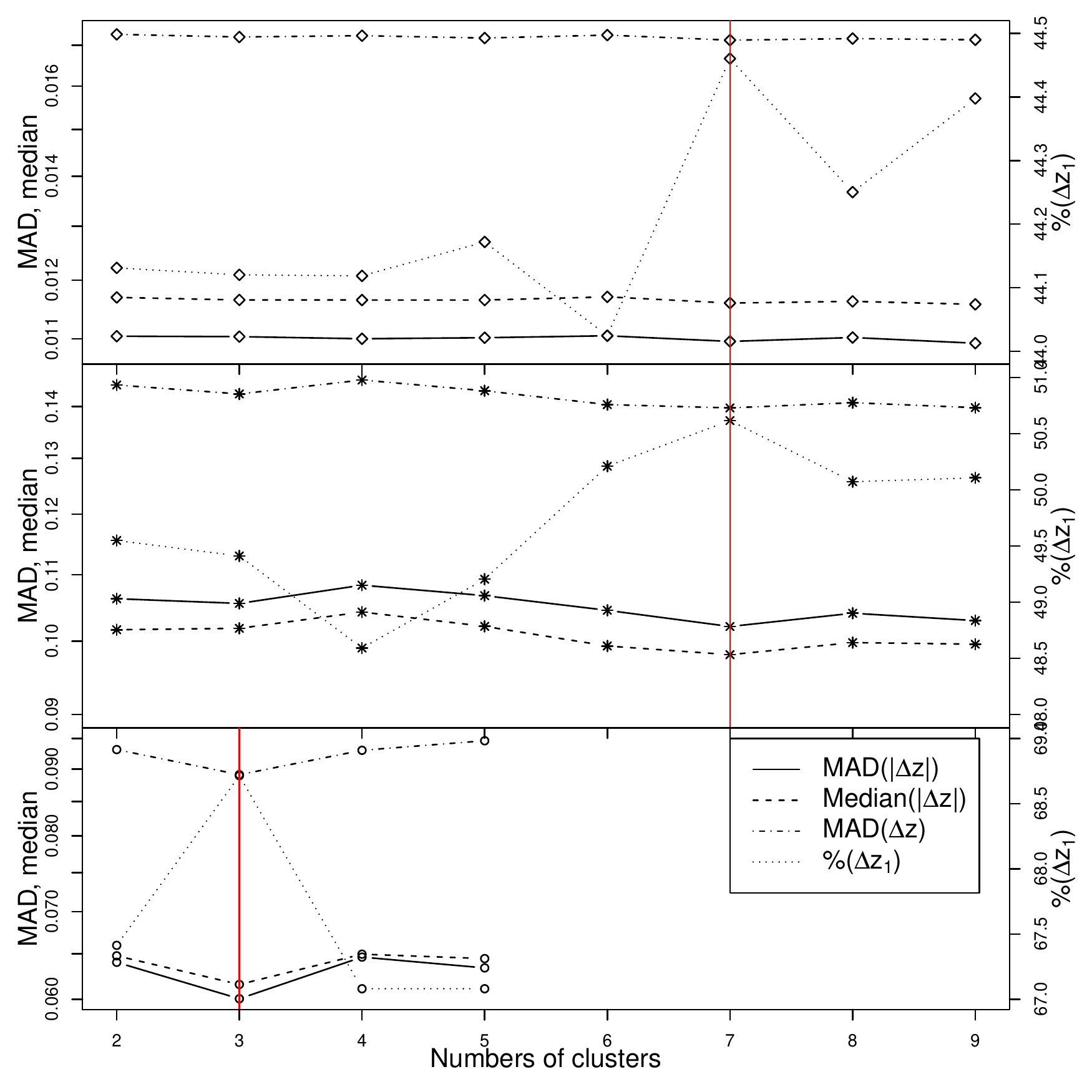} 
   \caption{Statistical diagnostics as a function of the number of fuzzy clusters for the three experiments described in this paper (from the
   		upper to lower panel, the experiments for the determination of the photometric redshifts of optical galaxies,  quasars with optical
		photometry and quasars with optical and ultraviolet photometry). The optimal number of clusters, as reported in table 
		\ref{table:experiments}, are marked with a red vertical line. In all cases, the optimal number of clusters are associated to the highest
		values of the $\%(\Delta z_{1})$ variable and to the lowest values of the three diagnostics MAD, MAD' and 
		${\mathrm median}(\|\Delta z\|)$.}
   \label{plot:diagnostics}
\end{figure*}

\noindent While the main criterion to select the best experiment is the first and the other two were used as tie-breakers in case of 
equal value of $\Delta z_{1}$, $\Delta z_{2}$ and $\Delta z_{3}$ (with a tolerance of 0.1\%), for all classes of experiments the best one
has been unambiguously selected by each of these criteria separately, as shown in figure \ref{plot:diagnostics}. In this plot, the values of the 
three diagnostics are shown respectively for all experiments of each class considered in this paper (optical galaxies, optical quasars and 
optical+ultraviolet quasars), as a function of the number of clusterings. 

\begin{table*}
\caption{Parameters of the best experiments for the evaluation of the photometric redshifts for optical galaxies, optical 
candidate quasars and optical plus ultraviolet candidate quasars.}             
\centering
{\small          
\begin{tabular}{l c c c}    
\hline\hline       
Parameters 			& Optical Galaxies  										&  Optical Quasars										& 	Optical+UV Quasars	 										\\ 
\hline
Params. clustering 		&	$\sigma_{u\!-\!g}$, $\sigma_{g\!-\!r}$, $\sigma_{r\!-\!i}$, $\sigma_{i\!-\!z}$ & $\sigma_{u\!-\!g}$, $\sigma_{g\!-\!r}$, $\sigma_{r\!-\!i}$, $\sigma_{i\!-\!z}$ 	&	$\sigma_{u\!-\!g}$,$\sigma_{g\!-\!r}$,$\sigma_{r\!-\!i}$,$\sigma_{i\!-\!z}$,			\\
					&													&													&	$\sigma_{fuv\!-\!nuv}$,$\sigma_{nuv\!-\!u}$							\\
Min. \# clusters			& 5													& 2													&	2														\\
Max. \# clusters			& 9													& 9													&	9														\\
Opt. \# clusters			& 7													& 7													&	3															\\	
Clusters threshold		&0.15												& 0.1													&	0.1														\\
Max. iterations clust.		&500												& 500												&	500														\\
Params. experts		&       $\sigma_{u\!-\!g}$,$\sigma_{g\!-\!r}$,$\sigma_{r\!-\!i}$,$\sigma_{i\!-\!z}$,	& $\sigma_{u\!-\!g}$,$\sigma_{g\!-\!r}$,$\sigma_{r\!-\!i}$,$\sigma_{i\!-\!z}$, 	&	$\sigma_{u\!-\!g}$,$\sigma_{g\!-\!r}$,$\sigma_{r\!-\!i}$,$\sigma_{i\!-\!z}$,			\\
					&	$(u\!-\!g)$,$(g\!-\!r)$,$(r\!-\!i)$,$(i\!-\!z)$							& $(u\!-\!g)$,$(g\!-\!r)$,$(r\!-\!i)$,$(i\!-\!z)$								&	$\sigma_{fuv\!-\!nuv}$,$\sigma_{nuv\!-\!u}$,							\\
					&													&													&	$(fuv\!-\!nuv)$,$(nuv\!-\!u)$,$(u\!-\!g)$,									\\	
					&													&													&	$(g\!-\!r)$,$(r\!-\!i)$,$(i\!-\!z)$											\\
Hid. neurons experts	& 30													& 20													&	20														\\
Max. epochs. experts	& 500												& 500												&	500														\\
Learning rate experts	& 0.01												& 0.01												&	0.01														\\
Steepness experts		& 1.0													& 1.0													&	1.0														\\
Hid. neurons gate		& 30													& 20													&	20														\\
Max. epochs. gate		& 500												& 500												&	500														\\
Learning rate gate		& 0.01												& 0.01												&	0.01														\\
Steepness gate		& 1.0													& 1.0													&	1.0														\\
\# training gates		& 30													& 20													&	50														\\
\hline                  
\label{table:experiments} 
\end{tabular}
}
\end{table*}

\noindent A first set of experiments were performed in order to set the steepness and the learning rate for all the 
experts in the whole features space. Once set, these values have not been treated as parameters of the WGE 
training but are considered fixed. Moreover, different values of the two parameters for the gating network have 
been explored, leading to a negligible variation in the final estimates of the photometric redshifts and associated
errors. For this reason, the values determined for the experts were used for all experiments.

\subsection{Photometric redshifts of galaxies with optical photometry}
\label{subsec:expgal}

The best experiment for the evaluation of the photometric redshifts of optical galaxies, retrieved 
from the SDSS photometric database, has been performed using the four SDSS colors and 
the corresponding errors (obtained by propagating the errors on the single magnitudes) as \emph{features} and 
the spectroscopic redshifts measured by the SDSS spectroscopic pipelines as target. The training of the WGE method, 
as described in detail in section \ref{sec:wge}, is obtained by first performing a clustering in the \emph{features} space 
and then training the single experts on each of the clusters, so that the final outcome of the method is evaluated 
by the gating network which combines the distinct outputs from the experts. For this experiment, the c-means 
clustering has been performed on the distribution of KB sources in the 4-dimensional \emph{features} space based 
on the uncertainties of the photometric colors $\sigma_{u\!-\!g}$, $\sigma_{g\!-\!r}$, $\sigma_{r\!-\!i}$ and $\sigma_{i\!-\!z}$, 
calculated by propagating the statistical uncertainties on the single magnitudes. The single experts have been 
trained on the different clusters determined by the fuzzy K-means algorithm in the 8-dimensional photometric \emph{features}
space obtained by adding the four colors $u\!-\!g$, $g\!-\!r$, $g\!-\!r$ and $i\!-\!z$ to their uncertainties $\sigma_{u\!-\!g}$, 
$\sigma_{g\!-\!r}$, $\sigma_{r\!-\!i}$ and $\sigma_{i\!-\!z}$. 

\begin{figure*} 
   \centering
   \includegraphics[width=6in, height=6in]{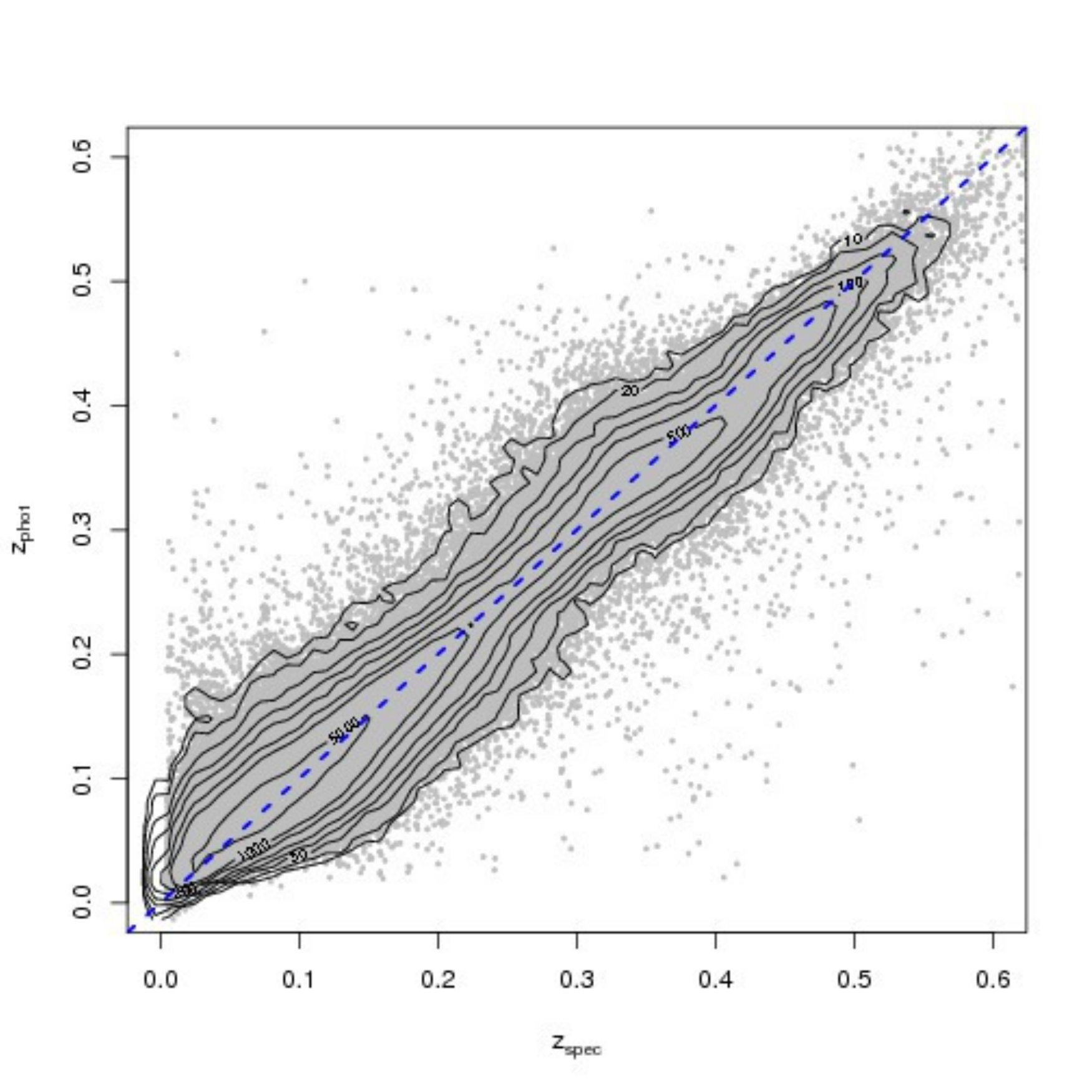} 
   \caption{Scatterplot of the spectroscopic redshifts vs photometric redshifts with isodensity contours for 
   		the sample of SDSS galaxies with optical photometry, belonging to the KB used to train the WGE in the first experiment.
		The isodensity contours are drawn for the following sequence of density values:
		$\{10, 20, 50, 100, 200, 500, 1000, 5000\}$.}
   \label{plot:zvsz_gal}
\end{figure*}

\noindent After multiple experiments performed with different values of the parameters of the WGE method,
the optimal value of the membership threshold on the fuzzy clustering has been fixed to $0.1$, so that each source has been 
considered member only of the clusters which accounts for at least $10\%$ of its total membership. 
The global MAD of the $\Delta z$ variable of this experiment is 0.017. The scatterplot showing the distribution of photometric redshifts against the 
corresponding spectroscopic redshifts for the members of the KB used for test the WGE method for the catalog of 
galaxies extracted from the SDSS DR7 database is shown in figure \ref{plot:zvsz_gal}. The histograms of the 
distributions of both photometric and spectroscopic redshifts for the test set 
of this experiment are shown in figure \ref{plot:zvsz_quasars_histo}.

\begin{figure*}
   \centering
   \includegraphics[width=6in, height=6in]{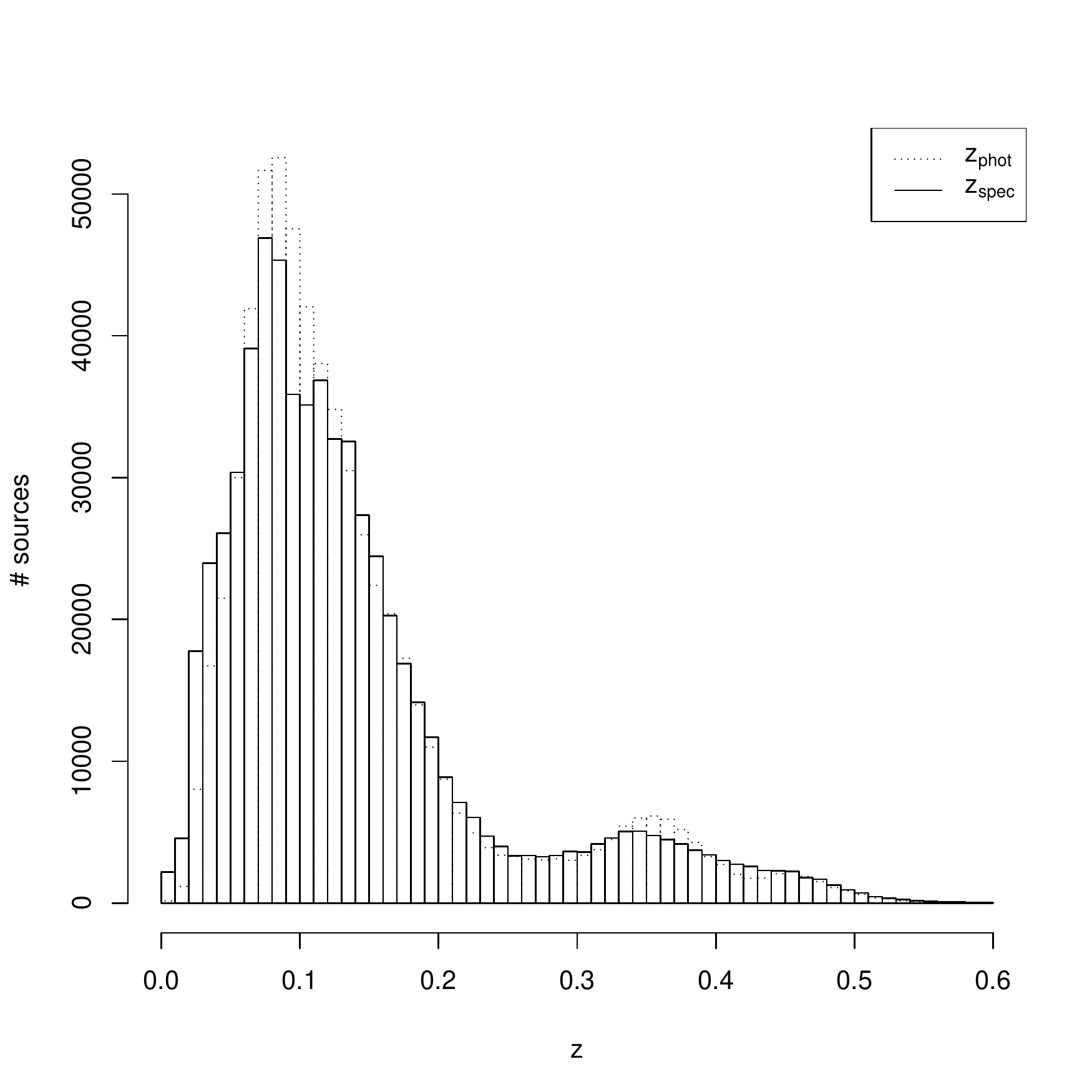} 
   \caption{Histograms of the distribution of spectroscopic and photometric redshifts for the sample of SDSS galaxies 
   		with optical photometry, belonging to the KB used to train the WGE in the first experiment.}
   \label{plot:zvsz_gal_histo}
\end{figure*}

\subsection{Photometric redshifts of quasars with optical photometry}
\label{subsec:expqua}

\begin{figure*}
   \centering
   \includegraphics[width=6in, height=6in]{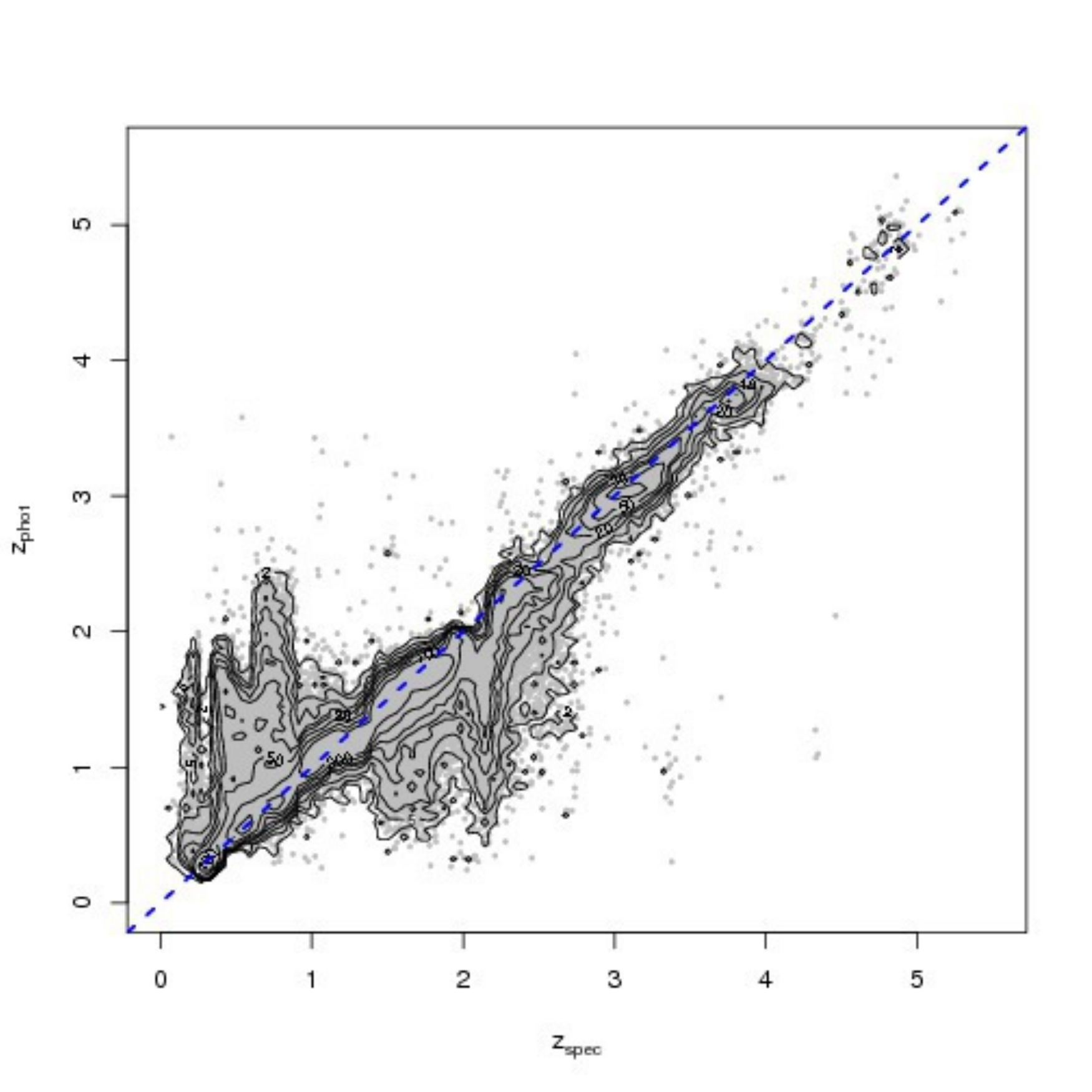} 
   \caption{Scatterplot of the spectroscopic redshifts vs photometric redshifts with isodensity contours for the sample of SDSS quasars 
   		with optical photometry, belonging to the KB used to train the WGE in the second experiment. The isodensity contours
		are drawn for the following sequence of density values: $\{2, 5, 10, 20, 30, 50, 100, 200\}$.}
   \label{plot:zvsz_quasars}
\end{figure*}

\begin{figure*}
   \centering
   \includegraphics[width=6in, height=6in]{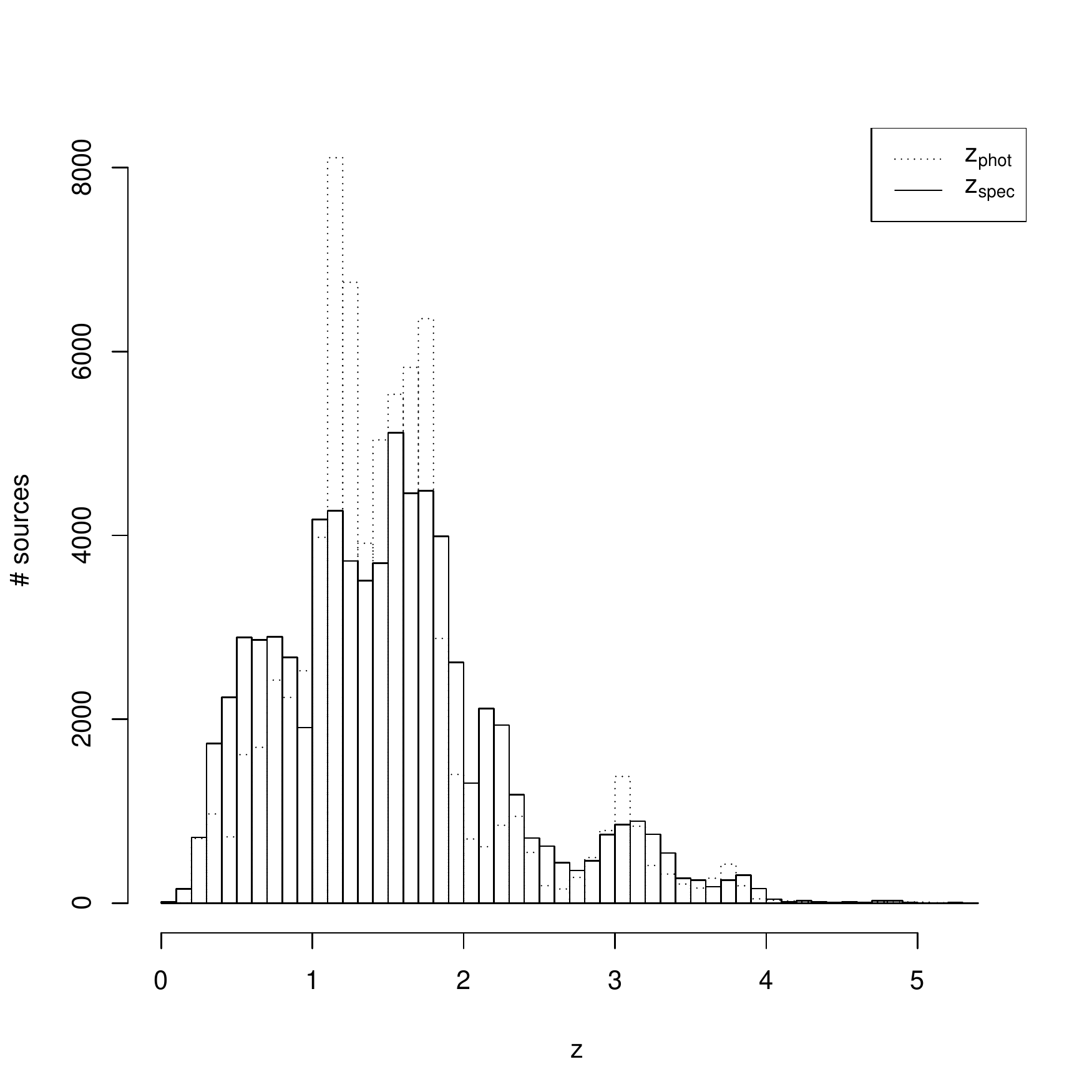} 
   \caption{Histograms of the distribution of spectroscopic and photometric redshifts for the sample of SDSS quasars 
   		with optical photometry, belonging to the KB used to train the WGE in the second experiment.}
   \label{plot:zvsz_quasars_histo}
\end{figure*}

The best experiment for the evaluation of the photometric redshifts of optical confirmed quasars 
extracted from the SDSS spectroscopic database made use of the four SDSS colors and associated 
uncertainties as features, and of the SDSS spectroscopic redshifts as targets. Similarly to what was 
described for the first experiment, the first step of the WGE training involved the determination of the 
optimal clustering of the KB sources in the 4-dimensional feature space consisting of the errors of 
the colors $\sigma_{u\!-\!g}$, $\sigma_{g\!-\!r}$, $\sigma_{r\!-\!i}$ and $\sigma_{i\!-\!z}$. 
On the other hand, the experts and the gating expert have been trained on the whole 8-dimensional 
feature space generated by the 4 optical colors and their uncertainties. After multiple
runs of the WGE method with different values of the parameters, the optimal value of the threshold on the fuzzy 
clustering has been fixed to 0.15. The clustering of the experiment for the determination of the errors on the 
photometric redshifts was carried out using, as features, the whole set of 8 photometric \emph{features} 
mentioned above in addition to the photometric redshifts $z_{\mathrm{phot}}$ and the variable $\Delta z$. 
The global MAD of the $\Delta z$ variable of this experiment is 0.14. The scatterplot of the distribution 
of photometric redshifts against the spectroscopic redshifts for the KB used to train the WGE method in 
this experiment is shown in figure \ref{plot:zvsz_quasars}, while the histograms of both spectroscopic 
and photometric redshifts distribution are shown in figure \ref{plot:zvsz_quasars_histo}.

\subsection{Photometric redshifts of quasars with optical and ultraviolet photometry}
\label{subsec:expquauv}

The most accurate reconstruction of the photometric redshifts for the quasars with SDSS optical 
and GALEX ultraviolet photometric data was achieved using, as \emph{features} for the clustering, 
the 6 uncertainties of the colors obtained by combining the 5 SDSS optical filters and the 2 
ultraviolet filters and by propagating the statistical errors on the magnitudes. 

\begin{figure*}
   \centering
   \includegraphics[width=6in, height=6in]{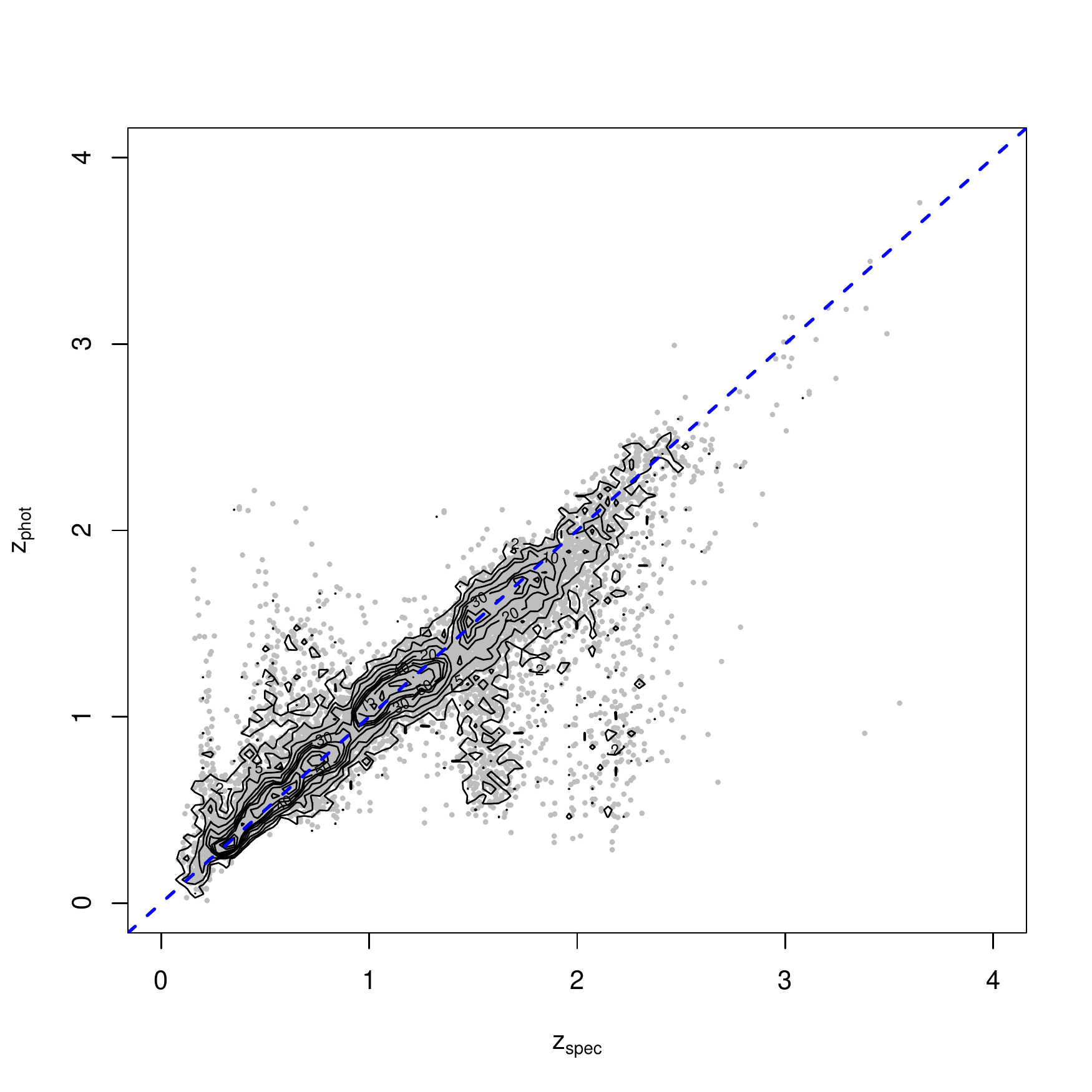} 
   \caption{Scatterplot of the spectroscopic redshifts vs photometric redshifts with isodensity contours for sample of quasars with 
   		optical and ultraviolet photometry, belonging to the KB used to train the WGE in the third experiment. The isodensity contours
		are drawn for the following sequence of density values: $\{2, 5, 10, 20, 30, 40, 50, 75, 100, 150, 200\}$.}
   \label{plot:zvsz_quasarsuv}
\end{figure*}

\begin{figure*}
   \centering
   \includegraphics[width=6in, height=6in]{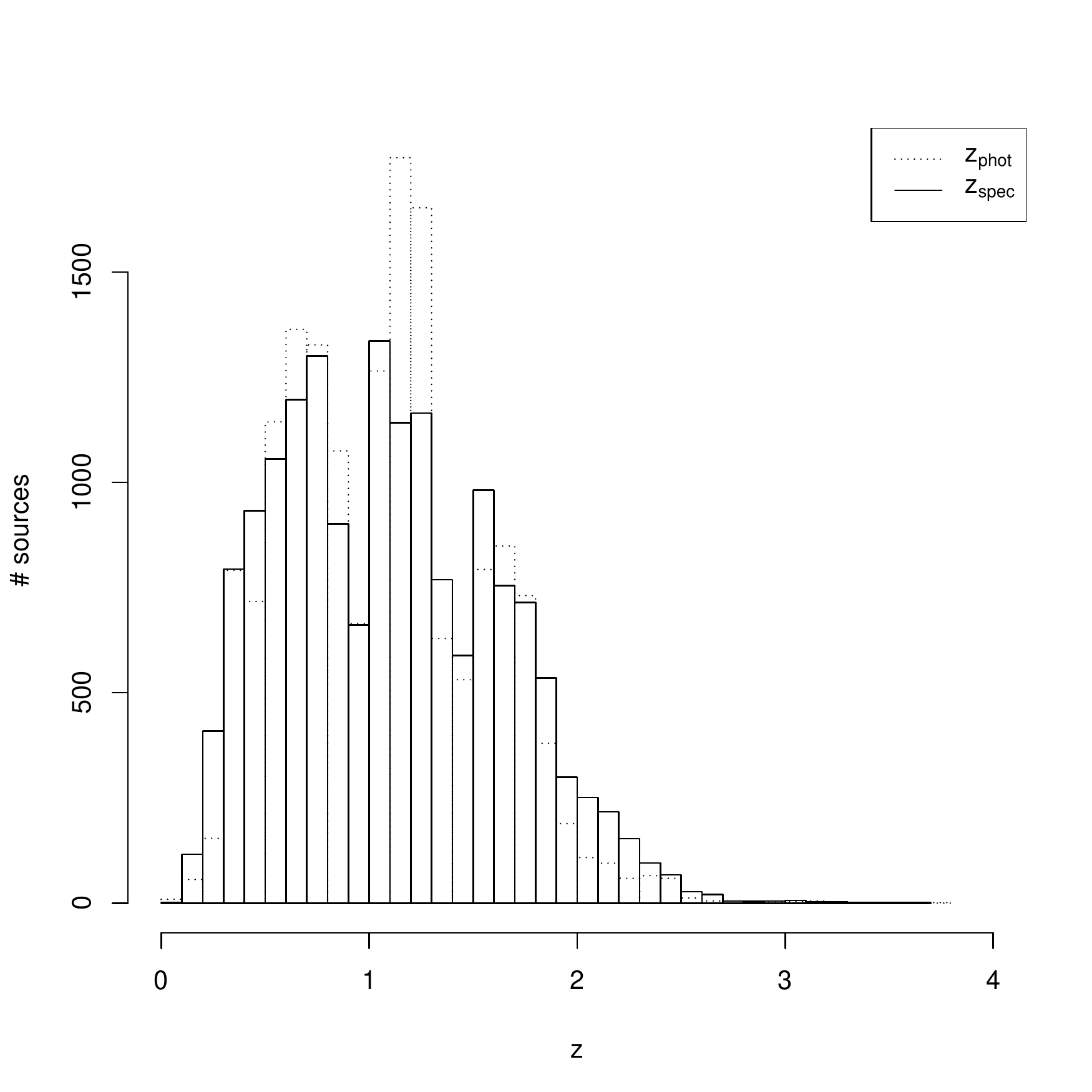} 
   \caption{Histograms of the distribution of spectroscopic and photometric redshifts for the sample of SDSS quasars 
   		with optical and ultraviolet photometry, belonging to the KB used to train the WGE in the third experiment.}
   \label{plot:zvsz_quasarsuv_histo}
\end{figure*}

The training of the experts and the gating expert was therefore carried out on the whole set of photometric 
\emph{features} available, i.e. the errors $\sigma_{u\!-\!g}$, $\sigma_{g\!-\!r}$, $\sigma_{r\!-\!i}$, $\sigma_{i\!-\!z}$, 
$\sigma_{fuv\!-\!nuv}$, $\sigma_{nuv\!-\!u}$ and the colors $(u\!-\!g)$,$(g\!-\!r)$,$(r\!-\!i)$,$(i\!-\!z)$,$(fuv\!-\!nuv)$,$(nuv\!-\!u)$. 
Also in this experiment, the clustering 
for the determination of the errors on the photometric redshifts was performed inside the feature space 
generated by the whole set of photometric \emph{features} used for the estimation of the photometric redshifts in 
addition to the photometric redshift $z_{\mathrm{phot}}$ itself and to the variable $\Delta z$. The MAD of 
the final $\Delta z$ variable in this experiment is 0.09, improving noticeably the accuracy of the photometric 
redshifts reconstruction obtained with the optical photometry only. As in the previous two experiments, the 
scatterplot of the distribution of photometric redshifts against the spectroscopic redshifts for the sources of 
the KB used to test the WGE method in this experiment is shown in figure \ref{plot:zvsz_quasarsuv}, while 
the histograms of both photometric and spectroscopic redshifts are shown in figure \ref{plot:zvsz_quasarsuv_histo}.

\section{The catalogs}

\subsection{The catalog of photometric redshifts for SDSS galaxies}
\label{sec:catgal}

A catalog of photometric redshifts for a sample of galaxies extracted from the SDSS-DR7 
database has been produced using the model obtained by training the WGE as described in section 
\ref{subsec:expgal}. The photometric galaxies were extracted, in a similar way to what done for 
the KB used for the training experiment, by querying the {\it Galaxy} table of the SDSS database 
for all primary extended sources with clean photometry in all filters $(u, g, r, i, z)$, and brighter 
than 21.0 in the $r$ band (the SQL query is shown in the appendix \ref{app:sqlgal}). 



\noindent In total, the catalog contains photometric redshifts for $\sim 3.2\cdot 10^{7}$ sources. The set of specific 
\emph{features} used for the evaluation of photometric redshifts, the estimated photometric redshifts values, errors and 
diagnostics flag together with 
some of the most common observational parameters retrieved directly from the SDSS database and useful 
for the identification of the sources in the SDSS database, have been included in the catalog for the sake of completeness. 
More information about the 24 columns of the catalog format are given in table \ref{table:catgal}. 
The photometric redshifts and uncertainties from our catalogs will also be incorporated into the NASA/IPAC
Extragalactic Database (NED) services.

\begin{table*}
\caption{Columns of the catalog of galaxies extracted from the SDSS with photometric redshifts 
evaluated using optical photometry.}             
\centering     
\begin{tabular}{l c c c}
\hline\hline       
\# & Name & Type & Description \\ 
\hline
1 & objID				& Long 			&	unique SDSS object ID				\\
2 & ra				&Double 			& 	right ascension in degrees (J2000)		\\
3 & dec				& Double	 		& 	declination in degrees (J2000)			\\
4 & dered\_u			& Float		 	&	SDSS dereddened $u$ model mag		\\
5 & dered\_g			& Float  			& 	SDSS dereddened $g$ model mag		\\
6 & dered\_r			& Float 			& 	SDSS dereddened $r$ model mag		\\
7 & dered\_i                         & Float 			& 	SDSS dereddened $i$ model mag		\\
8 & dered\_z			& Float 			& 	SDSS dereddened $z$ model mag		\\
9 & modelmagerr\_u		& Float 		 	&	SDSS $u$ model mag error			\\
10 & modelmagerr\_g	& Float 			& 	SDSS $g$ model mag error			\\
11 & modelmagerr\_r	&Float 	 		&	SDSS $r$ model mag error			\\
12 & modelmagerr\_i	& Float  			& 	SDSS $i$ model mag error			\\
13 & modelmagerr\_z	& Float 			& 	SDSS $z$ model mag error			\\
14 & extinction\_u		& Float 			&	SDSS $u$ mag extinction				\\
15 & extinction\_g		& Float 			& 	SDSS $g$ mag extinction				\\
16 & extinction\_r		& Float 			& 	SDSS $r$ mag extinction				\\
17 &extinction\_i		& Float 			& 	SDSS $i$ mag extinction				\\
18 & extinction\_z		& Float 			& 	SDSS $z$ mag extinction				\\
19 & u-g				& Double			& 	$u-g$ color						\\
20 & g-r				& Double 			& 	$g-r$ color						\\
21 & r-i				& Double 			& 	$r-i$ color							\\
22 & i-z				& Double 			& 	$i-z$ color						\\
23 & photoz			& Double  		& 	photometric redshift 					\\
24 & photoz\_err		&Double 			& 	photometric redshift error				\\
\hline                  
\label{table:catgal}  
\end{tabular}
\end{table*}

\subsubsection{Contamination of the catalog of photometric redshifts for SDSS galaxies}
\label{subsec:contamination}

\begin{figure*} 
   \centering
   \includegraphics[width=6in, height=6in]{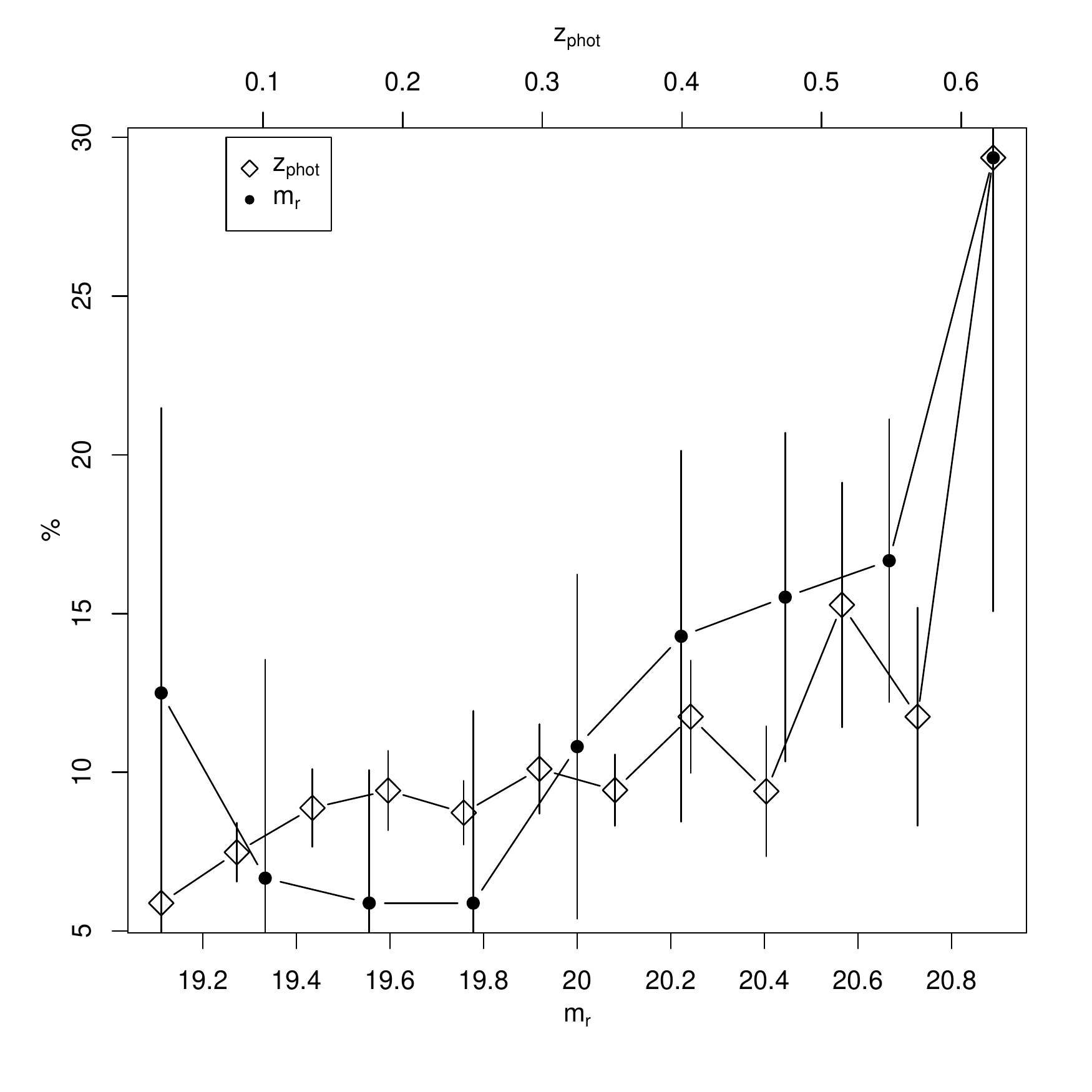} 
   \caption{Fraction of contaminants (galaxies with $z_{\mathrm{spec}} > 0.6$) in the catalog of SDSS DR7 photometric 
   		galaxies with photometric redshifts evaluated with the WGE method as function of the apparent magnitude in 
		the $r$ band (black symbols) and photometric redshifts (red symbols).}
   \label{plot:contamination}
\end{figure*}

The redshift distribution of the sources belonging to the KB used to train the WGE for the determination of the 
catalog of photometric redshifts for the galaxies extracted from the SDSS DR7, is shown in figure \ref{plot:zvsz_gal_histo}.
Even though no constraints on the redshift of the sources were explicitly required (as it is clear from the SQL 
version of the query in appendix \ref{app:sqlgal}), all galaxies belonging to this KB have spectroscopic 
redshift $z\!< 0.6$. A certain degree of contamination from galaxies at redshift $z\!> 0.6$ (and for this reason, 
not represented in the KB used for the WGE training) is expected in the catalog of photometric redshifts 
evaluated for the photometric galaxies extracted from the SDSS database. These galaxies could be mistakenly 
assigned a wrong value of their photometric redshift, in some case significantly lower than their real redshift.
The number and distribution of such galaxies, hereafter called contaminants, can be statistically evaluated either by 
using the luminosity function of the same galaxy population in the same band, similarly to what has been done in 
\cite{dabrusco2007}, or by employing a deeper catalog of galaxies with reliable measures of the redshifts. In the case 
of the catalog discussed in this section, the second method has been chosen to evaluate the contamination from high 
redshifts galaxies, using data from the DEEP2 survey \cite{davis2007}. DEEP2 is a 
spectroscopic survey that provides the most detailed census of the galaxy distribution at $z_{\mathrm{spec}}\!\sim\!1$, 
targeting $\sim\!5.0\!\cdot10^{5}$ galaxies in the redshift range $0\!<\!z\!<\!1.4$. The last data release (DR3) includes redshifts 
spanning four survey fields overlapping with the SDSS sky coverage. The SDSS galaxies with photometric redshifts estimated
with the WGE method have been positionally crossmatched with the catalog DEEP2 DR3 catalog of sources. The 
sample of cross-identified galaxies has been used to produce figure \ref{plot:contamination}, which shows the distribution of 
contaminants as functions of the apparent magnitude in the $r$ SDSS filter after correction for the extinction and the 
the photometric redshift $z_{\mathrm{phot}}$ of the galaxies. The fraction of contaminants is zero for $r$ magnitude smaller 
than 19 and is smaller than 20$\%$ for $r\!<\!20.5$. On the other hand, the fraction of contaminants as a function of the values
of the photometric redshifts assigned by the WGE method is consistently lower than 15$\%$ for $z_{\mathrm{phot}}\!<\!0.55$.
Uncertainties on the quantities plotted in the figure \ref{plot:contamination} have been evaluated applying 
poissonian statistics, and the large error bars for low magnitudes are caused by low statistics. 

\subsection{The catalog of photometric redshifts for SDSS optical candidate quasars}
\label{sec:catqua}

A catalog of photometric redshifts for the optical candidate quasars extracted from the SDSS-DR7 database 
is described in \cite{dabrusco2009}. The photometric redshifts for this sample of candidate quasars 
have been evaluated using the results of the WGE training experiment described in the section \ref{subsec:expqua}. 
The sample of point-like sources in the table {\it PhotoObjAll} of the SDSS-DR7 database from which the 
candidate quasars were extracted is composed of all the primary photometric stellar sources (using the SDSS 'type' flag, 
which provides a morphological classification of the sources by classifying them as extended or point-like) 
with clean photometry in all the filters $(u, g, r, i, z)$ and brighter than 21.3 in the $i$ band, for consistency with the 
sample of sources selected in \cite{richards2009}. 
\noindent The SQL query used to retrieve the data is given in the appendix \ref{app:sqlqsos}. The catalog retains 
the same basic structure of the catalog of photometric redshifts of galaxies, with few changes. 
This catalog contains $\sim 2.1\cdot 10^6$ candidate quasars, and consists of the list of candidate quasars 
with a small set of photometric \emph{features} used for the extraction process, with additional quantities derived by the 
method for the extraction of the candidates and the 
evaluation of photometric redshifts. Also in this case, some of the most common observational parameters available 
in the SDSS database were retrieved and added to the catalog to allow easier cross-matching with the original SDSS
database. More detailed information about the 31 columns of the catalog of photometric redshifts for the optical candidate 
quasars extracted from the SDSS-DR7 database are presented in table \ref{table:catqsos}. For this catalog a cone search service 
compliant with the VO standards will be made available as well.

\begin{table*}
\caption{Columns of the catalog of candidate quasars with photometric redshifts 
evaluated using optical photometry}             
\centering     
\begin{tabular}{l c c c}
\hline\hline       
\# & Name & Type & Description \\ 
\hline
1 & catjID				& Long 		& unique catalog object ID					\\
2 & objID				& Long 		& unique SDSS object ID						\\
3 & ra				& Double 		& right ascension in degrees (J2000)			\\
4 & dec				& Double 		& declination in degrees (J2000)				\\
5 & psfMag\_u			& Float 		& SDSS PSF $u$ model mag					\\
6 & psfMag\_g			& Float  		& SDSS PSF $g$ model mag					\\
7 & psfMag\_r			& Float 		& SDSS PSF $r$ model mag					\\
8 & psfMag\_i			& Float 		& SDSS PSF $i$ model mag					\\
9 & psfMag\_z			& Float 		& SDSS PSF $z$ model mag					\\
10 & psfmagerr\_u		& Float  		& SDSS $u$ PSF mag error					\\
11 & psfmagerr\_g		& Float 		& SDSS $g$ PSF mag error					\\
12 & psfmagerr\_r		& Float  		& SDSS $r$  PSF mag error					\\
13 & psfmagerr\_i		& Float  		& SDSS $i$  PSF mag error					\\
14 & psfmagerr\_z		& Float 		& SDSS $z$ PSF mag error					\\
15 & extinction\_u		& Float 		& SDSS $u$ mag extinction					\\
16 & extinction\_g		& Float 		& SDSS $g$ mag extinction					\\
17 & extinction\_r		& Float 		& SDSS $r$ mag extinction					\\
18 &extinction\_i		& Float 		& SDSS $i$ mag extinction					\\
19 & extinction\_z		& Float 		& SDSS $z$ mag extinction					\\
20 & strID 				& Long 		& SDSS stripe ID							\\
21 & u-g				& Double		& $u-g$ color								\\
22 & g-r				& Double 		& $g-r$ color								\\
23 & r-i				& Double 		& $r-i$ color								\\
24 & i-z				& Double 		& $i-z$ color								\\
25 & cluID 			& Integer 		& cluster ID								\\
26 & densKDEqsos 		& Double 		& KDE estimated p.d.f. relative to quasars distr.	\\
27 & densKDEnotqsos 	& Double 		& KDE estimated p.d.f. relative to not-quasars distr.	\\
28 & densKDEratio 		& Double 		& KDE estimated p.d.f. for quasars distr. to KDE 	\\
     & 					& 			& estimated p.d.f. for not quasars distr. ratio		\\
29 & photoz			& Double  	& photometric redshift (opt.+UV) 				\\
30 & photoz\_err		& Double 		& photometric redshift error					\\
31 & photoz\_flag		& Short  		& photometric redshift flag					\\
\hline                  
\label{table:catqsos}      
\end{tabular}
\end{table*}

\subsubsection{Candidate quasars}
\label{subsec:candidates}

The WGE method has been used to estimate photometric redshifts for the members of an updated version of the SDSS 
catalog of optical candidate quasars described in \cite{dabrusco2009}. While referring to the original work for a 
detailed description of the statistical method employed for the extraction of the candidate quasars, here we shall shortly summarize 
its basic facts in order to introduce some additional parameters included in the catalog. The method used to produce
the catalog of candidate quasars relies on the geometrical characterization of the distribution of spectroscopically 
confirmed quasars in the optical photometric features space and employs a combination of clustering techniques 
to achieve the best possible separation between regions of the \emph{features} space dominated by stars and 
quasars respectively. The method is based on the combination of different DM algorithms since it includes 
a dimensionality reduction phase obtained via Probabilistic Principle Surfaces (PPS) followed by a clustering performed 
using the Negative Entropy Clustering (NEC) respectively. The method allows to determine the salient correlations between the distribution of 
confirmed quasars in the photometric \emph{features} space and to use this information to extract new photometric candidate quasars.  
Given the original KB (a sample of point-like sources with spectroscopic classification), the extraction 
of the candidate quasars is performed by associating each photometric source to the closest cluster 
and retaining as candidates only those sources associated to clusters dominated by confirmed quasars. 
In the revised version of the catalog, the information provided for each candidate quasar has been completed 
by three parameters, namely the probabilities of each candidate quasar of being extracted from the underlying 
distributions of confirmed quasars or stars, and the ratio of these two probabilities. The first two values have been 
extracted from the probability density functions (pdf) associated to the two distinct distributions of stars and quasars, 
obtained by applying the Kernel Density Estimation (KDE) method. These parameters can be used to further refine 
the efficiency of the selection, at the cost of reducing the completeness of the sample. The catalog has been extracted 
from the DR7 SDSS database, thus yielding $\sim\!15\%$ more sources than the first version of the catalog. 

\subsection{The catalog of photometric redshifts for SDSS optical and ultraviolet candidate quasars}
\label{sec:catquauv}

A third catalog containing photometric redshifts estimates for a subsample of optical candidate quasars described 
in \ref{subsec:candidates} for which ultraviolet photometry from GALEX is available has been produced by using the 
results of the WGE training experiment described in the section \ref{subsec:expquauv}. The photometric redshifts
for quasars with both optical and ultraviolet photometry are significantly more accurate that those evaluated using
optical photometry only, and the fraction of catastrophic outliers is reduced as well (as will be described in detail in 
section \ref{sec:accuracy}). This catalog contains $\sim 1.6\cdot 10^5$ sources. The query used to retrieve the 
ultraviolet photometry of the sources with reliable GALEX counterparts is shown in appendix \ref{app:sqlqsosuv}.
The columns contained in the catalog are described in table \ref{table:catqsosuv}. Also in this case, the catalog 
will be available through a cone search service.

\begin{table*}
\caption{Columns of the catalog of candidate quasars with photometric redshifts 
evaluated using optical and ultraviolet photometry.}             
\centering     
\begin{tabular}{l c c c}
\hline\hline       
\# & Name & Type & Description \\ 
\hline
1 & catjID				& Long 		& unique catalog object ID					\\
2 & objIDsdss			& Long 		& unique SDSS object ID						\\
3 & objIDgal			& Long 		& unique GALEX object ID					\\
4 & ra				& Double 		& right ascension in degrees (J2000)			\\
5 & dec				& Double 		& declination in degrees (J2000)				\\
6 & nuv				& Float 		& GALEX $nuv$ mag						\\
7 & fuv				& Float 		& GALEX $fuv$ mag							\\
8 & psfMag\_u			& Float 		& SDSS PSF $u$ model mag					\\
9 & psfMag\_g			& Float  		& SDSS PSF $g$ model mag					\\
10 & psfMag\_r			& Float 		& SDSS PSF $r$ model mag					\\
11 & psfMag\_i			& Float 		& SDSS PSF $i$ model mag					\\
12 & psfMag\_z			& Float 		& SDSS PSF $z$ model mag					\\
13 & magerr\_nuv		& Float 		& GALEX $nuv$ mag error					\\
14 & magerr\_fuv		& Float 		& GALEX $fuv$ mag	 error					\\
15 & psfmagerr\_u		& Float  		& SDSS $u$ PSF mag error					\\
16 & psfmagerr\_g		& Float 		& SDSS $g$ PSF mag error					\\
17 & psfmagerr\_r		& Float  		& SDSS $r$  PSF mag error					\\
18 & psfmagerr\_i		& Float  		& SDSS $i$  PSF mag error					\\
19 & psfmagerr\_z		& Float 		& SDSS $z$ PSF mag error					\\
20 & extinction\_u		& Float 		& SDSS $u$ mag extinction					\\
21 & extinction\_g		& Float 		& SDSS $g$ mag extinction					\\
22 & extinction\_r		& Float 		& SDSS $r$ mag extinction					\\
23 &extinction\_i		& Float 		& SDSS $i$ mag extinction					\\
24 & extinction\_z		& Float 		& SDSS $z$ mag extinction					\\
25 & strID 				& Long 		& SDSS stripe ID							\\
26 & fuv-nuv			& Double		& $fuv-nuv$ color							\\
27 & nuv-u			& Double		& $nuv-u$ color							\\
28 & u-g				& Double		& $u-g$ color								\\
29 & u-g				& Double		& $u-g$ color								\\
30 & g-r				& Double 		& $g-r$ color								\\
31 & r-i				& Double 		& $r-i$ color								\\
32 & i-z				& Double 		& $i-z$ color								\\
33 & cluID 			& Integer 		& cluster ID								\\
34 & densKDEqsos 		& Double 		& KDE estimated p.d.f. relative to quasars distr.	\\
35 & densKDEnotqsos 	& Double 		& KDE estimated p.d.f. relative to not-quasars distr.	\\
36 & densKDEratio 		& Double 		& KDE estimated p.d.f. for quasars distr. to KDE 	\\
      & 				& 			& estimated p.d.f. for not quasars distr. ratio		\\
37 & photoz			& Double  	& photometric redshift (opt.+UV)				\\
38 & photoz\_err		& Double 		& photometric redshift error					\\
39 & photoz\_flag		& Short  		& photometric redshift flag					\\
\hline                  
\label{table:catqsosuv}      
\end{tabular}
\end{table*}

\section{Accuracy of the photometric redshift reconstruction}
\label{sec:accuracy}

Many different statistical diagnostics have been used in the literature to characterize the 
reconstruction of photometric redshifts as a function of the observational \emph{features} used to 
evaluate the quality of the redshifts. In this paragraph, a thorough statistical description of the 
performance of the WGE method will be given, in terms of the accuracy of the reconstruction, 
the biases of the reconstructed distribution of photometric redshifts and the fraction of outliers. 
A comparison of our results with others drawn from the literature is also provided in table 
\ref{table:diagnostics}, along with a comprehensive set of statistical diagnostics evaluated 
for the three different classes of experiments performed with the WGE method. All statistics have 
been calculated for the variables $\Delta z$ and $\Delta z_{\mathrm{norm}} = \frac{\Delta z}{1 + z_{\mathrm{spec}}} 
= \frac{z_{\mathrm{phot}}-z_{\mathrm{spec}}}{1 +  z_{\mathrm{spec}}}$. 
 
\noindent The statistical diagnostics evaluated for the results of the three experiments are the following:
\begin{itemize}
\item the averages $<\Delta z>$ and $<\Delta z_{\mathrm{norm}}>$ of both $\Delta z$ and 
$\Delta z_{\mathrm{norm}}$ variables, which accounts for the overall bias of the photometric redshifts distribution;
\item the Root Mean Square (RMS) of both variables $\Delta z$ and $\Delta z_{\mathrm{norm}}$, 
defined respectively as:
\begin{eqnarray}
RMS(\Delta z) = \sqrt{\sum(\Delta z)^{2}/N}\\
RMS(\Delta z_{\mathrm{norm}}) = \sqrt{\sum(\Delta z_{\mathrm{norm}})^{2}/N}
\end{eqnarray}
\noindent where N is the total number of values. The RMS accounts for the overall variation of the 
photometric redshifts distribution compared to the spectroscopic redshifts distribution;
\item the variances $\sigma^{2}(\Delta z)$ and $\sigma^{2}(\Delta z_{\mathrm{norm}})$ and the MAD 
of both $\Delta z$ and $\Delta z_{\mathrm{norm}}$ variables, accounting for the accuracy of the 
reconstruction measured as the spread of the two different variables; 
\item the values of the $MAD'$ for both $\Delta z$ and $\Delta z_{\mathrm{norm}}$ variables;
\item the percentage of sources with $\Delta z\!<\!\{\Delta z_{1} = 0.01, \Delta z_{2} = 0.02, \Delta z_{3} = 0.03\}$ and 
$\Delta z\!<\!\{\Delta z_{1} = 0.1, \Delta z_{2} = 0.2, \Delta z_{3} = 0.3\}$ for the 
experiments involving galaxy and quasars respectively (hereafter $\Delta z_1$, $\Delta z_2 $ and $\Delta z_3$ 
will be used for both galaxies and quasars, while $\Delta z_{\mathrm{norm},1}$, $\Delta z_{\mathrm{norm},2} $ and 
$\Delta z_{\mathrm{norm},3}$ will be used with the same meaning for the $\Delta z_{\mathrm{norm}}$ variable), which 
provide estimates of the performances of the reconstruction process at different levels of accuracy; 
\item the variance for the sources at $\Delta z_1$, $\Delta z_2 $ and $\Delta z_3$ ($\Delta z_{\mathrm{norm},1}$, 
$\Delta z_{\mathrm{norm},2} $ and $\Delta z_{\mathrm{norm},3}$), that represents an alternative measure of the 
performance of the reconstruction at three different levels of the accuracy;
\end{itemize}

\noindent In table \ref{table:diagnostics} we show the values of such diagnostics for the three experiments described in 
this paper and for a few other relevant papers in the literature that apply different methods to similar KBs and photometric datasets
(wide band photometry from ground based surveys in the optical and ultraviolet surveys). Namely, the results from 
\cite{ball2008, richards2009} for quasars with either optical or optical+ultraviolet photometry, and \cite{dabrusco2007} for optical 
galaxies are reported in the table. The WGE method noticeably improves over the accuracy achieved by \cite{dabrusco2007} in the 
reconstruction of the photometric redshifts for SDSS galaxies according to all the diagnostics, with only slightly smaller fractions 
of sources within $\Delta z_1$, $\Delta z_2 $ and $\Delta z_3$. 
In the case of the determination of the photometric redshifts for optical quasars, the kNN method used in \cite{ball2008} (column 
(2)) achieves a much larger variance for the $\Delta z$ variable while performing very similarly at the WGE method in terms of
$\Delta z_1$, $\Delta z_2$ and $\Delta z_3$, bias and variance of the distribution of $\Delta z_{\mathrm{norm}}$ variable. 
Similar results are achieved by the two methods also for the reconstruction of the photometric redshifts of quasars extracted 
from the SDSS with both optical and ultraviolet photometry, except for the fact that kNN achieves a much better variance 
for the distribution of the variable $\Delta z_{\mathrm{norm}}$. A different approach, not based on machine learning 
techniques, but similarly aimed at the determination of the empirical correlation between the colors and redshifts of 
the sources for the evaluation of the photometric redshifts is adopted in \cite{richards2009} (CZR method). Some of 
the diagnostics available for the application of this method
to SDSS quasars with both optical and optical+ultraviolet photometry show that such mok ethod achieves consistently lower accuracy
relative to both WGE and kNN methods (with the exception of the normalized variance for optical+UV experiment), while providing 
slightly larger fraction of sources within $\Delta z_1$, $\Delta z_2 $ and $\Delta z_3$ in the case of optical quasars. 

\begin{table*}
\caption{Statistical diagnostics of photometric redshifts reconstruction for all the experiments discussed in this paper
and for relevant papers in the literature. The first column (Exp. 1) contains the diagnostics for the experiment for the determination of the 
photometric redshifts of the optical galaxies from the SDSS catalog described in paragraph \ref{subsec:expgal}, while the columns 
(Exp. 2) and (Exp. 3) describe the diagnostics for the experiments concerning the determination of the photometric redshifts for 
quasars with optical and optical+ultraviolet photometry respectively (the details can be found in paragraphs \ref{subsec:expqua} and 
\ref{subsec:expquauv}). The same statistical diagnostics are shown for some papers from the literature, respectively
\protect\cite{dabrusco2007} for optical galaxies in column (1) and both \protect\cite{ball2008} and \protect\cite{richards2009} 
for optical and optical+ultraviolet quasars in the columns (2) and (3) respectively (as reported in \protect\cite{ball2008}). The 
definitions of the statistical diagnostics and other relevant} results of the literature are discussed in section \ref{sec:accuracy}.            
\centering     
\begin{tabular}{l c c c c c c c c}
\hline\hline       
Diagnostic 									& Exp. 1 				& (1)					& Exp. 2 				& (2)		& (3)		& Exp. 3				& (2)	 	& (3)		\\ 
\hline                                                                             
$\left\langle \Delta z \right\rangle$ 					& 0.015  				& 0.021				& 0.21				& -		& -		& 0.13 				& - 		& -     	\\
RMS$(\Delta z)$                                   				         	& 0.021  				& 0.074				& 0.35				& -		& -		& 0.25				& -		& -		\\
$\sigma^2(\Delta z)$ 							& $2.3\!\cdot\!10^{-4}$ 	& $5.0\!\cdot\!10^{-4}$	& 0.08 				& 0.123	& 0.27	& 0.044 				& 0.054	& 0.136	\\
MAD$(\Delta z)$ 								& 0.011   				& 0.012				& 0.11 				& - 		& - 		& 0.061 				& - 		& -		\\
MAD'$(\Delta z)$							& $\mathbf{0.012}$   	& - 					& $\mathbf{0.098}$	 	& - 		& - 		& $\mathbf{0.062}$	 	& - 		& -		\\
$\%(\Delta z_1)$ 								& 43.4    				& 41.1				& 50.7 				& 54.9	& 63.9	& 68.1 				& 70.8	& 74.9	\\
$\%(\Delta z_2)$ 								& 72.4   				& 68.4				& 72.3 				& 73.3	& 80.2	& 86.5 				& 85.8	& 86.9	\\
$\%(\Delta z_3)$		 						& 86.9   				& 83.4				& 80.5 				& 80.7	& 85.7	& 91.4 				& 90.8	& 91.0	\\
$\sigma^{2}(\Delta z_1)$			 				& $8.2\!\cdot\!10^{-6}$ 	& $8.2\!\cdot\!10^{-6}$	& $7.9\!\cdot\!10^{-4}$	& -		& -		& $7.6\!\cdot\!10^{-4}$	& - 		& -		\\	
$\sigma^{2}(\Delta z_2)$ 							& $3.0\!\cdot\!10^{-5}$ 	& $3.1\!\cdot\!10^{-5}$	& 0.003 				& -		& -		& 0.023 				& - 		& -		\\
$\sigma^{2}(\Delta z_3)$ 							& $6.1\!\cdot\!10^{-5}$ 	& $6.3\!\cdot\!10^{-5}$	& 0.005 				& -		& -		& 0.039 				& - 		& -		\\
$\left\langle \Delta z_{\mathrm{norm}} \right\rangle$ 		& 0.014 				& 0.017				& 0.095				& 0.095	& 0.115	& 0.058		 		& 0.06	& 0.071	\\	
RMS$(\Delta _{\mathrm{norm}})$                       			& 0.019 				& 0.037				& 0.19				& -		& -		& 0.11				& -		& - 		\\
$\sigma^2(\Delta z_{\mathrm{norm}})$ 				& $1.8\!\cdot\!10^{-4}$ 	& $1.1\!\cdot\!10^{-3}$	& 0.025				& 0.034	& 0.079	& 0.086				& 0.014	& 0.031	\\
MAD$(\Delta z_{\mathrm{norm}})$ 					& 0.009 				& 0.011				& 0.041				& -		& -		& 0.029				& -		& -		\\
MAD'$(\Delta z_{\mathrm{norm}})$ 				& $\mathbf{0.010}$  		& -					& $\mathbf{0.040}$	 	& - 		& - 		& $\mathbf{0.031}$	 	& - 		& -		\\
$\%(\Delta z_{\mathrm{norm},1})$ 					& 48.3   				& 45.6				& 77.3				& - 		& -		& 87.4				& -		& -		\\
$\%(\Delta z_{\mathrm{norm},2})$ 					& 77.2   				& 73.5				& 87.3				& - 		& -		& 94.0				& -		& -		\\
$\%(\Delta z_{\mathrm{norm},3})$ 					& 90.1   				& 87.0				& 91.8				& - 		& -		& 96.4				& -		& -		\\
$\sigma^{2}(\Delta z_{\mathrm{norm},1})$ 			& $8.3\!\cdot\!10^{-6}$ 	& $8.2\!\cdot\!10^{-6}$	& $6.2\!\cdot\!10^{-4}$	& -		& -		& $5.6\!\cdot\!10^{-4}$	& -		& -		\\
$\sigma^{2}(\Delta z_{\mathrm{norm},2})$ 			& $3\!\cdot\!10^{-5}$	   	& $3.0\!\cdot\!10^{-5}$	& 0.002				& -		& -		& 0.001				& -		& -		\\
$\sigma^{2}(\Delta z_{\mathrm{norm},2})$ 			& $5.8\!\cdot\!10^{-5}$ 	& $6.0\!\cdot\!10^{-5}$	& 0.004				& -		& -		& 0.002				& -		& -		\\
\hline
\label{table:diagnostics} 
\end{tabular}
\end{table*}

\begin{figure*}
   \centering
   \includegraphics[width=6in, height=6in]{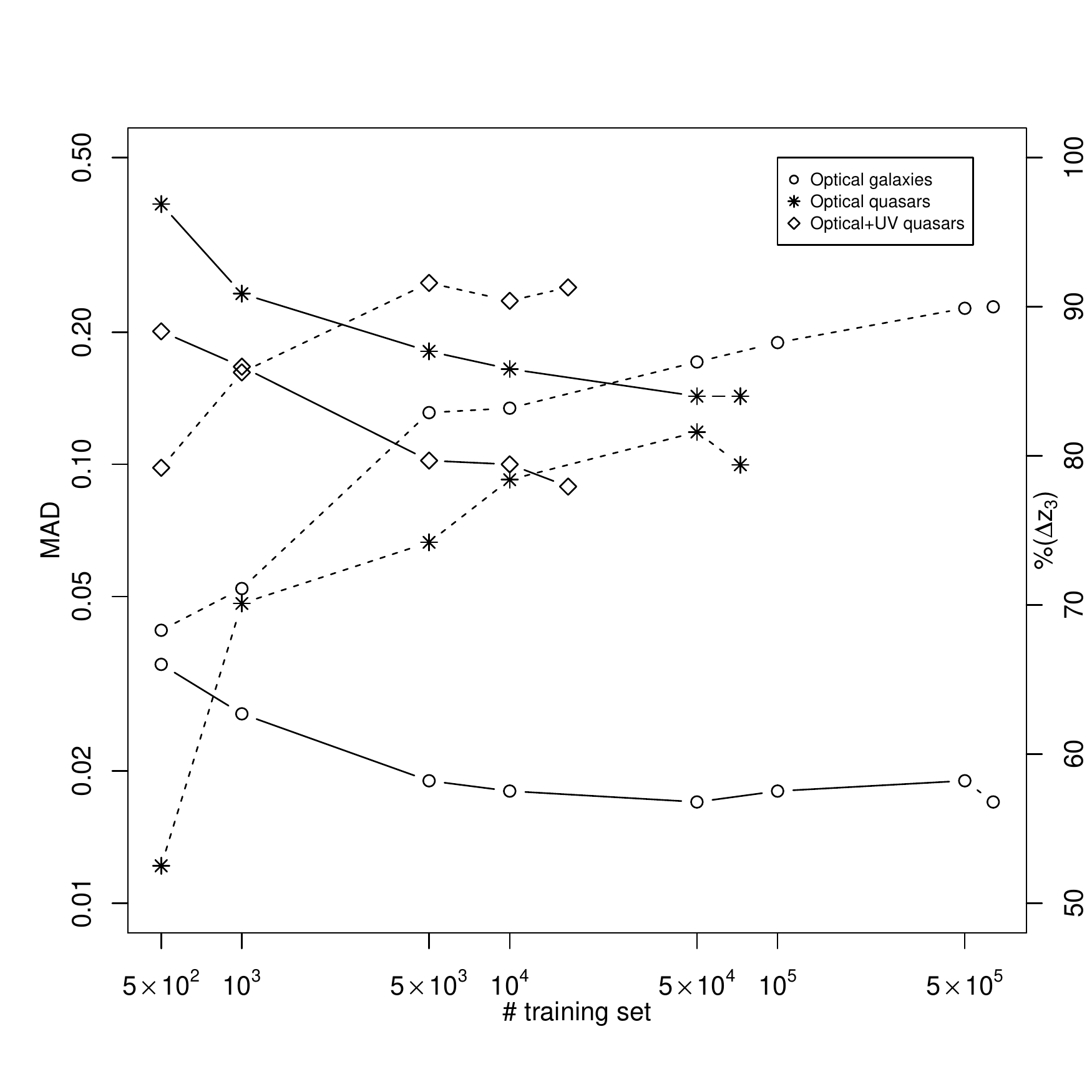} 
   \caption{Accuracy of the reconstruction of the photometric redshifts for the three experiments described in this paper as a 
	function of the number of sources composing the training set, randomly drawn from the whole KBs. In this plot are 
	shown the MAD of the $\Delta z$ variable (on the left y axis), and the percentage of sources with $\Delta z\!<\!0.3$ (or 
	$\Delta z\!<\!0.03$ for quasars) as the variable ($\%(\Delta z_{3})$ variable).}
   \label{plot:accuracy_training}
\end{figure*}

\noindent The accuracy of the reconstruction of the photometric redshifts depends on the number of sources belonging 
to the KB and on how well the KB samples the \emph{features} space defined by the photometric features. 
As a general statement, it is possible to state that the larger is the sample and the more homogeneous is the coverage of the 
\emph{features} space, the more accurate is the reconstruction  of the \emph{target} values. Plot \ref{plot:accuracy_training} 
shows the dependence of the robust sigma of the $\Delta z$ variable for all experiments discussed in this paper as a function 
of the number of sources of the KB. In more details, the plot \ref{plot:accuracy_training} shows (on the left y axis) the 
MAD of the $\Delta z$ variable and the percentage of sources of the KB with $\Delta z\!<\!\Delta z_{3}$ as functions of the 
number of sources of the training sets for the three experiments involving optical galaxies and quasars and optical+ultraviolet
quasars. The members of the training sets are extracted randomly from the whole KBs of the three experiments. The WGE 
method has been trained on such randomly drawn subsample of the original KBs in order to minimize the effects of all 
the other possible sources of variance. 
Both diagnostics of the performance of the WGE method considered show a common behavior, reaching a plateau after 
some characteristic threshold which apparently depends on the number of \emph{features} and the complexity of the experiment. 
The $\Delta z_{3}$ variable shows a steep increase at low cardinalities for all 
experiments, while the accuracy of the reconstruction appears to improve much more slowly with the number of sources in the 
training set.

\noindent The data used to create the plot in figure \ref{plot:accuracy_training} are presented in table \ref{table:accuracy_training}. 

\begin{table*}
\caption{Accuracy of the reconstruction of the photometric redshifts for the three experiments described in this paper as a 
function of the number of sources composing the KBs. Robust estimates of the robust standard deviation of the $\Delta z$ variable, 
obtained with the MAD algorithm are provided together with the percentages of sources with $\Delta z < 0.3$ and $\Delta z < 0.03$
for the experiments involving the quasars and the galaxy respectively.}             
\centering     
\begin{tabular}{c c c c c c c}
\hline\hline       
				& \multicolumn{3}{|c|}{$\sigma_{\mathrm{rob}}$}	&  \multicolumn{3}{|c|}{$\%(\Delta z_{3})$}			\\
$\#$ sources KB 	&	 Exp. 1 	& 	Exp. 2 	& 		Exp. 3	& 	Exp. 1 	& 	Exp. 2 	& Exp. 3				\\ 
\hline
$5\!\cdot\!10^{2}$	& 	0.035 	& 	0.392	&	0.201 		& 	68.3	 	& 	60.3	 	& 79.2				\\ 
$10^{3}$	 		& 	0.027	& 	0.245	& 	0.167		& 	71.1	 	& 	70.1	 	& 85.6				\\ 			
$5\!\cdot\!10^{3}$	& 	0.019	& 	0.181	& 	0.102		& 	82.9 		& 	74.2	 	& 91.6				\\ 
$10^{4}$	 		& 	0.018	& 	0.165	& 	0.100		& 	83.2 		& 	78.4	 	& 90.4				\\ 
$\!5\cdot\!10^{4}$	& 	0.017	& 	0.143	& 	-			& 	86.3 		& 	81.6	 	& -					\\ 			
$10^{5}$	 		& 	0.018	& 	-	 	& 	-			& 	87.6 		& 	-	 	& -					\\ 
$5\!\cdot\!10^{5}$	& 	0.018	 & 	-	 	& 	-			& 	88.9 		& 	-	 	& -					\\ 	
Whole KB	 		& 	0.017	& 	0.143 	& 	0.089		& 	90.1 		& 	79.4	 	& 91.3				\\ 	
\hline                  
\label{table:accuracy_training} 
\end{tabular}
\end{table*}

\section{Photometric redshifts errors and catastrophic outliers}
\label{sec:errors}

The determination of the uncertainty affecting the photometric redshifts has always been an open issue
\cite{quadri2010}. For instance, some methods in the past have provided a unique value of the error for all redshifts, 
based on the global evaluation of the accuracy of the evaluated redshifts themselves (see \cite{dabrusco2007}). A further 
advantage of the WGE algorithm over other methods is the ability to evaluate errors for each individual 
photometric redshift, based on the same \emph{features} used to train the WGE and on the value of the redshifts. 
While the evaluation of the statistical error is quite difficult and would not provide useful information for 
the scientific applications of the photometric redshifts, an estimate of the maximum error affecting each 
photometric redshift is represented by the value of the associated variable $\Delta z$, i.e. the 
difference between the photometric redshift and the corresponding value of the spectroscopic redshifts. 
The WGE has been trained to evaluate the uncertainty $\sigma_{z}$ for each photometric redshift as:

\begin{equation}
\mathrm{WGE}_{train}: (\mathbf{p}, z_{\mathrm{phot}})\!\rightarrow\!{\mathbf \|\Delta z\|}
\label{eq:wgeerrtrain}
\end{equation}  

\noindent where, as in equation \ref{eq:wgetrain}, $\mathbf{p}$ is the vector associated to a given 
collection of \emph{feature} values (i.e., a given set of colors or magnitudes), $z_{\mathrm{phot}}$ is the 
photometric redshift evaluated by the WGE in the first phase, and $\|\Delta z\|$ 
is the absolute value of the $\Delta z$ variable. 
Once trained, the WGE provides an estimated value of the error as a function of the \emph{features} and of 
the reconstructed targets, i.e. of the photometric \emph{features} and redshifts:

\begin{equation}
\sigma_{z_{\mathrm{phot}}} = \mathrm{WGE}(\mathbf{p}, z_{\mathrm{phot}})
\label{eq:wgeerrrun}
\end{equation}

\begin{figure*}
   \centering
   \includegraphics[width=6in, height=6in]{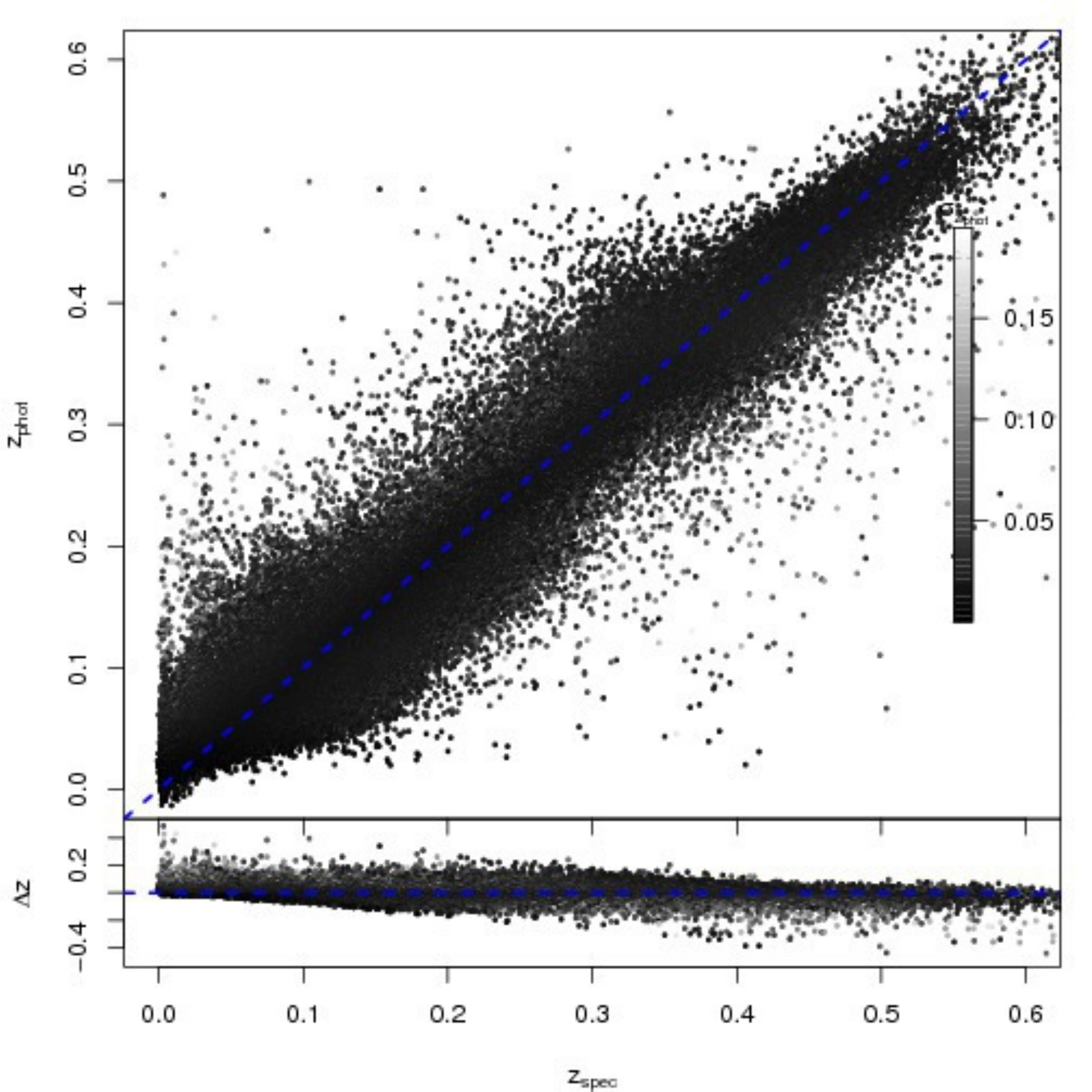} 
   \caption{In the upper panel, it is shown the scatterplot of the spectroscopic vs photometric redshifts evaluated with 
   		the WGE method for the members of the KB of the experiment for the SDSS galaxies 
		with optical photometry, while in the lower panel the scatterplot of the spectroscopic 
		redshift $z_{\mathrm{spec}}$ vs $\Delta z$ variable is shown for the same sources. All points are color-coded 
		according to the value of the errors $\sigma_{z_{\mathrm{phot}}}$ as evaluated but the WGE.}
   \label{plot:zvsz_gal_deltazerrbars}
\end{figure*}

\begin{figure*}
   \centering
   \includegraphics[width=6in, height=6in]{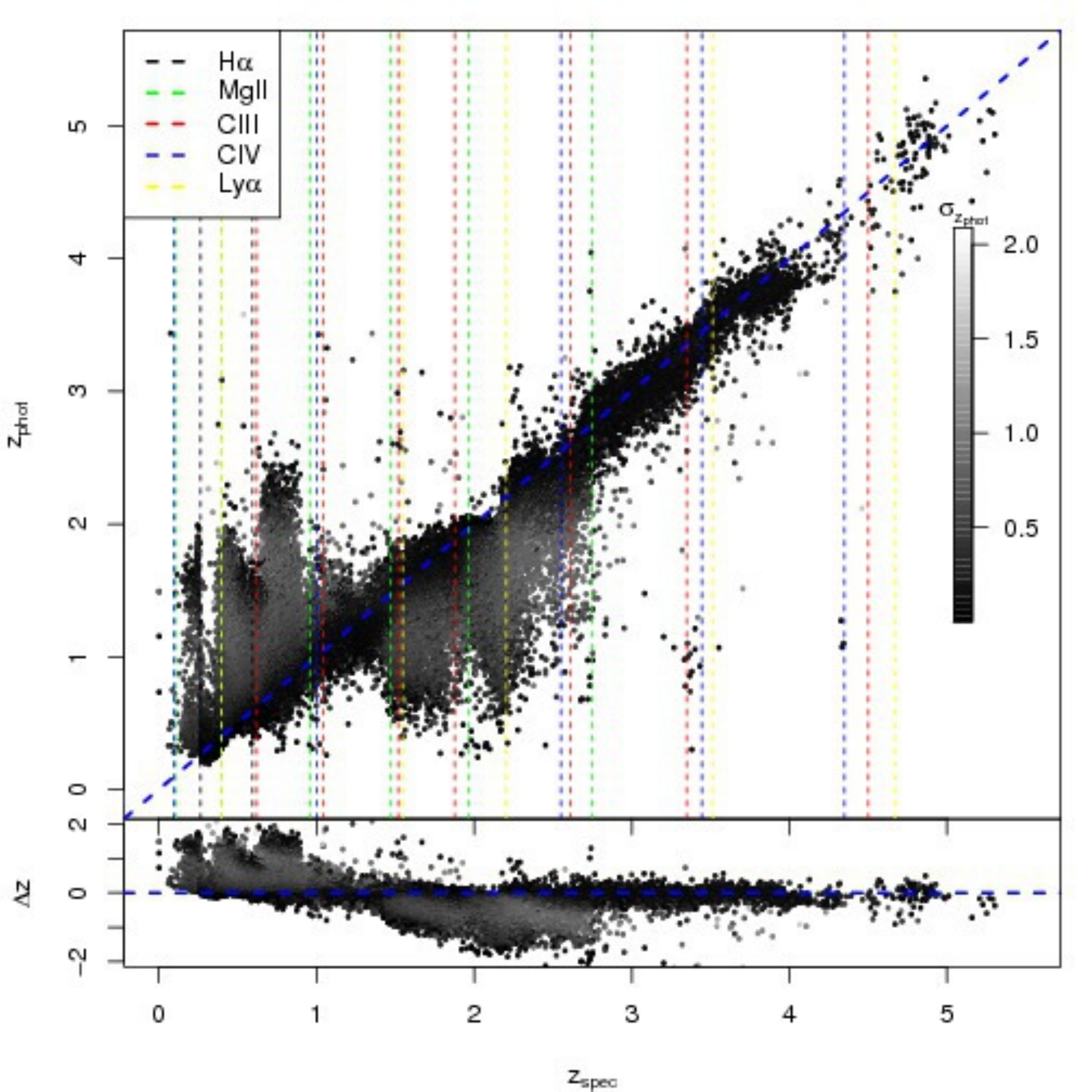} 
   \caption{In the upper panel, it is shown the scatterplot of the spectroscopic vs photometric redshifts evaluated with 
   		the WGE method for the members of the KB of the experiment for the quasars extracted from the SDSS catalog 
		with optical photometry, while in the lower panel the scatterplot of the spectroscopic 
		redshift $z_{\mathrm{spec}}$ vs $\Delta z$ variable is shown for the same sources. All points are color-coded
		according to the value of the errors $\sigma_{z_{\mathrm{phot}}}$ as evaluated but the WGE. The vertical dashed lines
		represent the redshift at which the most luminous emission lines characterizing quasars spectra
		shift off the SDSS photometric filters due to redshift. Most of the 
		features of the plot are associated to one or more of these lines.}
   \label{plot:zvsz_quasars_deltazerrbars}
\end{figure*}

\begin{figure*}
   \centering
   \includegraphics[width=6in, height=6in]{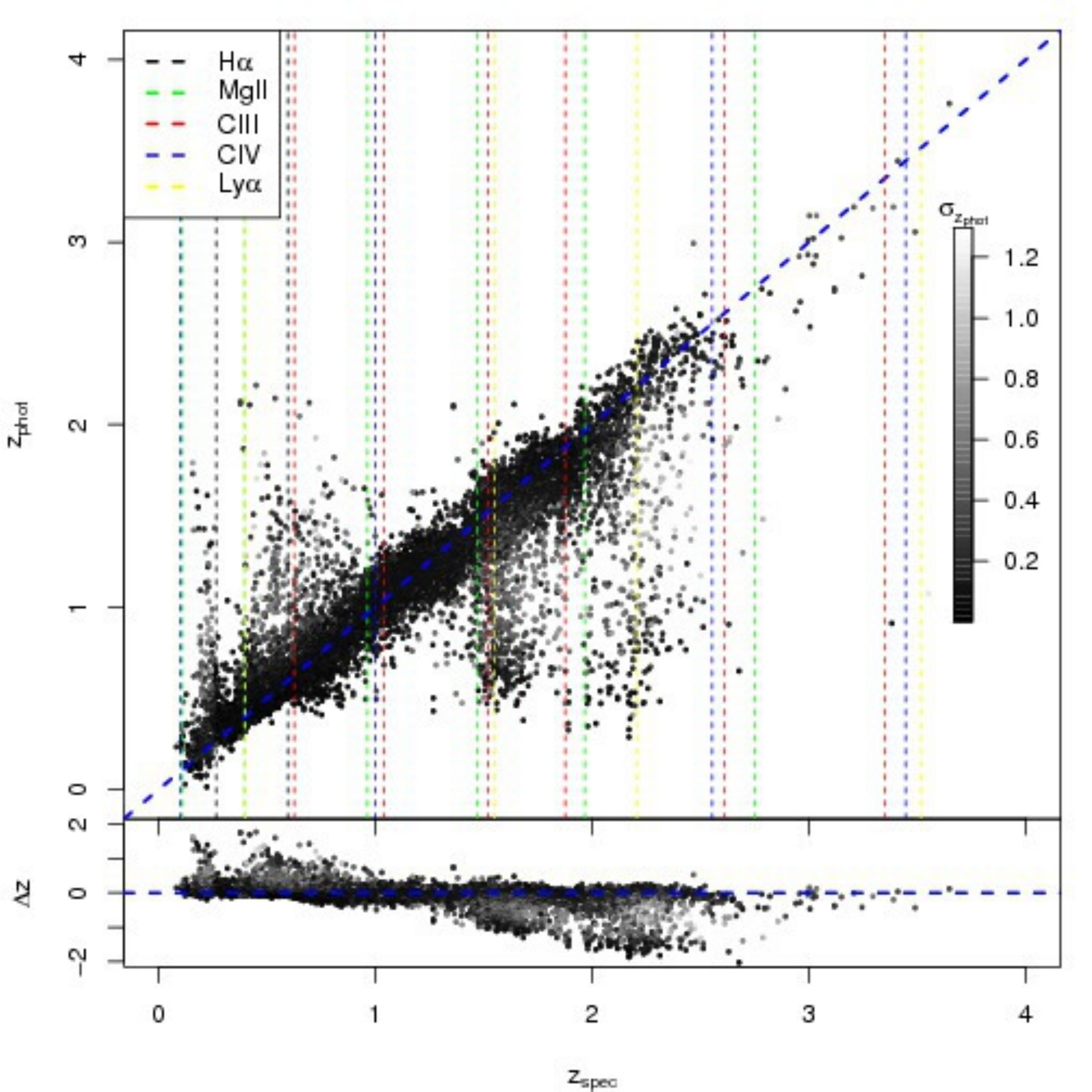} 
   \caption{In the upper panel, it is shown the scatterplot of the spectroscopic vs photometric redshifts evaluated with 
   		the WGE method for the members of the KB of the experiment for the quasars extracted from the SDSS catalog 
		with optical and ultraviolet photometry, while in the lower panel the scatterplot of the spectroscopic 
		redshift $z_{\mathrm{spec}}$ vs $\Delta z$ variable is shown for the same sources. All points are color-coded
		according to the value of the errors $\sigma_{z_{\mathrm{phot}}}$ as evaluated but the WGE. The vertical dashed lines
		represent the redshift at which the most luminous emission lines characterizing quasars spectra
		shift off the SDSS and GALEX photometric filters due to redshift. Similarly 
		to what is shown in figure \ref{plot:zvsz_quasars_deltazerrbars}, most of the features of the plot are associated to one or 
		more of these lines. Moreover, the lines associated to the GALEX filters resolve some of the degeneracies at low redshift.}
   \label{plot:zvsz_quasarsuv_deltazerrbars}
\end{figure*}

\noindent The evaluation of the errors on the photometric redshifts estimates with the WGE for the experiments 
described in sections \ref{subsec:expgal}, \ref{subsec:expqua} and \ref{subsec:expquauv}, has been carried out 
with a similar approach to the one described in the above sections for the evaluation of the photometric redshifts, except for 
the slightly different choice of the \emph{features}. For all three classes of experiments, the photometric \emph{features} 
used for the evaluation of the photometric redshifts, the photometric redshifts $z_{\mathrm{phot}}$  and the difference 
between photometric and spectroscopic redshifts $\Delta z$ have been used as \emph{features} for the clustering. 
The training of the experts has been performed on the same set of \emph{features}, except for the $\Delta z$ 
variable that has been employed as target of the training. A detailed list of the WGE parameters for the experiments
for the evaluation of the errors on the photometric redshifts is shown in table \ref{table:experimentserr}. 
The plots \ref{plot:zvsz_gal_deltazerrbars}, \ref{plot:zvsz_quasars_deltazerrbars} and 
\ref{plot:zvsz_quasarsuv_deltazerrbars} show the distribution of errors for the reconstructed 
photometric redshifts of the sources belonging to the KBs of the three distinct experiments. 
In these plots, the scatterplots of the variable $\Delta z$ and the 
spectroscopic redshifts $z_{\mathrm{spec}}$ are shown in the lower panels. Points in both 
panels are colored according to the value of the $\sigma_{z_{\mathrm{phot}}}$ variable. 

\begin{table*}
\caption{Parameters of the best experiments for the evaluation of the error on the photometric redshifts for optical galaxies, optical 
candidate quasars and optical plus ultraviolet candidate quasars.}             
\centering
{\small          
\begin{tabular}{l c c c}    
\hline\hline       
Params. clustering ($\sigma_{z}$)	& $\sigma_{u\!-\!g}$,$\sigma_{g\!-\!r}$,$\sigma_{r\!-\!i}$,$\sigma_{i\!-\!z}$ 	&$\sigma_{u\!-\!g}$,$\sigma_{g\!-\!r}$,$\sigma_{r\!-\!i}$,$\sigma_{i\!-\!z}$ 	&	$\sigma_{u\!-\!g}$,$\sigma_{g\!-\!r}$,$\sigma_{r\!-\!i}$,$\sigma_{i\!-\!z}$, 				\\
							& $(u\!-\!g)$,$(g\!-\!r)$,$(r\!-\!i)$,$(i\!-\!z)$,								&$(u\!-\!g)$,$(g\!-\!r)$,$(r\!-\!i)$,$(i\!-\!z)$,								&	$\sigma_{fuv\!-\!nuv}$,$\sigma_{nuv\!-\!u}$,									\\
							& $z_{\mathrm{phot}}$,$(z_{\mathrm{phot}}\!-\!z_{\mathrm{spec}})$	&$z_{\mathrm{phot}}$,$(z_{\mathrm{phot}}\!-\!z_{\mathrm{spec}})$	&	$(fuv\!-\!nuv)$,$(nuv\!-\!u)$, $(u\!-\!g)$,$(g\!-\!r)$,									\\
							&													&													& 	$(r\!-\!i)$,$(i\!-\!z)$,$z_{\mathrm{phot}}$,$(z_{\mathrm{phot}}\!-\!z_{\mathrm{spec}})$		\\
Min. \# clusters ($\sigma_{z}$)		& 2													& 2													&	2																\\
Max. \# clusters ($\sigma_{z}$)		& 9													& 9													&	9																\\
Opt. \# clusters ($\sigma_{z}$)		& 2													& 3													&	7																\\	
Clusters threshold ($\sigma_{z}$)	& 0.1													& 0.1													&	0.1																\\
Max. iterations clust. ($\sigma_{z}$)	& 500												& 500												&	500																\\
Params. experts ($\sigma$)		& $\sigma_{u\!-\!g}$,$\sigma_{g\!-\!r}$,$\sigma_{r\!-\!i}$,$\sigma_{i\!-\!z}$	& $\sigma_{u\!-\!g}$,$\sigma_{g\!-\!r}$,$\sigma_{r\!-\!i}$,$\sigma_{i\!-\!z}$, 	&	$\sigma_{u\!-\!g}$,$\sigma_{g\!-\!r}$,$\sigma_{r\!-\!i}$,$\sigma_{i\!-\!z}$,					\\
							& $(u\!-\!g)$,$(g\!-\!r)$,$(r\!-\!i)$,$(i\!-\!z)$,								& $(u\!-\!g)$,$(g\!-\!r)$,$(r\!-\!i)$,$(i\!-\!z)$,								&	$\sigma_{fuv\!-\!nuv}$,$\sigma_{nuv\!-\!u}$,									\\
							& $z_{\mathrm{phot}}$									& $z_{\mathrm{phot}}$									&	$(fuv\!-\!nuv)$,$(nuv\!-\!u)$, $(u\!-\!g)$, $(g\!-\!r)$,									\\
							&													&													&	$(r\!-\!i)$,$(i\!-\!z)$,$z_{\mathrm{phot}}$										\\
Hid. neurons experts ($\sigma_{z}$)	& 30												 	&20													&	20																\\
Max. epochs. experts ($\sigma_{z}$)& 500											 	&500												&	500																\\
Learning rate experts ($\sigma_{z}$)& 0.01											 	&0.01												&	0.01																\\
Steepness experts ($\sigma_{z}$)	& 1.0												 	&1.0													&	1.0																\\
Hid. neurons gate ($\sigma_{z}$)	& 30												 	&20													&	20 																\\
Max. epochs. gate ($\sigma_{z}$)	& 500											 	&500												&	500 																\\
Learning rate gate ($\sigma_{z}$)	& 0.01											 	&0.01												&	0.01																\\
Steepness gate ($\sigma_{z}$)		& 1.0													&1.0													&	1.0																\\
\# training gates ($\sigma_{z}$)	& 20												 	&20													&	20																\\
MAD ($\sigma$)				& 0.01												&0.086												&	0.053															\\
\hline                  
\label{table:experimentserr} 
\end{tabular}
}
\end{table*}

\begin{figure*}
	\centering
	\begin{tabular}{c c c}
	\includegraphics[width=3in, height=2.5in]{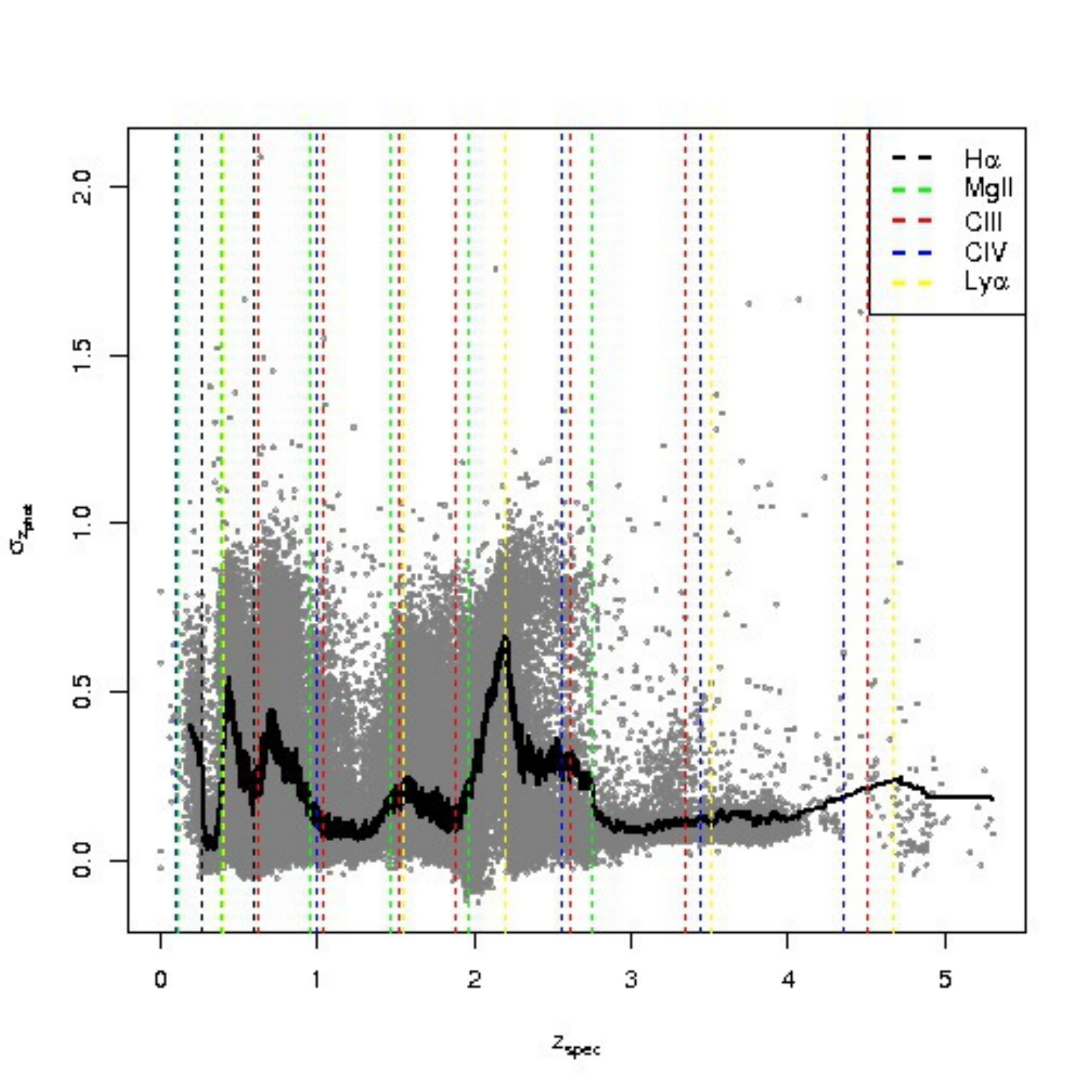} &
	\includegraphics[width=3in, height=2.5in]{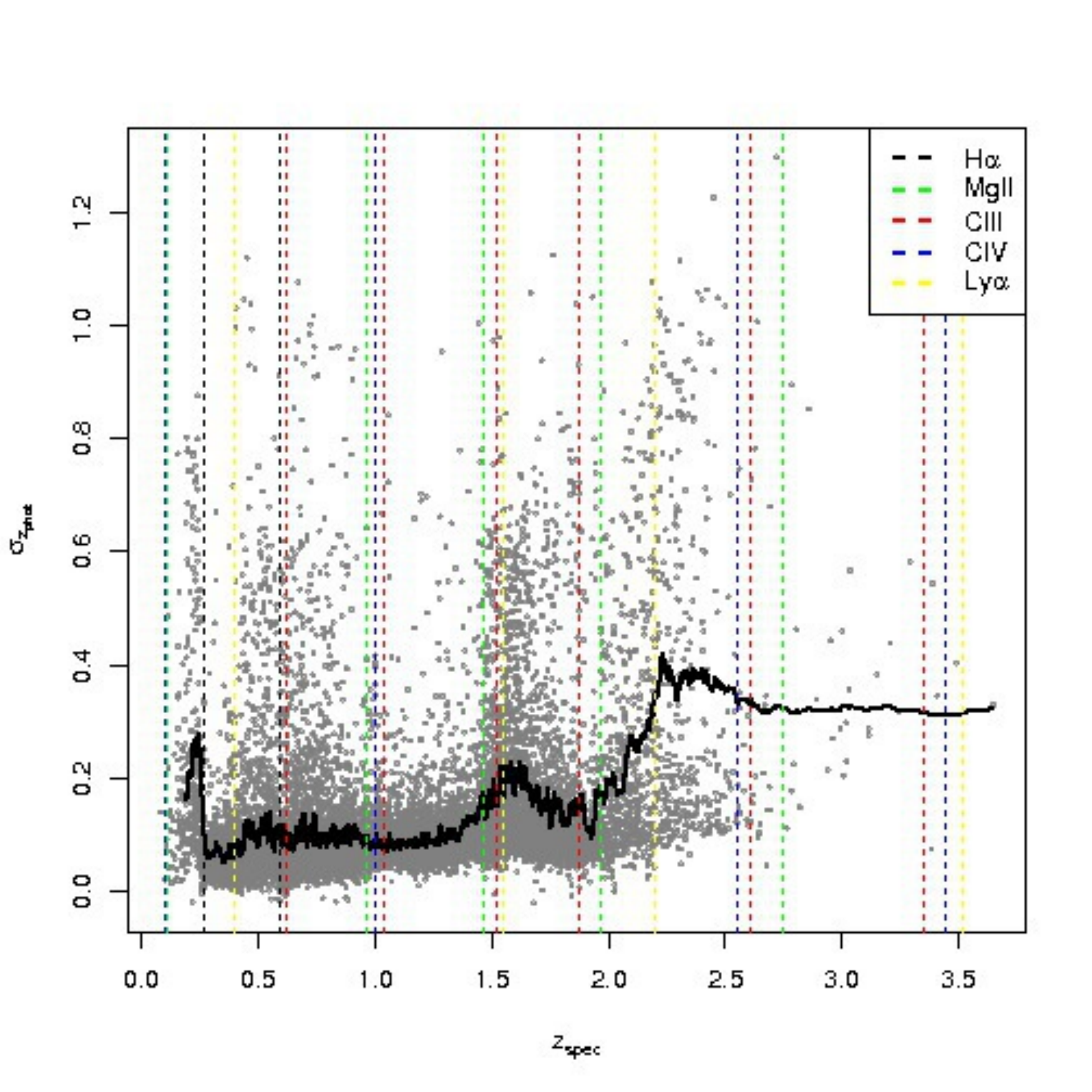}\\
	\includegraphics[width=3in, height=2.5in]{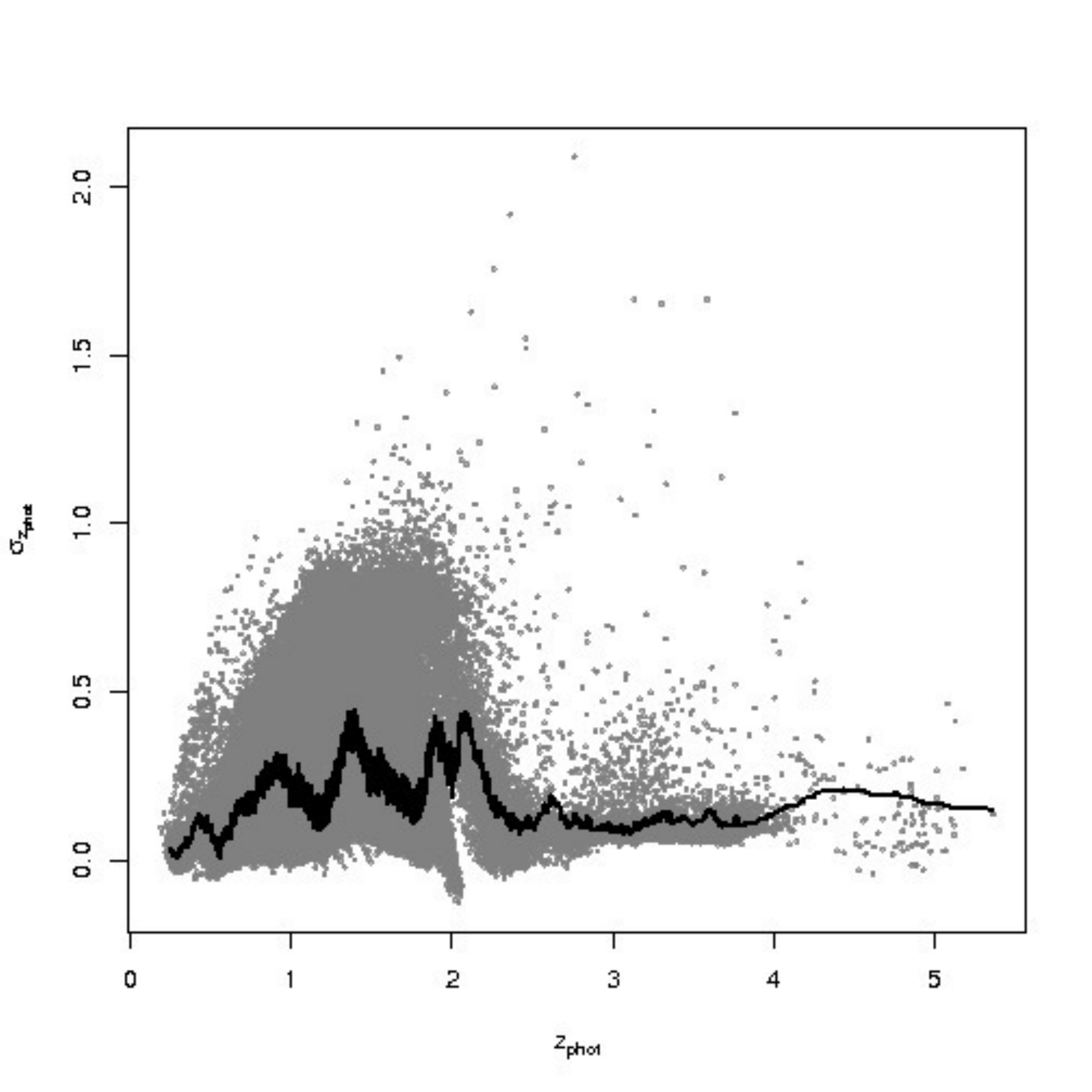} &
	\includegraphics[width=3in, height=2.5in]{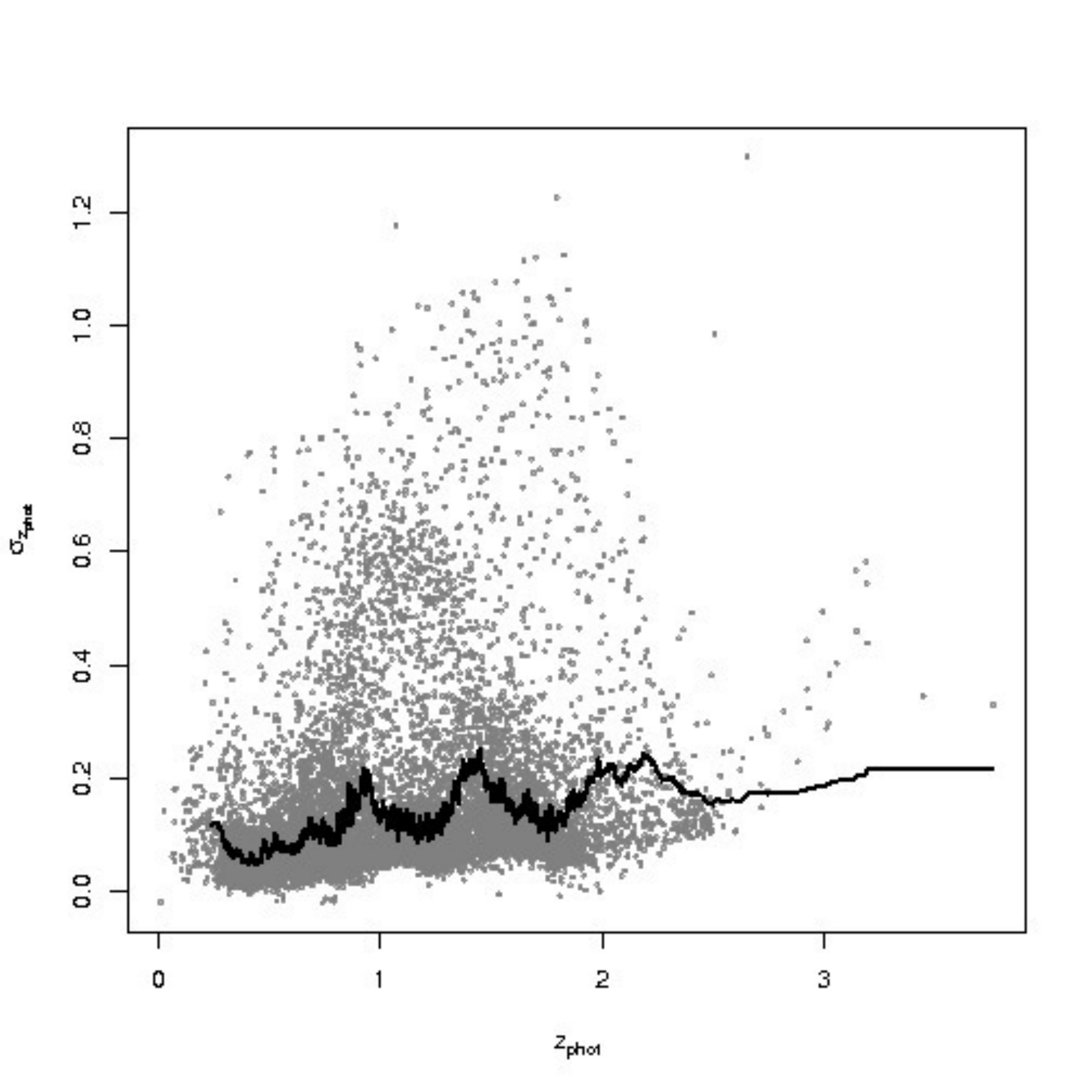}\\
	\includegraphics[width=3in, height=2.5in]{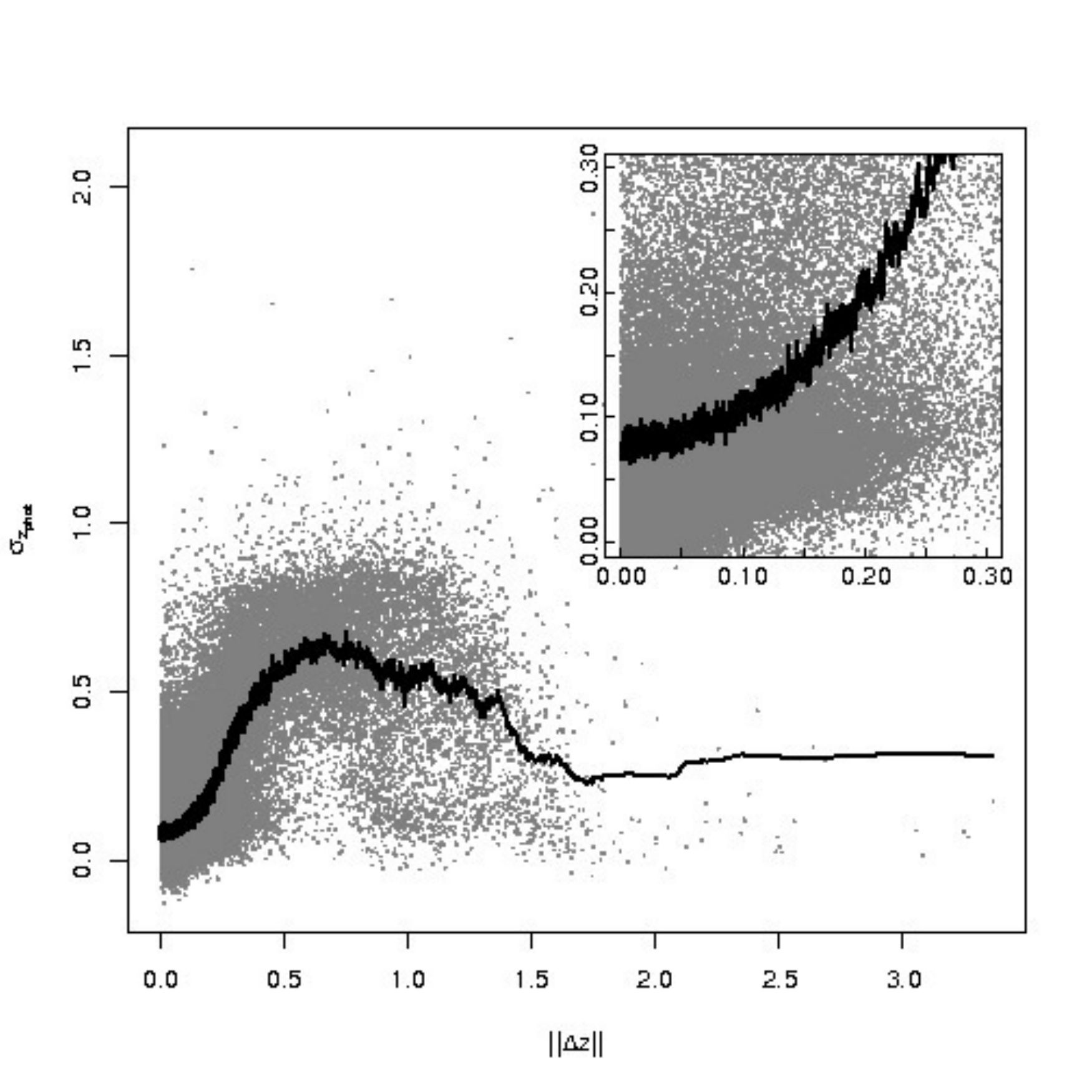} &
	\includegraphics[width=3in, height=2.5in]{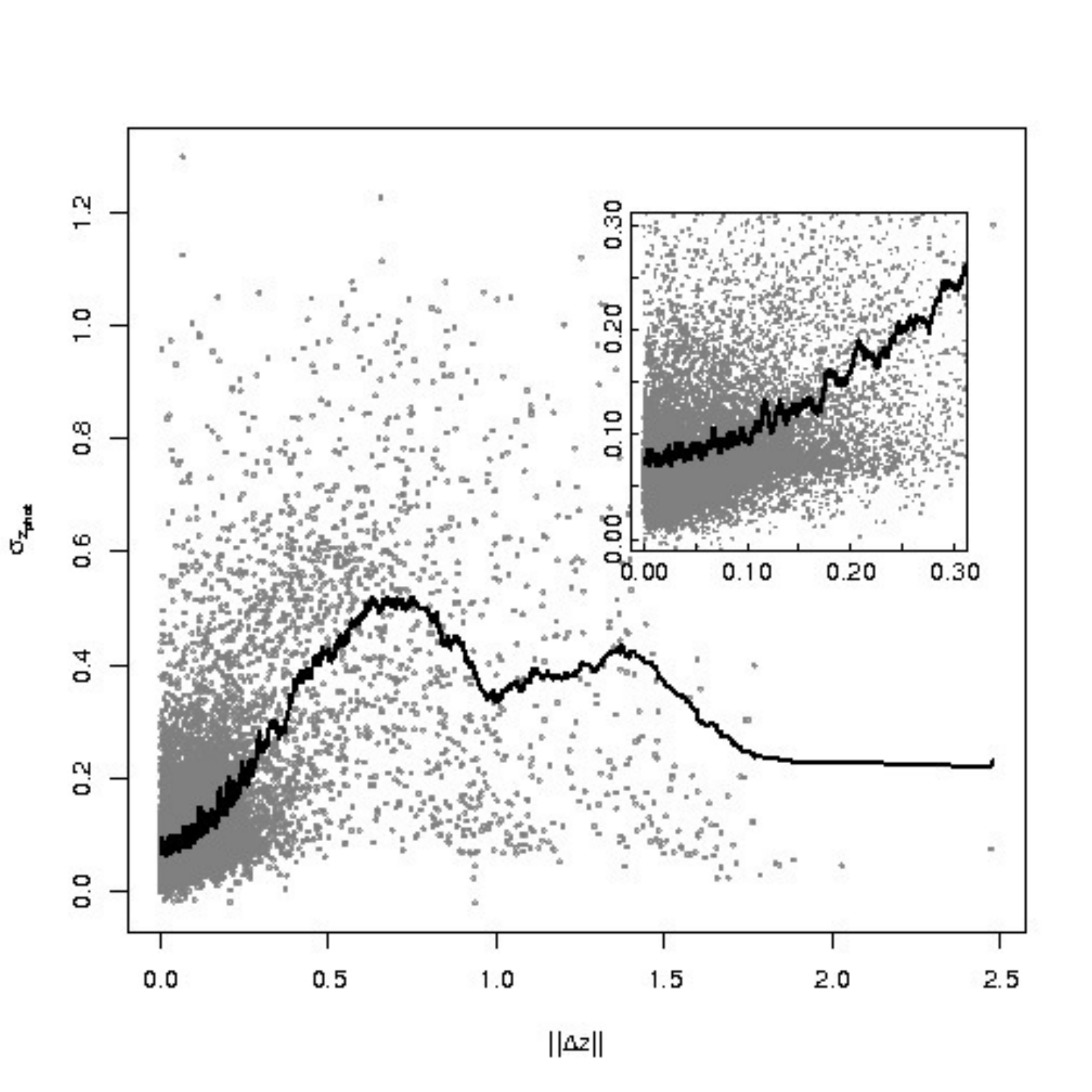}\\
	\end{tabular}
	\caption{From the upper to the lower plots, the distributions of the errors on the photometric redshifts 
	$\sigma_{z_{\mathrm{phot}}}$ as function 
	of the spectroscopic redshifts $z_{\mathrm{spec}}$, the photometric redshift
	$z_{\mathrm{phot}}$ and the variable $\|\Delta z\|$ respectively are
	shown for the two experiments regarding
	quasars with optical only photometry (left column) and quasars with optical and ultraviolet photometry (right column)
	discussed in this paper. The average profiles of the distribution of error 
	on the photometric redshifts are shown as a black line in all plots. In the upper plots, the redshifted emission lines
	are shown similarly to what is done in figures \ref{plot:zvsz_quasars_deltazerrbars} and \ref{plot:zvsz_quasarsuv_deltazerrbars}, 
	as lines over-plotted to the $z_{\mathrm{spec}}$ vs $\sigma_{z_{\mathrm{phot}}}$ scatterplots. Also in these cases, most of the 
	features in these two plot can be associated to one or more of the lines. In the lower two plots, the insets show the densest 
	regions of the plots. For the optical quasars (lower left plot), $\sim\!82\%$ of the sample is contained in the inset, while for the 
	optical and ultraviolet quasars (lower right plot), $\sim\!90\%$ of the sample is contained in the zoomed region.}
	\label{plot:sigmasigmarel}
\end{figure*}

\noindent The distribution of the errors on the photometric redshifts $\sigma_{z_{\mathrm{phot}}}$ as function of the spectroscopic redshifts, the 
photometric redshift and the variable $\Delta z$ are shown for the two experiments involving the samples of quasars in the figure \ref{plot:sigmasigmarel}.
As it was to be expected, in general, the WGE produces larger error estimates for the photometric redshifts of the sources lying inside the 
high degeneracy regions of the $z_{\mathrm{spec}}$ vs $z_{\mathrm{phot}}$ plots. As shown by the vertical dashed lines (upper panels), 
most of these regions occur at redshifts at which the most luminous emission lines characterizing SDSS quasars spectra shift off the filters of the 
SDSS or GALEX photometric systems. 
The shape of the average distribution of the error $\sigma_{z_{\mathrm{phot}}}$ as a function of the variable $\Delta z$, while not globally linear
as should have been expected in the case of perfect reconstruction of the errors by the WGE method, is compatible with a linear relation close to 
the diagonal of the plot for $\Delta z \lesssim 0.3$ for both experiments and represent an acceptable approximation since in this range lies a very high
percentage of the total number of sources (from $\sim\!80\%$ to $\sim\!90\%$ of the points). 

\noindent In the case of the reconstruction of the photometric redshifts for the candidate quasars using both the optical or the 
optical plus ultraviolet photometry, the characterization of the accuracy of the reconstruction of the photometric redshifts provided by 
the errors is not complete since, similarly to what happens for the $z_{\mathrm{phot}}$ values, the errors on such values are statistical 
estimates of the real uncertainty and are affected, to some extent, by the same degeneracies and systematic biases found in the 
$z_{\mathrm{phot}}$ reconstruction. This effect is noticeable in the scatterplots in figures \ref{plot:zvsz_quasars} and 
\ref{plot:zvsz_quasarsuv}, where consistent features of the plot deviate heavily from the ideal diagonal distribution. The degeneracies 
yielding such large effects cannot be completely resolved by the WGE during the phase of photometric redshifts estimation, but the 
same WGE generates information useful to flag the sources located in these regions of the plot (which cannot be recognized exactly 
in absence of spectroscopic redshifts, i.e. for all the sources belonging to the catalogs of photometric redshifts). For this reason, another 
measure of the reliability of the redshifts, hereafter called quality flag $q$, is provided for each object belonging to the catalog of photometric 
redshifts for optical candidate quasars. Unlike the photometric redshift value itself $z_{\mathrm{phot}}$ and the error on such 
value $\sigma_{z_{\mathrm{phot}}}$, the quality flag $q$ is evaluated on the basis of the global distributions of both photometric 
redshifts and photometric redshift errors, i.e. after the evaluation of photometric redshifts and of the corresponding errors for all 
sources in a given sample. The steps for the evaluation of the quality flags are the following:

\begin{itemize}
\item The distribution of photometric redshifts evaluated by the WGE for the training set is binned, inside the interval 
covered by the distribution of spectroscopic redshifts of the KB, in $n_{\mathrm{bin}}(z_{\mathrm{phot}})$ equally spaced intervals;
\item For each bin in the distribution of photometric redshifts, the associated set of errors on the estimates of the 
$z_{\mathrm{phot}}$ is binned in $n_{\mathrm{bin}}(\sigma_{z_{\mathrm{phot}}})$ equally spaced intervals;
\item The value of the quality flag of a given photometric redshift $\tilde{z}_{phot}$ is assigned according to the position 
of its uncertainty relatively to the overall presence of peaked features of the distribution: if the error $\sigma_{\tilde{z}_{\mathrm{phot}}}$ 
lies inside a bin belonging to the most prominent feature of the histogram (i.e. the component of the histogram containing 
the highest peak of the overall distribution), the quality flags of the corresponding photometric redshift estimate $q$ is set 
to 1, otherwise to 0;
\item The sources with $q\!=\!1$ are considered reliable, while the sources flagged by $q\!=\!0$ are considered unreliable, i.e. 
potential catastrophic outliers. 
\end{itemize}

\begin{figure*}
	\centering
	\begin{tabular}{c c}
	\includegraphics[width=3in, height=3in]{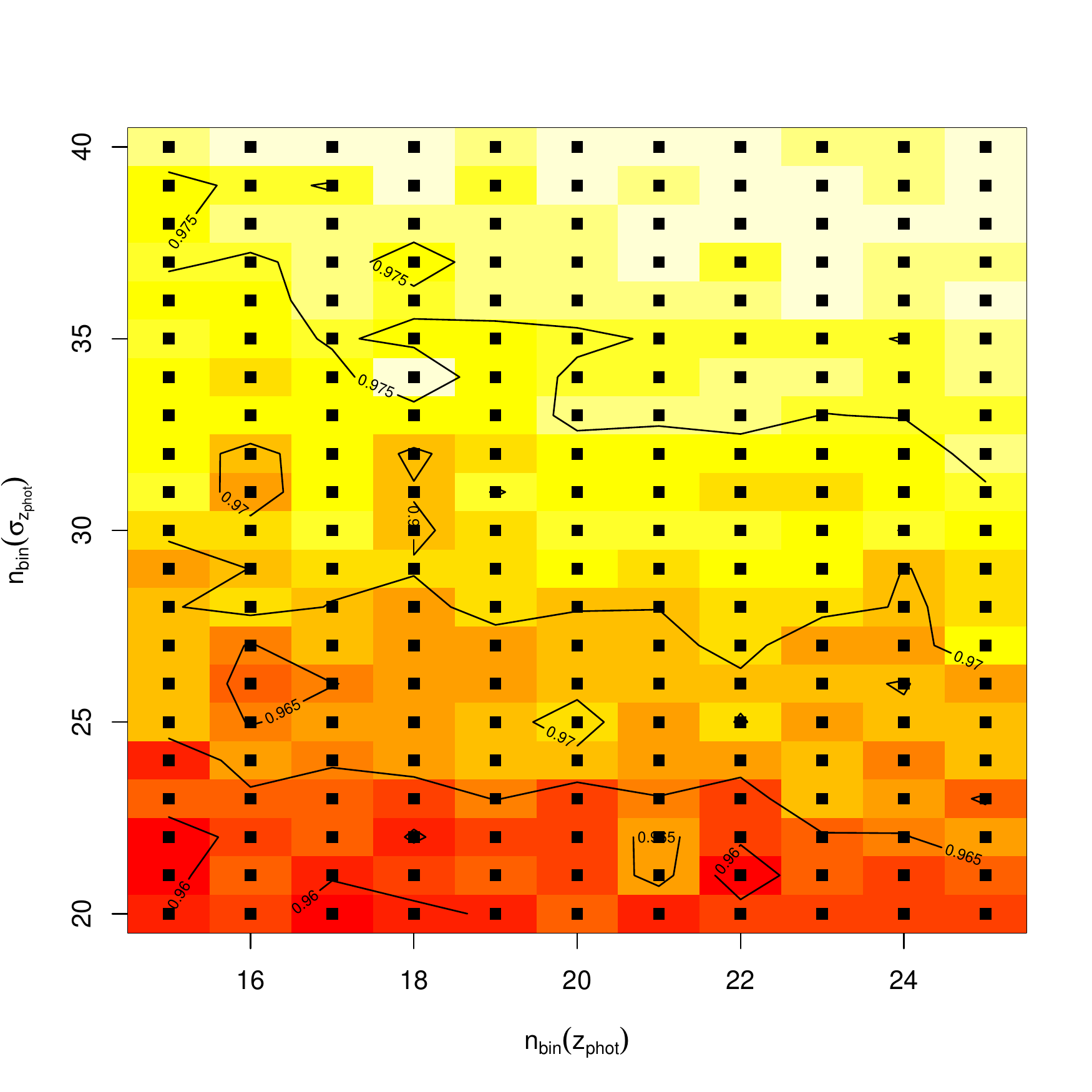} &
	\includegraphics[width=3in, height=3in]{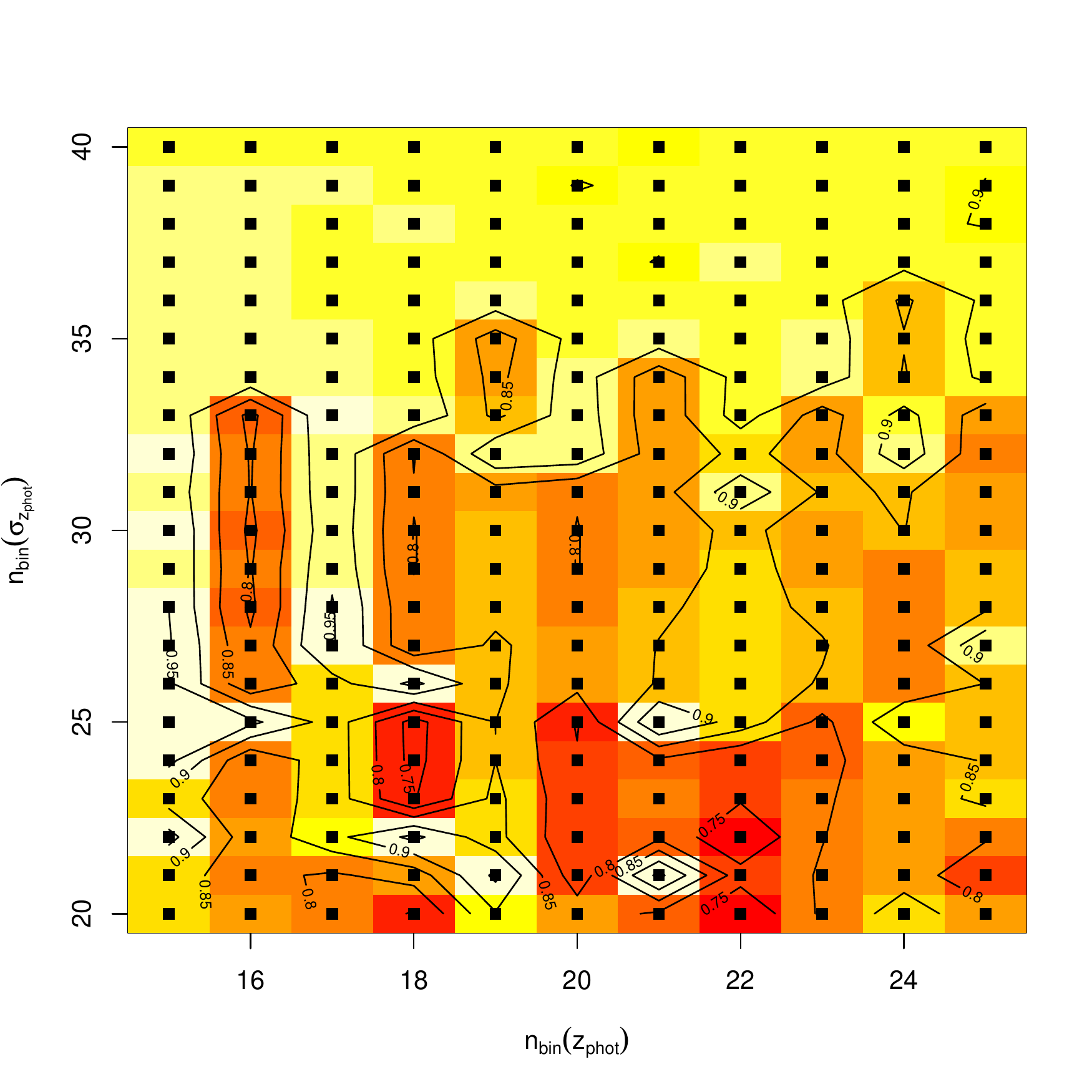}\\
	\includegraphics[width=3in, height=3in]{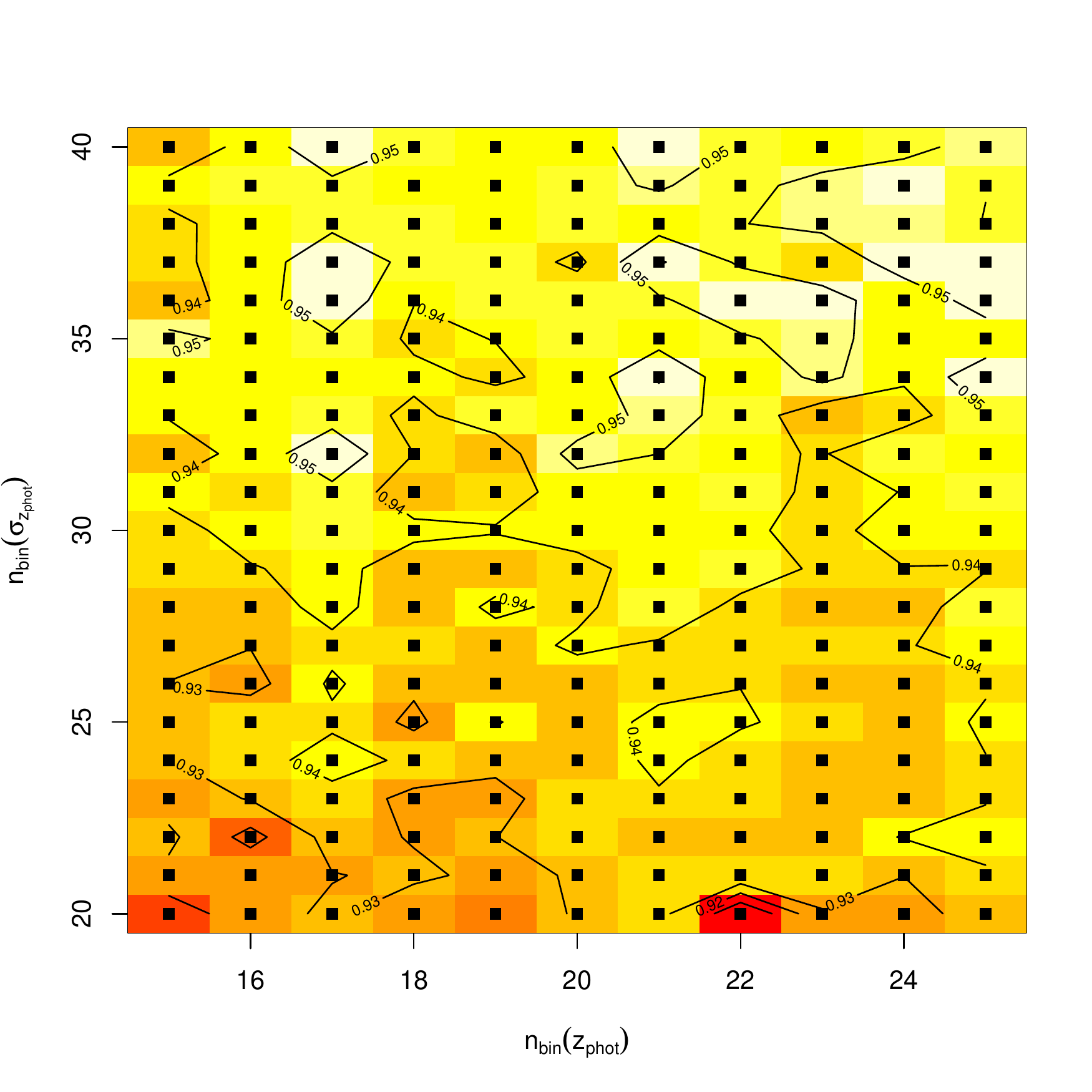}&
	\includegraphics[width=3in, height=3in]{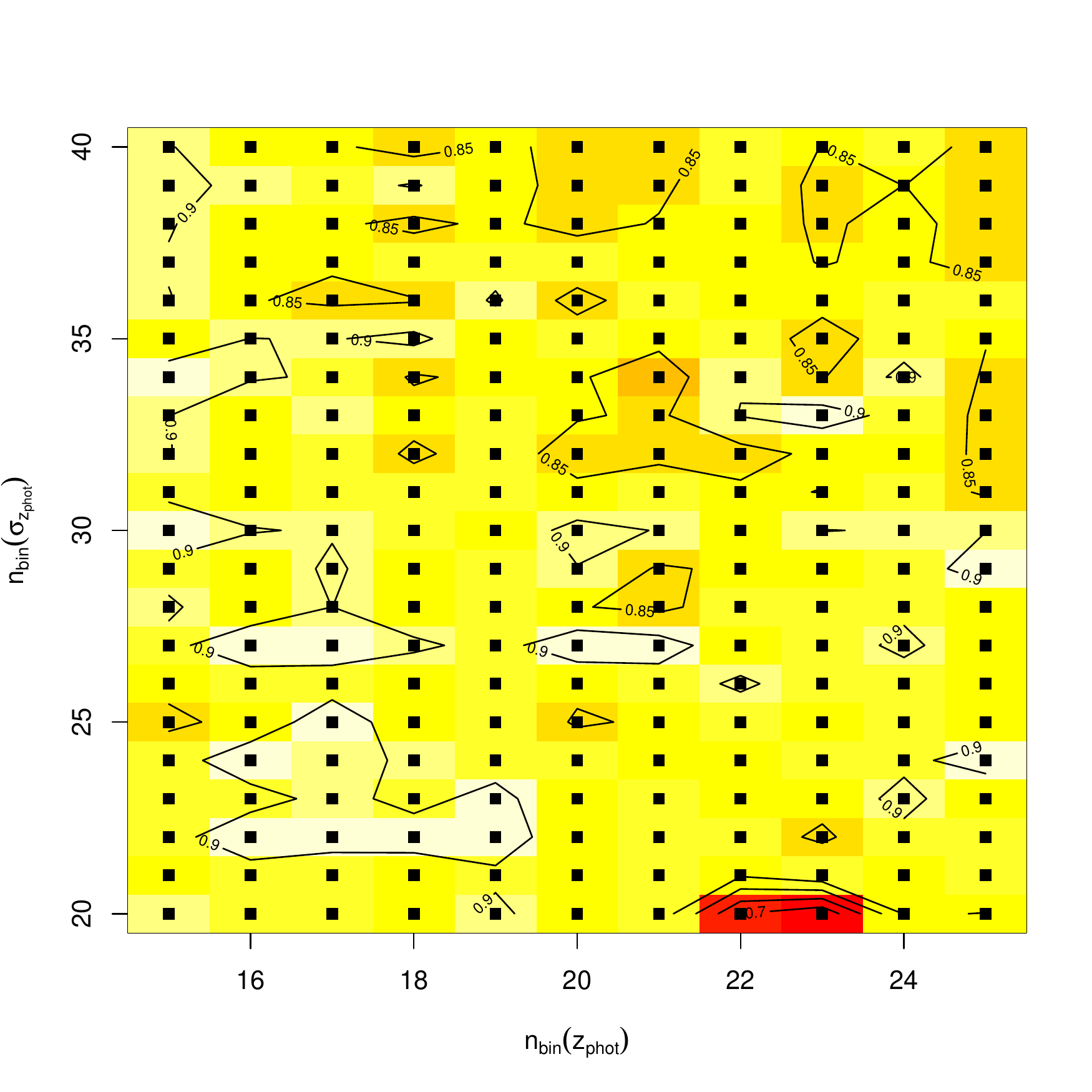}\\
	\end{tabular}
	\caption{Plots of the efficiency (left column) and of the completeness (right column) of the process of selection of the 
		       catastrophic outliers as functions of the two parameters $n_{\mathrm{bin}}(z_{\mathrm{phot}})$ and 
		       $n_{\mathrm{bin}}(\sigma_{z_{\mathrm{phot}}})$ involved in the procedure for the determination of the 
		       quality flag $q$. The upper plots are associated to the experiment for the evaluation of the photometric 
		       redshifts for the optical SDSS quasars, while the lower plots are associated to the third experiment for 
		       the estimation of the photometric redshifts of the SDSS quasars with optical and ultraviolet photometry.}
	\label{plot:effcomplquality}
\end{figure*}

\noindent The effectiveness of the quality flag in selecting the outliers of the photometric redshifts reconstruction depends 
critically on the value of the two parameters $n_{\mathrm{bin}}(z_{\mathrm{phot}})$ and $n_{\mathrm{bin}}(\sigma_{z_{\mathrm{phot}}})$ 
associated to the total number of bins for the photometric redshifts and the error on the photometric redshifts distribution respectively 
of the process described above. The optimal values of these two parameters have been determined by exploring the 
$n_{\mathrm{bin}}(z_{\mathrm{phot}})$ vs $n_{\mathrm{bin}}(\sigma_{z_{\mathrm{phot}}})$ space. Two different empirical 
diagnostics, based on the knowledge of the spectroscopic and photometric redshifts of the sources of the KBs, of the 
accuracy of the determination of the quality flags have been used, namely the efficiency and the completeness of the separation 
between reliable sources and unreliable sources. The efficiency $e$ is defined as the ratio of ``reliable" sources ($q\!=\!1$) with 
$\Delta z\!<\!0.3$ to the total number of ``reliable" sources ($q\!=\!1$), while the completeness $c$ is defined as the ratio of  
``reliable" sources ($q\!=\!1$) with $\Delta z\!<\!0.3$ to the total number of sources, independently from the value of the quality flag, 
with $\Delta z\!<\!0.3$.

\begin{figure*}
   \centering
   \includegraphics[width=6in, height=6in]{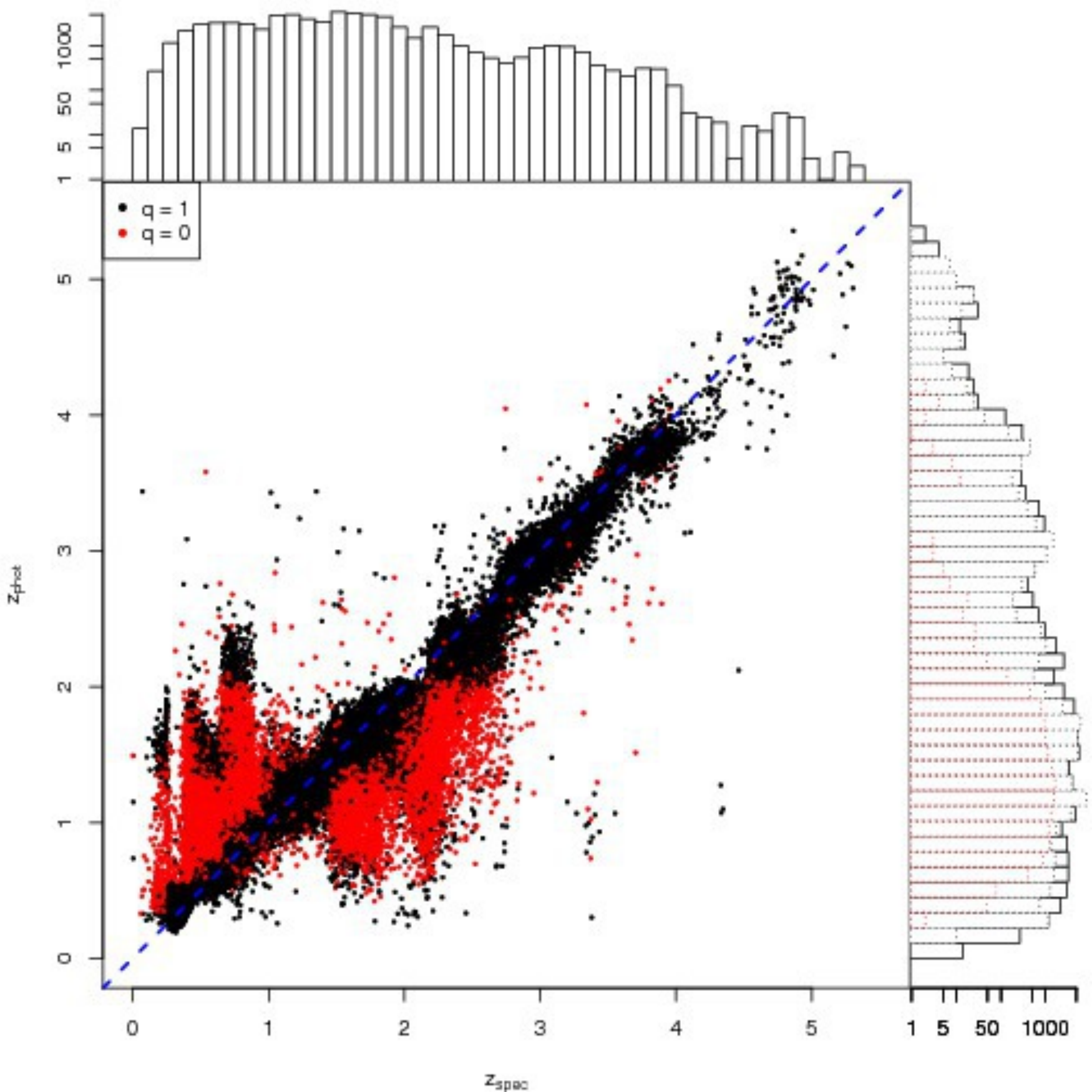} 
   \caption{Scatterplot of the spectroscopic vs photometric redshifts for the KB of the second experiment (quasars with 
   		optical photometry), with marginal histograms for reliable ($q\!=\!1$) and unreliable ($q\!=\!0$) photometric redshift 
		estimates according to the quality flag $q$. In the vertical marginal panel, the histograms of the distributions of 
		reliable and unreliable photometric redshifts
		are respectively plotted with black and red dotted lines, while the histogram of the spectroscopic redshifts distribution is 
		shown as a solid black line in both marginal panels.}
   \label{plot:zvsz_quasars_flags}
\end{figure*}

\noindent The efficiency and completeness for the second and third experiments as functions of the $n_{\mathrm{bin}}(z_{\mathrm{phot}})$ and 
$n_{\mathrm{bin}}(\sigma_{z_{\mathrm{phot}}})$ parameters are shown in the figures \ref{plot:effcomplquality}. The optimal 
values of the two parameters $n_{\mathrm{bin}}(z_{\mathrm{phot}})$ and $n_{\mathrm{bin}}(\sigma_{z_{\mathrm{phot}}})$ 
have been chosen to maximize at the same time the efficiency and completeness, i.e. the product of the efficiency 
and completeness $t=e\cdot c$, and in the case of equal values, priority has been given to the couple of values associated
to the larger efficiency. The optimal values of the parameters for the second experiment, i.e. the determination 
of the photometric redshifts of the optical SDSS quasars, are $n_{\mathrm{bin}}(z_{\mathrm{phot}})\!=\!18$ and 
$n_{\mathrm{bin}}(\sigma_{z_{\mathrm{phot}}})\!=\!34$ respectively. For the third experiment, involving the 
evaluation of the photometric redshifts for SDSS quasars with optical and ultraviolet photometry, the optimal parameters are
$n_{\mathrm{bin}}(z_{\mathrm{phot}})\!=\!17$ and $n_{\mathrm{bin}}(\sigma_{z_{\mathrm{phot}}})\!=\!32$.
The values of the flags associated to the high redshift quasars $(z_{\mathrm{spec}} \geq 4.5)$ have 
all been fixed to 1 (reliable photometric redshifts estimates) since, because of low total number of sources in such redshift interval, 
the method described above for the evaluation of the determination of the outliers based on the overall
shape of the binned $z_{\mathrm{phot}}$ distribution in bins of spectroscopic redshifts cannot be applied.
The decision to retain all such sources as reliable is based on the eye inspection of the $z_{\mathrm{spec}}$ vs 
$z_{\mathrm{phot}}$ scatterplot in figure \ref{plot:zvsz_quasars_flags}.

\begin{figure*}
	\centering
	\begin{tabular}{c c}
	\includegraphics[width=3in, height=3in]{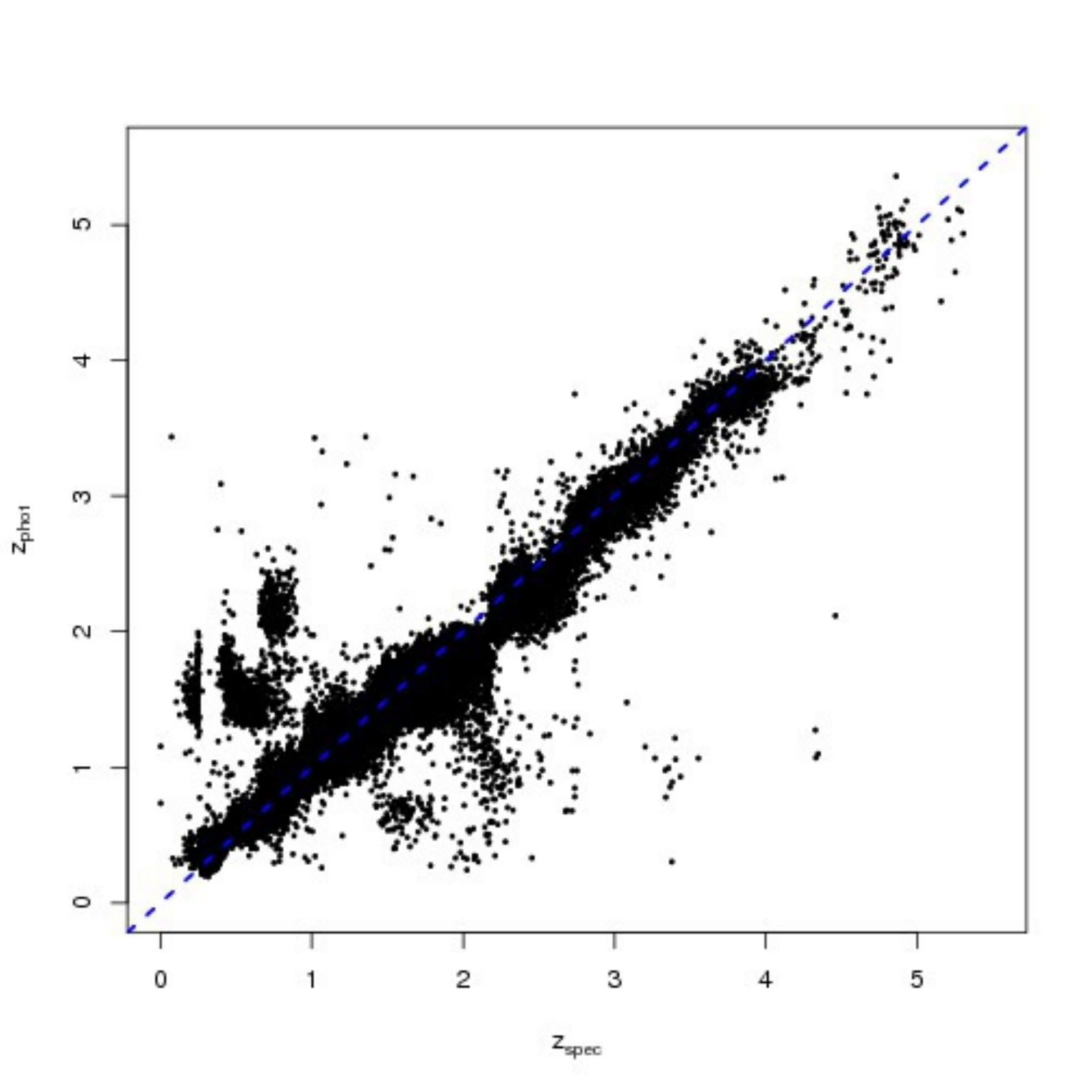} &
	\includegraphics[width=3in, height=3in]{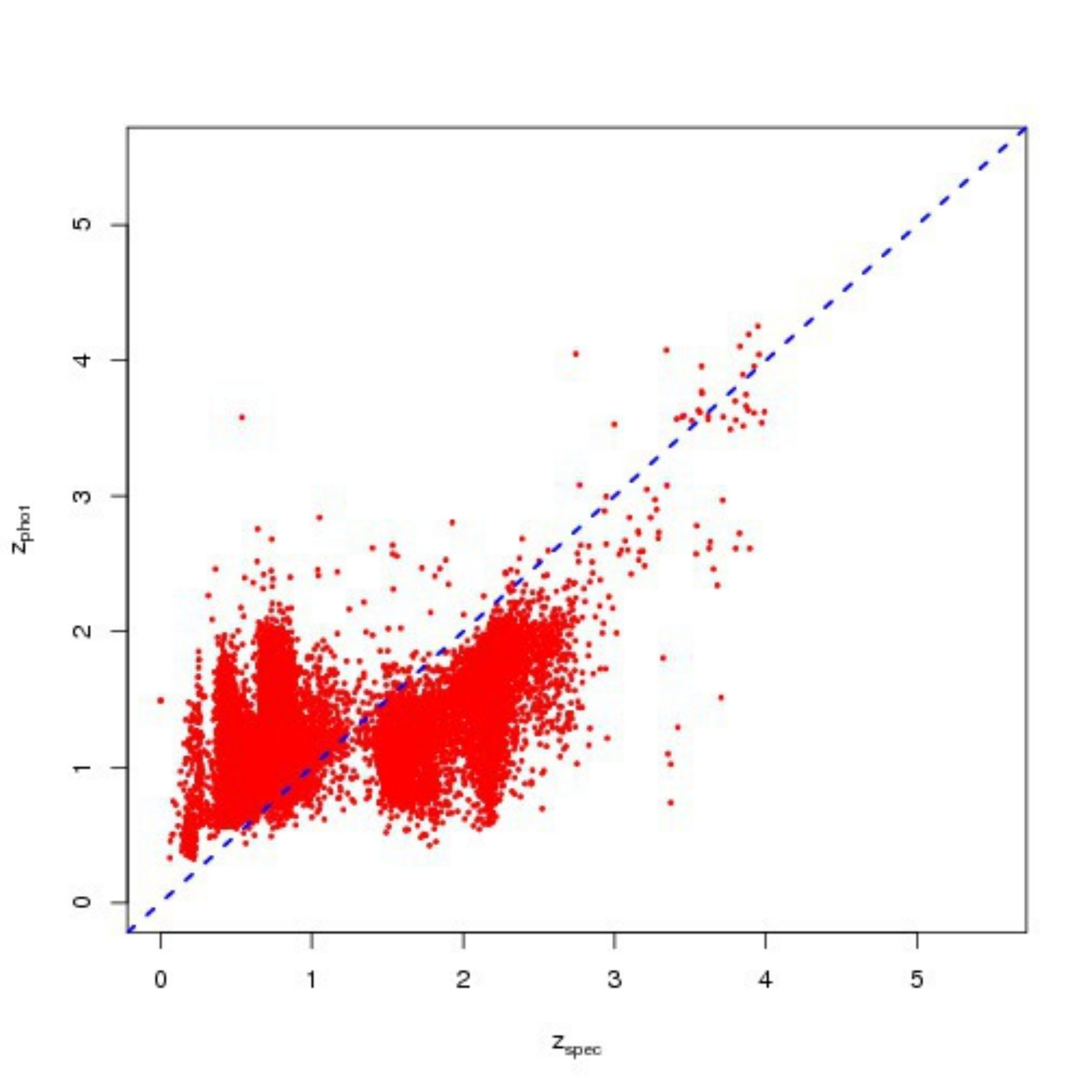}\\
	\end{tabular}
	\caption{Scatterplots of the spectroscopic vs photometric redshifts distributions for the KB of the second 
		       experiment (quasars with optical photometry) separately for reliable and unreliable estimations of the 
		       photometric redshifts according to the quality flag $q$. The sources with reliable $z_{\mathrm{phot}}$ 
		       values ($q\!=\!1$) are shown in the plot on the left, while sources with unreliable
		       $z_{\mathrm{phot}}$ values ($q\!=\!0$) are shown in the plot on the right.}
	\label{plot:qualitygoodbad}
\end{figure*}

\begin{figure*}
   \centering
   \includegraphics[width=6in, height=6in]{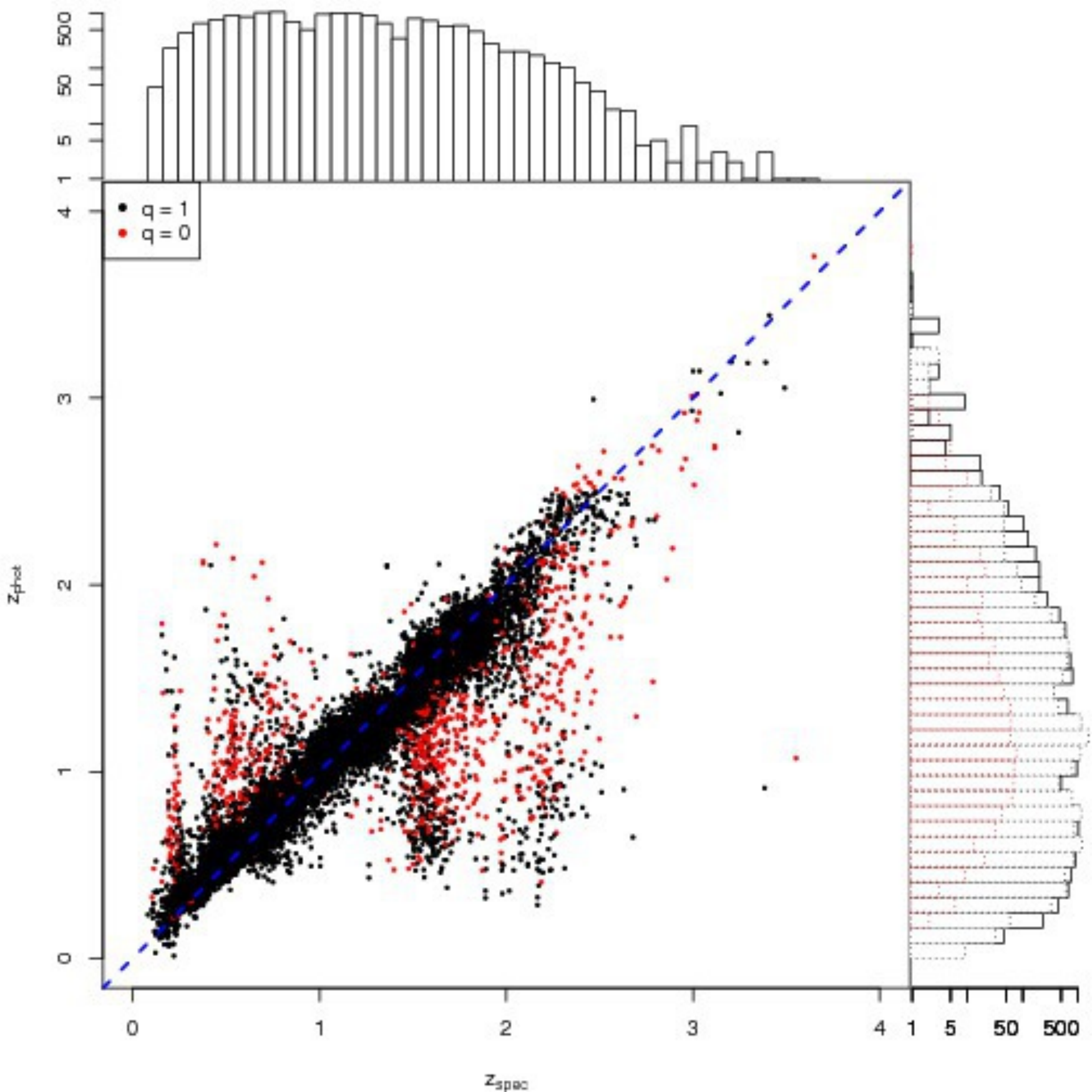} 
   \caption{Scatterplot of the spectroscopic vs photometric redshifts for the KB of the third experiment (quasars with optical 
   		and ultraviolet photometry), with marginal histograms for reliable and unreliable photometric redshift estimates 
		according to the quality flag $q$. In the vertical marginal panel, the histograms of the distributions of 
		reliable and unreliable photometric redshifts
		are respectively plotted with black and red dotted lines, while the histogram of the spectroscopic redshifts distribution is 
		shown as a solid black line in both marginal panels.}
   \label{plot:zvsz_quasarsuv_flags}
\end{figure*}

\begin{figure*}
	\centering
	\begin{tabular}{c c}
	\includegraphics[width=3in, height=3in]{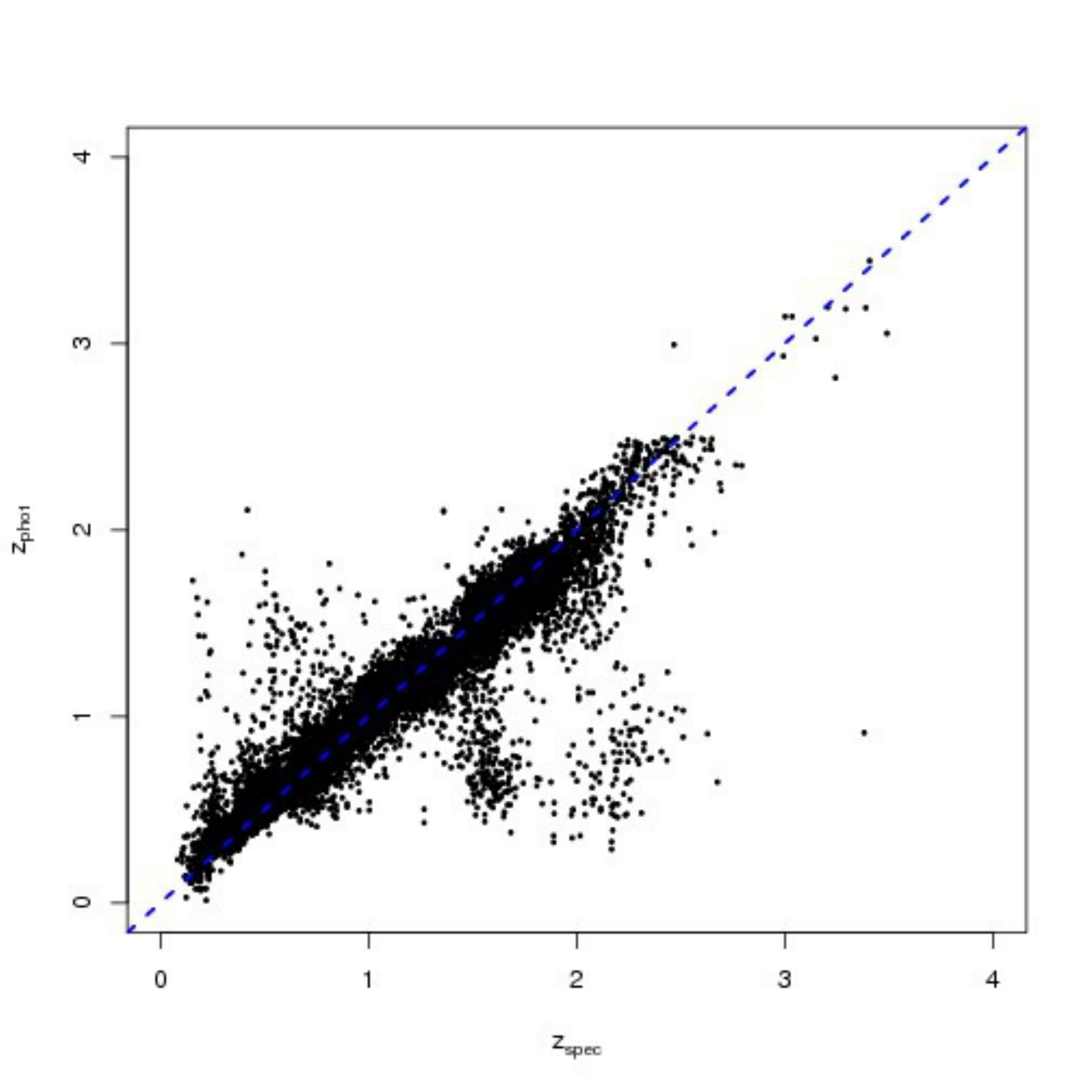} &
	\includegraphics[width=3in, height=3in]{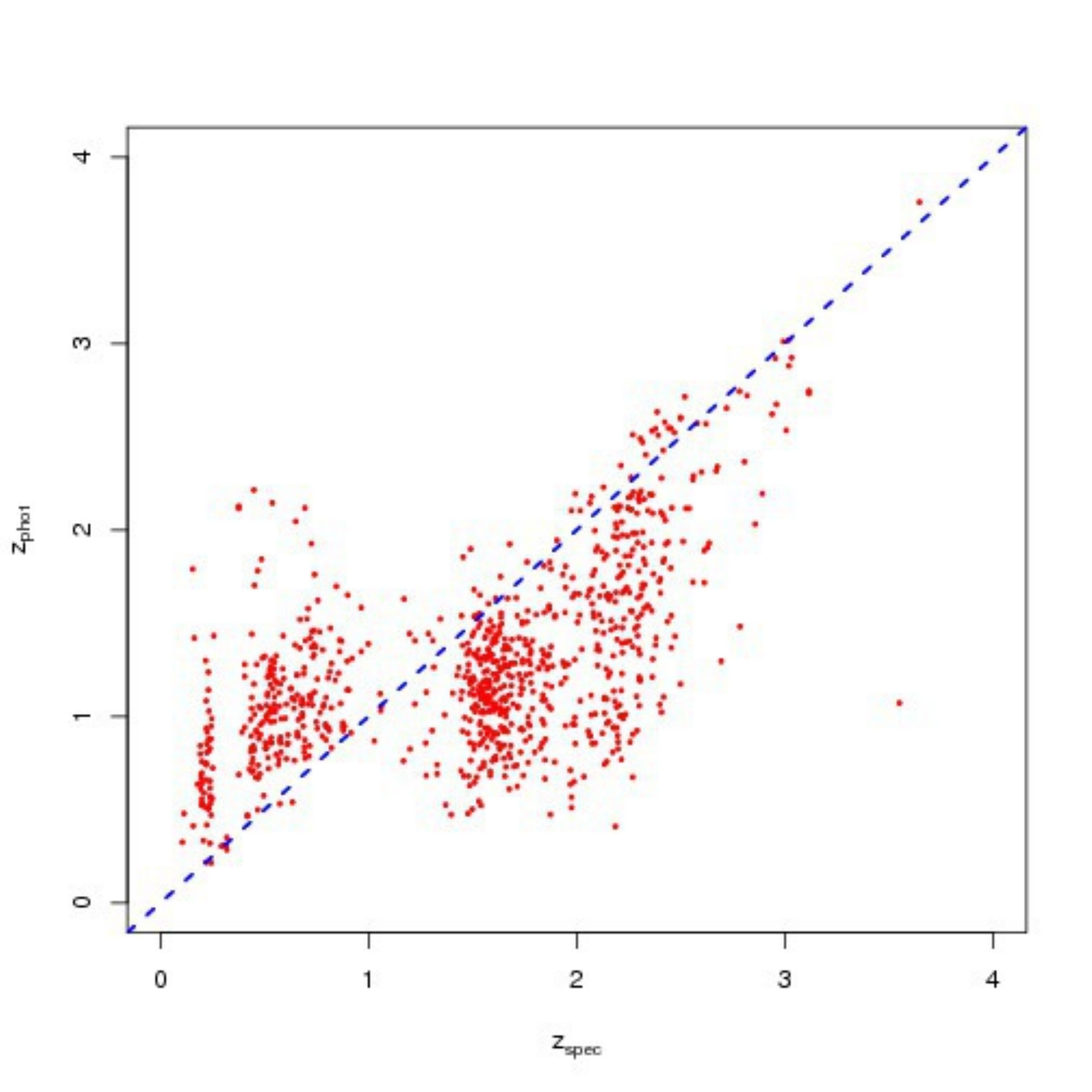}\\
	\end{tabular}
	\caption{Scatterplots of the spectroscopic vs photometric redshifts distributions for the KB of the third 
		       experiment (quasars with optical and ultraviolet photometry) separately for reliable and unreliable estimations of the 
		       photometric redshifts according to the quality flag $q$. The sources with reliable $z_{\mathrm{phot}}$ 
		       values ($q\!=\!1$) are shown in the plot on the left, while sources with unreliable
		       $z_{\mathrm{phot}}$ values ($q\!=\!0$) are shown in the plot on the right.}
	\label{plot:qualitygoodbaduv}
\end{figure*}

\noindent The scatterplot of the distribution of photometric redshifts as function of the spectroscopic redshifts for the KB associated to the 
second experiment performed by the WGE (quasars with optical photometry) with different color of the symbol associated to 
the two different values of the quality flags is shown in figure \ref{plot:zvsz_quasars_flags}, with marginal histograms of the 
distribution of the different subsets according to $q$. In order to highlight the differences in the distributions of 
the sources with reliable or unreliable photometric redshifts values, the $z_{\mathrm{phot}}$ vs $z_{\mathrm{spec}}$ scatterplots 
for the two samples with $q\!=\!1$ and $q\!=\!0$ respectively are shown in figure \ref{plot:qualitygoodbad}. These same plots 
for the experiment concerning the estimation of the photometric redshifts of quasars with optical and ultraviolet photometry, are 
shown in \ref{plot:zvsz_quasarsuv_flags} and \ref{plot:qualitygoodbaduv} respectively. The set of statistical diagnostics calculated 
for the whole KBs of the three experiments discussed in this paper and shown in table \ref{table:diagnostics}, have been evaluated 
for the KBs of the second and third experiments separately for sources with $q\!=\!1$ and $q\!=\!0$ (see table \ref{table:diagnosticsquaflags}).  

\begin{table*}
\caption{Statistical diagnostics of the accuracy of the photometric redshifts reconstruction for the second 
and third experiments, evaluated for reliable and unreliable $z_{\mathrm{phot}}$ estimates according to 
the quality flag $q$. For the definition of the statistical diagnostics see section \ref{sec:accuracy}.}             
\label{table:diagnosticsquaflags}      
\centering     
\begin{tabular}{l c c c c c c}
\hline\hline       
											& \multicolumn{3}{|c|}{Exp. 2}							&  \multicolumn{3}{|c|}{Exp. 3}							\\
Diagnostic 									& All 				& $q=1$ 			&  $q=0$			& All				& $q=1$ 			&  $q=0$			\\ 
\hline
$\left\langle \Delta z \right\rangle$ 					& 0.21			& 0.13			& 0.53			& 0.13		 	& 0.10			& 0.52			\\
RMS$(\Delta z)$                                   				         	& 0.35			& 0.24			& 0.62			& 0.25			& 0.20			& 0.63			\\
$\sigma^2(\Delta z)$ 							& 0.08 			& 0.04 			& 0.11			& 0.044 			& 0.031 			& 0.12			\\
MAD$(\Delta z)$ 								& 0.11 			& 0.07  			& 0.32			& 0.061   			& 0.056 			& 0.34 			\\
MAD'$(\Delta z)$ 							& $\mathbf{0.098}$	& $\mathbf{0.064}$  	& $\mathbf{0.41}$ 	& $\mathbf{0.062}$ 	& $\mathbf{0.047}$	& $\mathbf{0.29}$	\\
$\%(\Delta z_1)$ 								& 50.7 			& 61.9 			& 5.9				& 68.1 			& 71.4 			& 8.5				\\
$\%(\Delta z_2)$ 								& 72.3 			& 86.6			& 15.2			& 86.5 			& 90.4 			& 18.2 			\\
$\%(\Delta z_3)$ 								& 80.5 			& 90.6 			& 27.5			& 91.4 			& 95.0 			& 28.6			\\
$\sigma^{2}(\Delta z_1)$ 							& $7.9\!\cdot\!10^{-4}$	& $7.9\!\cdot\!10^{-4}$	& $8.3\!\cdot\!10^{-4}$	& $7.6\!\cdot\!10^{-4}$	& $7.6\!\cdot\!10^{-4}$	& $8.4\!\cdot\!10^{-4}$	\\
$\sigma^{2}(\Delta z_2)$ 							& 0.003 			& 0.003 			& 0.003			& 0.023 			& 0.002			& 0.003 			\\
$\sigma^{2}(\Delta z_3)$ 							& 0.005 			& 0.004 			& 0.007			& 0.039 			& 0.004 			& 0.007 			\\
$\left\langle \Delta z_{\mathrm{norm}} \right\rangle$ 		& 0.095			& 0.056			& 0.25			& 0.058		 	& 0.049  			& 0.23			\\
RMS$(\Delta _{\mathrm{norm}})$                       			& 0.19			& 0.13			& 0.32			& 0.11			& 0.09 			& 0.29			\\
$\sigma^2(\Delta z_{\mathrm{norm}})$ 				& 0.025			& 0.014			& 0.036			& 0.086			& 0.006 			& 0.03			\\
MAD$(\Delta z_{\mathrm{norm}})$ 					& 0.041			& 0.028			& 0.14			& 0.029			& 0.027			& 0.16			\\
MAD'$(\Delta z_{\mathrm{norm}})$ 				& $\mathbf{0.04}$	& $\mathbf{0.030}$   & $\mathbf{0.19}$ 	& $\mathbf{0.031}$ 	& $\mathbf{0.029}$	& $\mathbf{0.204}$	\\
$\%(\Delta z_{\mathrm{norm},1})$ 					& 77.3			& 92.2			& 17.5			& 87.4			& 91.0			& 23.3			\\
$\%(\Delta z_{\mathrm{norm},2})$ 					& 87.3			& 96.8			& 49.3			& 94.0			& 96.6			& 49.1			\\
$\%(\Delta z_{\mathrm{norm},3})$ 					& 91.8			& 97.1			& 70.3			& 96.4			& 97.8 			& 71.2 			\\
$\sigma^{2}(\Delta z_{\mathrm{norm},1})$ 			& $6.2\!\cdot\!10^{-4}$	& $5.7\!\cdot\!10^{-4}$	& $8.4\!\cdot\!10^{-4}$	& $5.6\!\cdot\!10^{-4}$	& $5.5\!\cdot\!10^{-4}$	& $8.2\!\cdot\!10^{-4}$	\\
$\sigma^{2}(\Delta z_{\mathrm{norm},2})$ 			& 0.002			& $9.8\!\cdot\!10^{-4}$	& 0.003		& 0.001			& 0.001			& 0.003			\\
$\sigma^{2}(\Delta z_{\mathrm{norm},3})$ 			& 0.004			& 0.001			& 0.006			& 0.002			& 0.002			& 0.007			\\
\hline                  
\end{tabular}
\end{table*}

\noindent The accuracy of the reconstruction of the photometric redshifts for the reliable sources ($q = 1$) increases with a factor 
from 1.2 to 2 in terms of both the variables RMS($\Delta z$) and MAD($\Delta z$) for both experiments involving the determination
of the photometric redshifts of quasars. While a significant contamination from photometric redshifts with $\|\Delta z\| > 0.1$ is still present 
in the subsets of reliable sources in both experiments ($\%(\|\Delta z\| < 0.1) = 61.9$ and 71.4 for Exp. 2 and Exp. 3 respectively), the 
fraction of accurate $z_{\mathrm{phot}}$ ($\|\Delta z\| < 0.1$) selected as unreliable ($q = 0$) at the $\|\Delta z\| = 0.1$ level is very low 
($5.9\%$ and $8.5\%$ respectively).  

\section{Conclusions}
\label{sec:conclusions}

The Weak Gated Expert or WGE is an original method for the determination of the photometric redshifts capable 
of working on both galaxies and quasars. The WGE, which is based on a combination of clustering and regression 
techniques, is able to mitigate most of the degeneracies which arise from the distribution of KB templates in the 
\emph{features} space, and to derive accurate estimates of both the photometric redshifts values and of their errors. 
Besides giving a detailed description of how the WGE works, in this paper we have also presented an application 
of the WGE to the determination of photometric redshifts of optical galaxies and to the candidate quasars with optical
and ultraviolet photometry, both extracted from the SDSS-DR7 database. The accuracy of the reconstruction of the 
redshifts for optical galaxies, obtained by comparing photometric and spectroscopic redshifts, can be expressed a robust
estimate of the dispersion of the $\Delta z$ variable, which is equal to MAD($\Delta z$) = 0.011 with $\sim86.9\%$ of the 
sources within $\Delta z_{3}$. The same diagnostics for the estimation of $z_{\mathrm{phot}}$ 
for candidate quasars are MAD($\Delta z$) = 0.11 and $\%(\Delta z_{3}) = 80.5$ when only optical
photometry is used, reaching MAD($\Delta z$) = 0.061 and $\%(\Delta z_{3}) = 91.4$ when the 
photometric redshifts are evaluated using both optical and ultraviolet photometry.
A thorough discussion and a comparison of the WGE with several other methods applied to the same or similar data 
is also provided in the paper. To perform such comparison, a large set of statistical diagnostics shows that the 
WGE performs better than or similarly to all the other methods. The results of the best experiments with the WGE for 
optical galaxies and quasars have been used to produce the catalogs of photometric redshifts of $\sim 3.2\!\cdot\!10^7$ 
galaxies photometrically selected, a sample of $\sim 2.1\!\cdot\!10^6$ optical candidate quasars from 
\cite{dabrusco2009} with photometric redshifts estimated using optical only photometry and a smaller catalog 
of more accurate photometric redshifts derived from optical and ultraviolet photometry for a subset of $\sim 1.6\!\cdot\!10^5$ 
optical candidate quasars respectively. All catalogs will be publicly 
available and a complete description of the parameters associated to each photometric redshift estimates is available
(see \ref{sec:catgal}, \ref{sec:catqua} and \ref{sec:catquauv} respectively for details on the catalogs). 
In this paper, we have also shown the results of the application of the WGE method to a relatively small sample of 
spectroscopically selected optical SDSS quasars for which also the ultraviolet (GALEX) photometry was available. 
Since the largest computational load is in the training phase, once the WGE has been trained and has has achieved 
the required accuracy (either by matching some a priori constraint or by convergence), it can be ``frozen" and newly 
acquired data falling in the same region of the \emph{features} space sampled by the KB can be processed without 
the need for a re-training of the method. This implies that, regardless the rate at which data 
are acquired, the WGE can produce estimates of photometric redshifts in real-time. If needed, a new training of the method 
can be performed off-line when a larger/improved KB becomes available. This requirement is becoming of the utmost 
importance for data mining techniques in order for them to cope with the data streams foreseen for the current and future 
optical synoptic surveys (such as 
Pan-STARRS or the LSST) that will produce overnight an amount of data (images and catalog) similar or even larger 
than the total amount of data collected by the SDSS.
\noindent It is worth stressing that the WGE is part of the larger realms of Astroinformatics and Data Mining. As a data-driven discipline, 
through the application of Data Mining methods, Astroinformatics can provide Astronomy and Astrophysics with a framework for 
tackling new problems or old problems with a novel approach: in particular, where the traditional approach uses data from observations 
in order to prove or disprove an hypothesis, with Data Mining we want data itself to provide hypotheses that can be then proved or 
disproved with more accurate follow-up observation. For example, using a catalog of photometric redshifts for galaxies of the SDSS
DR7 survey, \cite{capozzi2009} put constraints on the nature of the so called Shakbazian groups by studying the properties of such 
groups as they appeared to be in the de-projected space. 
This data-driven approach is well described also by the fact that by using machine learning methods many assumptions can be dropped 
in favor of a more agnostic approach: for instance, by employing machine learning techniques to the photometric redshift problem, one can 
drop any assumptions on the form of the SED of the source, so that it is up to the model, for example a neural network, to find a 
representation of the highly non-linear relation between the photometric information and the spectroscopic redshift, instead of 
fitting the data with a set of SED templates. However, since the hypothesis driven approach of template fitting has noticeable advantages, 
it can be useful to underline an interesting feature offered by the WGE: it is possible, through the WGE, to link together different 
\emph{experts} employed in complex architectures, in which different 
predictors can be integrated to take advantage of the peculiar strengths of each of them. Even an algorithm which does not 
belong to the domain of the DM techniques could be consistently used together with machine learning experts. In this case, 
however, the predictors not based on DM techniques will not be trained in the first step of the training algorithm and 
will only participate to the training of the gate predictor. This feature was not exploited in this work, but a likely outcome of 
such a hybrid approach will be the creation of mixed WGE architectures in which empirical machine learning algorithms cooperate 
with more traditional algorithms based on physical knowledge, for instance neural networks and SED template fitting\footnote{For a 
review of the most used template fitting methods in the literature see \cite{hildebrandt2010}).}

\noindent One interesting feature of this approach is the generalization offered by the WGE: linking together different \emph{experts} 
can lead to complex architectures in which different 
predictors can be integrated to take advantage of the peculiar strengths of each of them. Even an algorithm which does not 
belong to the domain of the DM techniques can be consistently used together with machine learning experts. In this case, 
however, the predictors not based on DM techniques will not be trained in the first step of the training algorithm and 
will only participate to the 
training of the gate predictor. A likely outcome of such hybrid approach will be the creation of mixed WGE architectures in 
which empirical machine learning algorithms cooperate with more classical algorithms based on physical consideration 
or models typical of the specific domain. For instance, for the particular problem of the estimation of photometric redshifts, 
the WGE method could be used to integrate machine learning algorithms and the empirical methods based on SED 
template fitting (for a review of the most used template fitting methods in the literature see \cite{hildebrandt2010}). 

\section*{Acknowledgments}

This paper is based on work that took advantage of several technologies the authors would like to acknowledge.
The WGE code is mostly based on the Fast Artificial Neural Network library\footnote{Available here: http://leenissen.dk/}. 
Most of the statistical code is implemented in R \footnote{The official reference for the R programming language is 
\emph{R: A Language and Environment for Statistical Computing}, published by the R Foundation for Statistical 
Computing and available at the URL: http://www.R-project.org}, while for data retrieval, analysis and publication, 
multiple tools, services and protocols developed by the International Virtual Observatory Alliance\footnote{Home page at 
the URL: www.ivoa.net} were used. In particular, all the catalogs derived from this publication will be published as standard 
Cone Search services through the VODance service hosted at the Italian center for Astronomical Archives (IA2), Trieste 
Astronomical Observatory. TOPCAT \cite{taylor2005} was used extensively in both its desktop version and its command 
line counterpart STILTS \cite{taylor2006}. The authors thank the anonymous reviewer for insightful comments that have 
helped to improve the paper. 

\appendix
\section{SQL query for SDSS galaxies}
\label{app:sqlgal}

This is an example of the SQL queries used to retrieve the galaxies in the SDSS photometric dataset whose redshifts 
have been evaluated using the results of the WGE experiment described in \ref{sec:catgal}. The queries were run 
on the DR7 SDSS database through the SDSS Catalog Archive Server Jobs System (CASJobs).\footnote{The CASJobs 
system can be reached at the URL: http://cas.sdss.org/CasJobs.}
\\
\\
{\tt SELECT\\
\\
g.objID,\\
g.ra, g.dec,\\ 
g.dered\_u, g.dered\_g ,g.dered\_r, g.dered\_i, g.dered\_z,\\
g.modelmagerr\_u, g.modelmagerr\_g, g.modelmagerr\_r, \\
g.modelmagerr\_i, g.modelmagerr\_z,\\
g.extinction\_u, g.extinction\_g, g.extinction\_r, g.extinction\_i,\\ 
g.extinction\_z,\\
g.petroR50\_u,g.petroR90\_u,\\
g.petroR50\_g,g.petroR90\_g,\\
g.petroR50\_r,g.petroR90\_r,\\
g.petroR50\_i,g.petroR90\_i,\\
g.petroR50\_z,g.petroR90\_z,\\
g.lnLDeV\_u,g.lnLDeV\_r,\\
g.lnLExp\_u,g.lnLExp\_r,\\
g.lnLStar\_u,g.lnLStar\_r \\
\\
FROM \\
\\
Galaxy AS g, Segment AS seg, Field AS f\\
\\
WHERE\\
\\
g.mode = 1 AND\\
seg.segmentID = f.segmentID AND\\
f.fieldID = g.fieldID AND\\
seg.stripe = 16 AND\\
g.dered\_r < 21.5 AND\\ 
dbo.fPhotoFlags('PEAKCENTER') != 0 AND\\
dbo.fPhotoFlags('NOTCHECKED') != 0 AND\\
dbo.fPhotoFlags('DEBLEND\_NOPEAK') != 0 AND\\
dbo.fPhotoFlags('PSF\_FLUX\_INTERP') != 0 AND\\
dbo.fPhotoFlags('BAD\_COUNTS\_ERROR') != 0 AND\\
dbo.fPhotoFlags('INTERP\_CENTER') != 0\\
} 

\section{SQL query for optical SDSS stellar sources}
\label{app:sqlqsos}

This is an example of the SQL queries used to retrieve the stellar sources in the SDSS photometric dataset from which the 
candidate quasars have been extracted with the method described in \ref{subsec:candidates} and the photometric redshifts 
have been evaluated using the results of the WGE experiment described in \ref{sec:catqua}. The queries were run 
on the DR7 SDSS database through the SDSS Catalog Archive Server Jobs System (CASJobs).\footnote{The CASJobs 
system can be reached at the URL: http://cas.sdss.org/CasJobs.}
\\
\\
{\tt SELECT\\
\\
p.objID,\\
p.ra, p.dec,\\
p.psfMag\_u, p.psfMag\_g, p.psfMag\_r, p.psfMag\_i,p.psfMag\_z,\\ 
p.psfmagerr\_u,p.psfmagerr\_g,p.psfmagerr\_r,\\
p.psfmagerr\_i,p.psfmagerr\_z,\\ 
p.extinction\_u,p.extinction\_g,p.extinction\_r,\\
p.extinction\_i, p.extinction\_z\\ 
\\
FROM\\
\\
PhotoObjAll AS p, Segment AS seg, Field AS f\\
\\
WHERE\\
\\
p.mode = 1 AND\\
p.type = 6 AND\\
seg.segmentID = f.segmentID AND\\ 
f.fieldID = p.fieldID AND\\
seg.stripe = 11 AND\\
p.psfmag\_i > 14.5 AND\\
(p.psfMag\_i - p.extinction\_i) < 21.3 AND\\
p.psfmagErr\_i < 0.2 AND\\
dbo.fPhotoFlags('PEAKCENTER') != 0 AND\\
dbo.fPhotoFlags('NOTCHECKED') != 0 AND\\
dbo.fPhotoFlags('DEBLEND\_NOPEAK') != 0 AND\\
dbo.fPhotoFlags('PSF\_FLUX\_INTERP') != 0 AND\\
dbo.fPhotoFlags('BAD\_COUNTS\_ERROR') != 0 AND\\
dbo.fPhotoFlags('INTERP\_CENTER') != 0\\
}

\section{SQL query for ultraviolet GALEX counterparts of optical candidate quasars}
\label{app:sqlqsosuv}

This is an example of the SQL queries used to retrieve the ultraviolet GALEX counterparts of the optical 
candidate quasars composing the catalog described in \ref{sec:catquauv}, whose photometric redshifts 
have been evaluated using the results of the WGE experiment described in \ref{subsec:expquauv}.
\\
\\
{\tt SELECT\\
\\
p.objid AS galex$\_$objid,\\
my.objID AS sdss$\_$objid,\\
p.nuv$\_$mag as nuv, p.nuv$\_$magErr as nuv$\_$err,\\
p.fuv$\_$mag as fuv, p.fuv$\_$magErr as fuv$\_$err,\\
p.e$\_$bv,\\
x.distance,\\ 
x.distanceRank,\\ 
x.reverseDistanceRank,\\
x.multipleMatchCount,\\
x.reverseMultipleMatchCount\\ 
\\
FROM\\
\\
MYDB.candidate$\_$quasars$\_$objid AS my\\
INNER JOIN xSDSSDR7 AS x ON my.objID = x.SDSSobjid\\
INNER JOIN photoobjall AS p ON x.objid = p.objid\\
\\
WHERE\\
\\
x.distanceRank = 1\\
AND x.reverseDistanceRank = 1\\
AND x.distance < 2\\
AND p.nuv$\_$mag > 0\\ 
AND p.fuv$\_$mag > 0\\
}

\label{lastpage}
\end{document}